\documentclass[a4paper]{article}
\usepackage{typearea}
\usepackage{graphicx}  
\usepackage{hyperref}
\usepackage{makeidx}
\usepackage{amsmath}

\usepackage{mathtools}
\usepackage{slashed}
\usepackage{amssymb}
\usepackage{amsthm}
\usepackage{mathrsfs}
\usepackage{color}
\usepackage{bm}
\usepackage{booktabs}
\usepackage{multirow}
\usepackage{multicol}
\usepackage{ytableau}
\usepackage{ulem}

\usepackage{cite} 
\usepackage{pifont}
\DeclareSymbolFont{extraup}{U}{zavm}{m}{n}
\DeclareMathSymbol{\varheart}{\mathalpha}{extraup}{86}
\DeclareMathSymbol{\vardiamond}{\mathalpha}{extraup}{87}

\usepackage{mathabx}
\providecommand{\U}[1]{\protect\rule{.1in}{.1in}}

\include{settings}
\numberwithin{equation}{section}
\begin{document}
\typearea{18}
\title{
Quantum GraviElectro Dynamics
}
\author{Yoshimasa Kurihara\footnote{yoshimasa.kurihara@kek.jp}
\\
{\it\footnotesize High Energy Accelerator Organisation (KEK), 
Tsukuba, Ibaraki 305-0801, Japan}
}
\date{}
\maketitle
\begin{abstract} 
This report presents a possible attempt at renormalisable quantum gravity based on the standard BRST quantisation used for Yang--Mills theory.
We have provided the BRST invariant Lagrangian of the gravitationally interacting $U(1)$ gauge theory, namely the \textit{Quantum GraviElectro Dynamics} (\QGED),  including the gauge fixing and the ghost Lagrangian.
We have extracted a set of Feynman rules in the local inertial frame where gravity locally vanishes from this Lagrangian.
Using the Feynman rules of the \QGED constructed here, we have built all the renormalisation constants and show that the theory is perturbatively renormalisable, at least in the one-loop order.
All infinite naked objects in the naked Lagrangian can replace experimentally measured ones.
In addition to the standard $\QED$ parameters, we have shown that the gravitational coupling constant can be experimentally measured.

We have also discussed a running effect of the gravitational coupling constant and the perturbative estimation of the Hawking radiation as examples of perturbative \QGED.
The difference between our theory and the widely known un-renormalisable quantum general relativity is clarified.
\end{abstract}
\tableofcontents
\section{Introduction}
The standard theory of particle physics, based on quantum field theory, has much experimental support for the microscopic aspect of the universe.
On the other hand, general relativity is the fundamental theory of the large-scale structure of the Universe, and many experiments have confirmed its accuracy.
Our understanding of nature thus covers a wide range of length and time scales, from the Universe's large-scale structure to the microscopic behaviour of subatomic particles.
However, two fundamental theories are inconsistent; the first is quantum theory, and the second is classical theory.
Due to the uncertainty principle, the fundamental physical theory must be a quantum theory; thus, constructing a quantum theory of gravity is one of the essential goals of modern physics.

We understand the \textit{quantum theory of gravity} as a theory that describes the behaviour of four-dimensional spacetime (the metric tensor) in regions where the uncertainty principle is essential and a theory consistent with well-established general relativity at large spacetime scales.
Moreover, a theory that can provide experimentally measurable predictions is desirable.
The establishment of general relativity and quantum mechanics in the 1920s immediately followed the development of quantum gravity in the 1930s.
For a detailed history, see Refs.\!\cite{Rovelli:2000aw,doi:10.1142qg} and references therein.
In this article, we do not mention quantum gravity other than in four-dimensional spacetime, such as superstring theory, or other than the Einstein--Hilbert Lagrangian, such as the effective action method\! \cite{GAVilkovisky_1992,buchbinder2017effective}.

It is commonly recognised that perturbative expansion of general relativity in four-dimensional spacetime is not renormalisable.
Although trials to construct a perturbatively renormalised theory of general relativity have failed\!\cite{BERENDS197599,Goroff198581,GOROFF1986709}, its non-existence has yet to be proven.
The current author discussed the quantisation of four-dimensional general relativity using the Heisenberg picture in Ref$.$\!\cite{doi:10.1140/epjp/s13360-021-01463-3}.
Recently, the gauge theory of gravity resembling the formulation of the standard theory of particle physics appears\!\cite{Partanen_2025}, which has utilised the eight-spinor representation of the Lagrangian, enabling the extraction of four-dimensional space-time quantities from the eight-dimensional spinors. 

On the other hand, the quantisation of the Yang--Mills theory in flat spacetime is well established, and the theory is proven to be renormalisable in all orders of perturbation expansions.
However, it is not trivial in curved spacetime.
A primary objective of this report is to construct renormalisable quantum gravity and the Yang--Mills theory simultaneously.

General relativity is also the Yang--Mills type gauge theory, initially investigated in a pioneering work by Utiyama\!\cite{PhysRev.101.1597}.  
(see also Ref$.$ \!\cite{blagojevi2013gauge}).
The Yang--Mills theory incorporating the gravitational force as the gauge interaction is called the Yang--Mills--Utiyama theory in Ref$.$\!\cite{Kurihara:2025tro}.
The common aspects between general relativity and the Yang-Mills theory may provide insight into renormalisable perturbation theory for general relativity.
We summarise correspondences of mathematical ingredients between them in Table-\ref{table2}.
\begin{table}[b]
\begin{center}
\caption{\label{table2}\small
Comparison between general relativity and the Yang--Mills theory at a classical level.}
\vskip 2mm
\begin{tabular}{l||llll|c}
\hspace{2em}Property & \multicolumn{2}{c}{General Relativity} & \multicolumn{2}{c|}{Electromagnetism}&dimension \\
\hline
structural group & \multicolumn{2}{c}{$SO(1,3)$}& \multicolumn{2}{c|}{$(SO(1,3)\ltimes T^4)\otimes\SU\!(\!N\!)$} &~\\
connection & spin connection:& $\omega_\mu^{~ab}$ 
& gauge potential:& $\Aa^\mu$&$L^{-1}$ \\
curvature & curvature:& $R^{ab}_{~~cd}$ 
& field strength:& $\f^{\mu\nu}$ &$L^{-2}$\\
section & vierbein:& $\E_a^\mu/\lp$ 
& Dirac spinor:& $\psi^{2/3}$&$L^{-1}$ \\
%
%
coupling constant &gravitational charge:& $\cGR$ &electric charge:&$e$&${\it 1}\hspace{1em}$
\end{tabular}
\end{center}
\end{table}
Comparing two gauge theories, it is clear that the spin connection (gauge potential) and vierbein field (section) are appropriate targets of quantisation\!\cite{Kurihara_2020} in the \QEGD.
Moreover, the gravitational coupling constant must be included in the Lagrangian and renormalised, even if it is unity after renormalisation, which has not been considered in previous studies.
Renormalizability of the Yang--Mills theory must be ensured, including gravitational vierbein and spin connection fields.
In the curved spacetime, the Yang--Mills Lagrangian also has an interaction term of Dirac spinor, spin connection and vierbein\!\cite{fre2012gravity}.
In this report, we develop quantum general relativity quantised with the U(1)-gauge theory, namely the ``Quantum GraviElectroDynamics (\QGED)''.

Owing to the two keywords, \textit{Yang--Mills theory} and \textit{gravity}, we are afraid that one may recall, so-called, the \textit{double copy} method\cite{adamo2022} inspired by the string theory.
The double copy method is represented conceptually as $(\text{gravity})=(\text{Yang--MIlls theory})^2$ and pursues to develop the method to calculate amplitudes in one theory, general relativity, using amplitudes from one or two different simpler theories, $(\text{Yang--MIlls theory})^2$, as input.
Our approach treats both gravity and the Yang--Mills gauge force on equal footing, and thus it is entirely different from the double copy method.

The \QGED does not quantise spacetime itself; thus, the spacetime coordinates $x^\mu$ are not the quantum operator but the classical continuous object\!\cite{nakanishi1990covariant,NakanishiSK2009}.
The subject for quantisation is the vierbein field (equivalently, the metric tensor $g^{(c)}_{\hspace{.3em}\mu\nu}(x)$), which is obtained as a solution of the Einstein equation.
In classical general relativity, the geometrical metric tensor $g^{(g)}_{\hspace{.3em}\mu\nu}(x)$ is given by the solution of the classical Einstein equation, such as $g^{(c)}_{\mu\nu}(x)$=$g^{(g)}_{\hspace{.3em}\mu\nu}(x)$, i.e., it is Einstein's equivalence principle.
This relation is not simply true at the quantum level, and the geometrical metric tensor is given as the expected value of the quantum metric tensor $g^{(g)}_{\hspace{.3em}\mu\nu}(x)=\langle g^{(q)}_{\hspace{.3em}\mu\nu}(x)\rangle$.

Here, we briefly mention our theory's renormalisability compared with previously failed approaches\!\cite{BERENDS197599,Goroff198581,GOROFF1986709}.
The obstacle for the renormalisable perturbation arises at a self-interaction of dimensionless fields.
In the \QGED\!, we identify the spin connection, which has an inverse length dimension, with a graviton, so renormalisation has no obstacle.
Although the \QGED also has a dimensionless field of the vierbein as a section, there are no vierbein self-couplings in the \QGED Lagrangian density.
Thus,  the theory does not have the unrenormalisable vertices.
The detailed discussion is given in \textbf{Appendix \ref{appC}}.

This report is structured as follows:
After this introductory section, \textbf{section 2} provides mathematical preliminaries describing the geometrical background of general relativity and Yang-Mills theory based on Ref$.$\!\cite{Kurihara:2022sso}.
We introduce a gravitational coupling constant in a covariant differential concerning a local $SO(1,3)$ group.
Although this method is not standard in general relativity, it is indispensable in Yang--Mills theory.
When the Yang--Mills theory treats multiple gauge symmetries, they provide a relative strength of these couplings.
\textbf{Section 3} introduces the classical (unrenormalised) \QGED Lagrangian, which consists of the pure and interaction Lagrangians for general relativity and Yang--Mills theory.
The interaction Lagrangian includes a gravitational interaction between a fermion and a gravitational gauge boson.
In addition, we need to include a gauge fixing and ghost Lagrangian to quantise this system.
This section gives all the necessary Lagrangians and shows their BRST invariance, which ensures the quantisation of gauge theory.
From the \QGED Lagrangian given in \textbf{section 3}, \textbf{section 4} extracts Feynman rules for the perturbative calculation of amplitudes.
The canonical quantisation conditions for physical fields are also given in this section.
Subsequently, \textbf{section 5} investigates the renormalisability of \QGED using the Feynman rules given in \textbf{section 4}.
Before performing concrete one-loop calculations, we check the renormalisability of the theory using the power-counting test.
We prepare all the necessary renormalisation constants and show that all ultraviolet divergences are absorbed in a finite number of renormalisation constants at the one-loop level.
Consequently, the renormalisation group equation provides an energy scale dependence of the effective gravitational coupling (a running gravitational coupling).
Section 6 presents an application of the QGED, i.e. the Hawking radiation of the Schwarzschild black hole, similar to the Schwinger effect of QED.
Finally, the Summary section summarises the main results of this report and discusses essential aspects of the \QGED.
In addition, we provide two appendices, a proof of nilpotency for all fields appearing in the Lagrangian by \textbf{Appendix \ref{app1}} and the possible method to measure the gravitational coupling constant in particle physics experiments by \textbf{Appendix \ref{appB}}.
If the manifold is torsionless, we can obtain a useful relation, the soldering relation.
\textbf{Appendix \ref{Soldering}} describes the soldering of the local and global cotangent bundles used in this study.
Finally, in \textbf{Appendix \ref{appC}} we compare the perturbative expansion coefficients of gravity in the standard method and the QEGD, and explain why the QEGD is renormalisable while the standard method is not.

The physical units used in this study are as follows:
The speed of light and the vacuum permittivity are set to unity, $c=1$ and $\varepsilon^{~}_0=1$ respectively.
Einstein's gravitational constant, $\kE=\GN/8\pi$, and the reduced Planck constant, $\hbar=h/2\pi$, are written explicitly, where $\GN$ is the Newtonian gravitational constant.
We define the fine structure constant as $\alpha=e^2/4\pi\varepsilon^{~}_0c\hbar=e^2/4\pi\hbar$, which is a dimensionless constant.
In these units, the physical dimensions of the fundamental constants are $[\hbar\hspace{.1em}\kE]_\text{pd}=L^2=T^2$ and $[\hbar/\kE]_\text{pd}=E^2=M^2$, where $L$, $T$, $E$ and $M$ are the length, time, energy and mass dimensions respectively.
An electric charge is also dimensional as $[e]=[\hbar]^\hlf=\sqrt{LE}$.
Here $[\bullet]_\text{pd}$ gives a physical dimension of the quantity $\bullet$.
The Planck mass $\Pmass$ and Planck length $\lp$ are defined as $\Pmass:=\sqrt{\hbar/\kE}$ and $\lp:=\sqrt{\hbar\kE}$ respectively.

\section{Mathematical preliminaries}
\subsection{Space-time manifold}
We introduce a four-dimensional Riemannian manifold $(\MM,\bm{g})$, where $\MM$ is a smooth and oriented four-dimensional manifold, and $\bm{g}$ is a metric tensor with a negative signature in $\MM$\!.
We refer to $\MM$ as the global manifold.
In an open  neighbourhood $U_{\hspace{-.1em}p{\hspace{.1em}\in\hspace{-.1em}}\MM}\subset\MM$\!, we introduce the standard coordinate $x^\mu$\!. 
Orthonormal bases in $T\MM_{\!p}$ and $\TsMM_{\!p}$ are denoted as $\partial_\mu$ and $dx^\mu$\!, respectively.
We use the abbreviation $\partial_\mu:=\partial/\partial x^\mu$ throughout this report.
Two trivial vector bundles $\TMM:=\bigcup_{p} \TMM_{p}$ and $\TsMM:=\bigcup_{p}\TsMM_{p}$ are referred to as  a global tangent and cotangent bundles in $\MM$\!, respectively.
We introduce an inertial system in which the Levi-Civita connection locally vanishes at any point in $\MM$\!.
An inertial frame at point $p\!\in\!\MM$ denoted as $\M_{p}$ has an $SO(1,3)$ symmetry.
Suppose a trivial bundle $\TM:=\bigcup_{p\in\M}\TM_p$ and its dual space $\TsM:=\bigcup_{p}\TsMM_{p}$ exist, namely the local tangent and cotangent bundle, respectively.

We define the vierbein $\E^a_\mu(x)\in{C^\infty(\MM)}$ as a map transferring a vector in $\TsMM$ to that in $\TsM$ such that: 
\begin{align}
\E^a_\mu(x\in\!\TsMM)dx^\mu\big|_{p\in\MM_p}:=\eee^a\big|_{p\in\!\M_p}
\in\(\Omega^1\!(\TsM){\otimes}V^1\!(\TM)\)\otimes\sss\ooo(1,3)\implies
g_{\mu\nu}=\eta^{~}_{ab}\hspace{.2em}\E^a_\mu(x)\E^b_\nu(x),\label{gYEE}
\end{align}
where $\Omega^p(\bullet)$ ($V^q(\bullet)$) is a space of a $p$-form object (a q-tensor) defined in space $\bullet$, and $\eta^{~}_{ab}$ is the metric tensor in $\TsM$ given as $\bm{\eta}=\textup{diag}(1,-1,-1,-1)$.
In a trivial bundle $U_{\hspace{-.1em}p{\hspace{.1em}\in\hspace{-.1em}}\MM}\subset\M$, we introduce the standard coordinate $\xi^a$\!. 
Roman letter suffixes are used for components in $\M$ throughout this study; on the other hand, Greek letters are used for them in $\MM$.
The vierbein is a smooth and invertible function globally defined in $\MM$.
One-form object $\eee^a$ is referred to as the vierbein one-form.
The vierbein inverse 
\begin{align*}
[\E^{-1}\(\xi(x)\)]_a^\mu=\E_a^\mu(\xi\in\TsM)\in{C^\infty(\TsM)}, 
\end{align*}
which is also called the vierbein, is an inverse transformation such that
\begin{align*}
\E^a_\mu(x)\E_a^\nu\(\xi(x)\)=\delta^\nu_\mu\in\TsMM\!.
\end{align*}
Our index convention makes it possible to distinguish two abbreviated vectors, $\partial_\mu$ and $\partial_a$, as in
\begin{align*}
\partial_\mu&:=\frac{\partial}{\partial{x^\mu}}\in V(\TMM)
~~\text{and}~~
\partial_a:=\frac{\partial~}{\partial\xi^a}:=\E^\mu_a\frac{\partial}{\partial{x^\mu}}\in V(\TM).
\end{align*}

The $GL(4)$ invariant two-, three- and four-dimensional volume forms are, respectively, defined using vierbein forms as 
\begin{subequations}
\begin{align}
\SSS^{~}_{ab}&:=\frac{1}{2}\epsilon_{ab\bcdots}\hspace{.1em}
\eee^{\bcdot}\wedge\eee^{\bcdot}&\in\Omega^2\!,\\
\VVV^{~}_{a}\hspace{.2em}&:=\frac{1}{3!}\epsilon_{a\bcdot\bcdots}\hspace{.1em}
\eee^{\bcdot}\wedge\eee^{\bcdot}\wedge\eee^{\bcdot}&\in\Omega^3\!,\\
\vvv\hspace{.5em}&:=\frac{1}{4!}\epsilon_{\bcdots\bcdots}\hspace{.1em}\eee^\bcdot\wedge\eee^\bcdot\wedge\eee^\bcdot\wedge\eee^\bcdot
=\deteps\hspace{.2em}dx^0{\wedge}dx^1{\wedge}dx^2{\wedge}dx^3&\in\Omega^4\!,\label{vvv}
\end{align}
\end{subequations}
which give the standard orthonormal bases of  two-, three- and four-forms.
We note that $\deteps=\sqrt{-\text{det}[\bm{g}]}$ owing to (\ref{gYEE}).

\subsection{Connections and curvatures}
This section introduces the connection and curvature in the principal bundle given in the previous section.
In physics, a connection provides a gauge boson field that mediates a force between matter fields, and a curvature is called a field strength and provides, for example, electromagnetic fields in $U(1)$ gauge theory.
The Lagrangian consists of the gauge covariant differential of matter and gauge fields, which ensures the gauge invariance of the theory.
The covariant differential includes a \textit{coupling constant} in physics, which provides the relative strength among the fundamental forces in nature.
On the other hand, the standard treatment of connection and curvature in mathematics does not include a coupling constant in its definitions.
We introduce the gravitational coupling constant in order to discuss the gravitational force and other forces simultaneously.
We first give the standard definitions of connection and curvature in general relativity, and then introduce the gravitational coupling constant in our theory.
\subsubsection{Standard definition}
We introduce connection one-form $\hat\www$ in the spinor bundle, namely the spin connection-form. 
An object with a hat, $\hat\bullet$, represents one by the standard definition in this section.
The $SO(1,3)$-covariant differential of the $p$-form object $\aaa^{a_1a_2\cdots a_q}\in\Omega^p(\TsMM)\otimes{V}^q(\TM)$ is defined using the spin connection-form as
\begin{subequations}
\begin{align}
d_{\hat{\www}}\aaa^{a_1\cdots a_q}&:=d\aaa^{a_1\cdots a_q}+
\hat\www^{a_1}_{\hspace{.5em}\bcdot}\wedge\aaa^{\bcdot a_2\cdots a_p}+
\hat\www^{a_2}_{\hspace{.5em}\bcdot}\wedge\aaa^{a_1\bcdot\cdots a_p}+\cdots+
\hat\www^{a_q}_{\hspace{.5em}\bcdot}\wedge\aaa^{a_1\cdots\bcdot},
\label{dwwwST}
\intertext{The spin connection-form has a component representation as }
\hat\www^a_{~b}&=
\hat\omega^{\hspace{.3em}a}_{\mu\hspace{.3em}b}\hspace{.1em}dx^\mu=
\hat\omega^{\hspace{.3em}a}_{\bcdot\hspace{.3em}b}\hspace{.1em}\E^\bcdot_\mu\hspace{.1em}dx^\mu=
\hat\omega^{\hspace{.3em}a}_{\bcdot\hspace{.3em}b}\hspace{.1em}\eee^\bcdot
\in\(\Omega^1(\TsM)\otimes V^2(\TM)\)\otimes Ad(\sss\ooo(1,3)),
\label{wwwTsM}
\end{align}
\end{subequations}
concerning the standard bases in $\TsM$.
Dummy \textit{Roman} indices are often abbreviated to a small circle $\bcdot$ (and $\star$, $\times$, $\cdots$) when the dummy index pair of the Einstein convention is obvious as above.
When multiple circles appear in an expression, the pairing must be in a left-to-right order at the subscripts and superscripts.
Raising and lowering indices are done using metric tensor $\bm\eta$ in $\M$.
Two-form object 
\begin{subequations}
\begin{align}
\hat\TTT^a&:=d_{\hat\www}\eee^a=
d\eee^a+\hat\www^{a}_{\hspace{.3em}\bcdot}\wedge\eee^\bcdot
\in\(\Omega^2(\TsMM)\otimes V^1(\TM)\)\otimes\sss\ooo(1,3)\label{torsionFMST}
\intertext{
is referred to as a torsion form.
We define a component of the torsion two-form using the standard bases as
}
\text{(\ref{torsionFMST})}&=:
\frac{1}{2}\hat\TT^a_{\hspace{.3em}\mu\nu}\hspace{.1em}dx^\mu{\wedge}dx^\nu=
\frac{1}{2}\hat\TT^a_{\hspace{.3em}\bcdots}\hspace{.1em}\eee^\bcdot\wedge\eee^\bcdot
\in\(\Omega^2(\TsM)\otimes V^1(\TM)\)\otimes\sss\ooo(1,3).
\label{torsionFMST2}
\end{align}
\end{subequations}

Local $SO(1,3)$ group action $\Gso:\TsM\rightarrow\TsM$ is known as the Lorentz transformation.
That for the vierbein form and the spin connection-form are
\begin{subequations}
\begin{align}
\Gso&:\eee\mapsto\Gso(\eee)=\eee'=\LLambda\eee,\label{GsoST1}\\
\Gso&:\hat\www\mapsto\Gso(\hat\www)=\hat\www'=
\LLambda\hat\www\LLambda^{-1}+\LLambda\hspace{.1em}d\hspace{-.1em}\LLambda^{\hspace{-.2em}-1}.\label{GsoST2}
\end{align}
\end{subequations}
The section belongs to the fundamental representation, and the connection does to adjoint one. 
A spacetime curvature (or simply \textit{curvature}) is defined owing to the structure equation as
\begin{align}
\hat\RRR^{ab}&:=d\hat\www^{ab}+\hat\www^a_{~\bcdot}\wedge\hat\www^{\bcdot b}
\in\(\Omega^2(\TsMM)\otimes V^2(\TM)\)\otimes Ad(\sss\ooo(1,3)),\label{RRRST}
\end{align}
which is a two-form valued rank-$2$ tensor  represented using the trivial basis as
\begin{align*}
\hat\RRR^{ab}&=
\sum_{\mu<\nu}\hat{\Ri}^{ab}_{\hspace{.7em}\mu\nu}\hspace{.1em}dx^\mu\wedge dx^\nu =
\frac{1}{2}\hat{\Ri}^{ab}_{\hspace{.7em}\mu\nu}\hspace{.1em}dx^\mu\wedge dx^\nu=
\frac{1}{2}\hat{\Ri}^{ab}_{\hspace{.7em}\bcdots}\hspace{.1em}\eee^\bcdot\wedge\eee^\bcdot.
\end{align*}
Tensor coefficient $\hat{R}^{ab}_{\hspace{.7em}cd}$ is referred to as the Riemann-curvature tensor.
Ricci-curvature tensor and scalar curvature are defined, respectively, owing to the Riemann-curvature tensor as
\begin{align*}
\hat{R}^{ab}:=\hat\Ri^{\bcdot a}_{\hspace{.7em}\mu\nu}\hspace{.1em}
\E^\mu_\bcdot\E^\nu_\star\eta^{\star b}~~\textrm{and }~~
\hat{R}:=\hat\Ri^{\bcdot\star}_{\hspace{.7em}\mu\nu}\hspace{.1em}
\E^\mu_\bcdot\E^\nu_\star.
\end{align*}
The first and second Bianchi identities are
\begin{align}
d_{\hat\www}\hat\TTT^a&=d_{\hat\www}(d_{\hat\www}\eee^a)=
\eta_{\bcdots}\hat\RRR^{a\bcdot}\wedge\eee^{\bcdot}
~~\text{and}~~
d_{\hat\www}\hat\RRR^{ab}=0.\label{BianchiRST}
\end{align}
We define the spin connection-form, the curvature form and their coefficients intrinsically in $\TsM$.

\subsubsection{Coupling constant}\label{CC}
We redefine the connection and the curvature with the gravitational coupling constant $\cGR$.
Objects without a hat represent one defined with the gravitational coupling constant corresponding to those with a hat.
We redefine the covariant differential (\ref{dwwwST}) utilising the scaled spin connection,
\begin{align}
{\cGR}\hspace{.1em}\www^{ab}:=\hat\www^{ab}\implies
{\cGR}\hspace{.1em}\omega_c^{\hspace{.3em}ab}=\hat\omega_c^{\hspace{.3em}ab},\label{cgrwww}
\end{align}
as
\begin{align}
d_{{\www}}\aaa^{a_1\cdots a_q}&=d\aaa^{a_1\cdots a_q}+\cGR\(
\www^{a_1}_{\hspace{.5em}\bcdot}\wedge\aaa^{\bcdot\cdots a_p}+\cdots\)\!.
\tag*{(\ref{dwwwST})\texttt{'}}\label{dwww}
\end{align}
Accordingly, the scaled torsion form is provided as
\begin{align}
\TTT^a&=d_\www\eee^a=d\eee^a+\cG\hspace{.1em}\www^a_{~\bcdot}\wedge\eee^\bcdot\!=\hat\TTT^a\implies
\TT^a_{\hspace{.3em}bc}=\hat\TT^a_{\hspace{.3em}bc}.
\tag*{(\ref{torsionFMST})\texttt{'}}\label{torsionFM}
\end{align}
The structure equation owing to the scaled spin connection provides the scaled curvature as
\begin{align}
\RRR^{ab}&:=d\www^{ab}+\cG\www^a_{~\bcdot}\wedge\www^{\bcdot b}\!,
\tag*{(\ref{RRRST})\texttt{'}}\label{RRR}
~~
\text{yielding}
~~
\cGR\hspace{.1em}\RRR^{ab}=\hat\RRR^{ab}\implies
\cGR\hspace{.1em}\Ri^{ab}_{\hspace{.7em}cd}=\hat\Ri^{ab}_{\hspace{.7em}cd}.
\end{align}
The Lorentz transformation for the scaled spin connection scales as
\begin{align}
\text{(\ref{GsoST2})}&\implies
\Gso:\www\mapsto\Gso(\www)=\www'=
\LLambda\www\LLambda^{-1}+{\cG}^{\hspace{-.5em}-1}\LLambda\hspace{.1em}d\hspace{-.1em}\LLambda^{\hspace{-.2em}-1},
\tag*{(\ref{GsoST2})\texttt{'}}\label{Gso2}
\intertext{
and the first Bianchi identity as
}
\text{(\ref{BianchiRST})}&\implies
d_\www\TTT^a=d_\www(d_\www\eee^a)=\cG\hspace{.2em}
\eta_{\bcdots}\RRR^{a\bcdot}\wedge\eee^{\bcdot}
\tag*{(\ref{BianchiRST})\texttt{'}}~~\text{and}~~
d_{\www}\RRR^{ab}=0.\label{BianchiT}
\end{align}

As seen in (\ref{RRRST}) and \ref{RRR}, a simultaneous re-scaling of the connection and curvature eliminates the coupling constant in the system.
When treating only the gravitational force, the gravitational coupling constant has no physical meaning.
However, when considering the Yang--Mills gauge forces in curved spacetime, coupling constants provide a relative strength among forces, including the gravitational one.

%
%
\subsubsection{co-Poincar\'{e} bundle}
The Yang--Mills theory in the Lorentzian metric space has the Poincar\'{e} symmetry in addition to the gauge symmetry.
On the other hand, general relativity is not invariant under the four-dimensional translation.
The current author has introduced a modified Poincar\'{e} symmetry, keeping general relativity invariant, namely the co-Poincar\'{e} symmetry\cite{doi:10.1063/1.4990708,Kurihara_2020}.
This section introduces a principal bundle having the co-Poincar\'{e} group as the structure group.

The Poincar\'{e} group is the semidirect product of Lorentz group actions and the four-dimensional translation group as
\begin{align*}
I\hspace{-.15em}S\hspace{-.1em}O(1,3)=SO(1,3)\ltimes T^4,
\end{align*}
whose Lie algebra is given as 
\begin{align}
\left[P_{a},P_{b}\right]&=0,\notag\\
\left[J_{ab},P_{c}\right]&=-\eta_{ac}\hspace{.1em}P_{b}+\eta_{bc}\hspace{.1em}P_{a},\label{JJ}\\
\left[J_{ab},J_{cd}\right]&=
-\eta_{ac}\hspace{.1em}J_{bd}+\eta_{bc}\hspace{.1em}J_{ad}
-\eta_{bd}\hspace{.1em}J_{ac}+\eta_{ad}\hspace{.1em}J_{bc},\notag
\end{align}
where $P_a$ and $J_{ab}$ are generators of the $T^4$ group and the $\SO(1,3)$ group, respectively.
Although the Yang--Mills Lagrangian is invariant under the Poincar\'{e} group, the Einstein--Hilbert gravitational Lagrangian does not respect the symmetry\cite{0264-9381-29-13-133001}, obstructing quantisation of general relativity.
The current author has discovered a novel symmetry named the co-Poincar\'{e} symmetry\cite{doi:10.1063/1.4990708} which allows the construction of general relativity in four-dimensional space-time and defined the invariant quadratic in four-dimensional space-time owing to the co-Poincar\'{e} group\cite{doi:10.1140/epjp/s13360-021-01463-3}.

The co-Poincar\'{e} group, denoted as $\GcP$, is the Poincar\'{e} group that replaced the translation operator with the co-translation operator.
A principal co-Poincar\'{e} bundle is defined as tuple 
\begin{align}
\left(\MM\otimes\M,\pi^{~}_{\textsc{i}},\MM,\GcP\right),
\end{align}
which is the same as the inertial bundle except the structure group $\GcP$.
A generator of the co-translation is defined as $P_{ab}:=P_a\iota_b$, where $\iota_b$ is a contraction concerning trivial frame field $\partial_b$.
The generator of the co-Poincar\'{e} group can be represented by two ($\FxF$)-matrices as
\begin{subequations}
\begin{align}
\left[\Theta_I\right]_{ab}:=
\begin{cases}
P_{ab},&I=1\\
J_{ab},&I=2
\end{cases}\!,
\end{align}
whose the Lie algebra are provided as
\begin{align}
\left[P_{ab},P_{cd}\right]&=0,~~~
\left[J_{ab},P_{cd}\right]=-\eta_{ac}\hspace{.1em}P_{bd}+\eta_{bc}\hspace{.1em}P_{ad}
~~\text{and}~~
 \left[J_{ab},J_{cd}\right]=\text{(\ref{JJ})}.\label{coPoincare1}
\end{align}
\end{subequations}
We denote the structure constants of the co-Poincar\'{e} group obtained from the above Lie algebra as
\begin{subequations}
\begin{align}
&[\Theta_I,\Theta_J]:=\FF_{~IJ}^{K}\hspace{.1em}\Theta_K,\label{coPoincare2}
\intertext{yielding}
&\left\{
\begin{array}{l}
\FF_{~11}^{1}=\FF_{~11}^{2}=\FF_{~12}^{2}=
\FF_{~21}^{2}=\FF_{~22}^{1}=0,\\
\left[\FF_{~12}^{1}\right]_{ab;cd}^{ef}=
-\left[\FF_{~21}^{1}\right]_{cd;ab}^{ef}=
\eta_{ac}\hspace{.1em}\delta^e_b\delta^f_d-\eta_{bc}\hspace{.1em}\delta^e_a\delta^f_d,\\
\left[\FF_{~22}^{2}\right]_{ab;cd}^{ef}=
-\eta_{ac}\hspace{.1em}\delta_b^e\delta_d^f+\eta_{bc}\hspace{.1em}\delta_a^e\delta_d^f
-\eta_{bd}\hspace{.1em}\delta_a^e\delta_c^f+\eta_{ad}\hspace{.1em}\delta_b^e\delta_c^f.
\end{array}
\right.\label{FgrLie}
\end{align}
\end{subequations}

Unscaled Co-Poincar\'{e} connection form $\hat\AAA_\cP$ and curvature form $\hat\FFF_\cP$ are, respectively, introduced as Lie-algebra valued one- and two-form objects concerning the co-Poincar\'{e} group, and they are expressed using the trivial basis as
\begin{subequations}
\begin{align}
\hat\AAA^{I}_\cP&=
\begin{cases}
J_\bcdots\otimes\hat\www^\bcdots,&\hspace{1.2em}I=1,\\
P_\bcdots\otimes\SSS^\bcdots,&\hspace{1.2em}I=2,
\end{cases}
\hspace{2em}\in\Omega^1(\TsM)\otimes Ad(\ggg_\cP),\label{cPLieA}\\
\hat\FFF^{I}_\cP&=
\begin{cases}
J_\bcdots\otimes\hat\RRR^\bcdots,&I=1,\\
P_\bcdots\otimes d_{\hat\www}\SSS^\bcdots,&I=2,
\end{cases}
\hspace{2em}\in\Omega^2(\TsM)\otimes Ad(\ggg_\cP),\label{cPLieA2}
\end{align}
where $\ggg_\cP$ is a Lie algebra of the co-Poincar\'{e} symmetry.
\end{subequations}
Whereas $d_{\hat\www}\SSS_\bcdots$ is the three-form object, $P^\bcdots\otimes d_{\hat\www}\SSS_\bcdots\in\Omega^2(\TsM)$ is the two-form object due to $\iota_\bullet:\Omega^p\rightarrow\Omega^{p-1}$ yielding $\iota_a\eee^b=\delta^b_a$.

\subsection{Spinor-gauge bundle}
This section introduce the spinor-gauge bundle in which the \YMU theory is constructed.
Details of the spinor-gauge bundle are given in Ref$.$\!\cite{Kurihara:2025tro}.
\paragraph{spinor bundle:}
We introduce a spin structure into inertial manifold $\M$ owing to a spinor group $Spin(1,3)$.
The principal spinor bundle is a tuple such as
\begin{align*}
\(\M{\otimes}\VV^{~}_{\hspace{-.2em}\hlf\otimes\overline\hlf},\pi_\Sp,\M, G^{~}_\cP\hspace{.1em}{\otimes}Spin(1,3)\)\!,
\end{align*}
where $\VV^{~}_{\hspace{-.2em}\hlf\otimes\overline\hlf}$ is a space of the Dirac spinor.  
A total space is the trivially combined Dirac spinor space with the inertial manifold.
Projection map $\pi_\Sp$ is provided owing to the covering map,
\begin{align*}
&\tau_\cov:Spin(1,3)\rightarrow SO(1,3)\otimes\{R,L\},
\intertext{yeilding}
&\pi_\Sp:=\tau_\cov/\{R,L\}:
\M{\otimes}\VV^{~}_{\hspace{-.2em}\hlf\otimes\overline\hlf}
\rightarrow\M/\{R,L\}:
\left.{\psi}\right|_p\hspace{.1em}\mapsto p\in\M,
\end{align*}
where $\psi$ is the Dirac spinor.
The structure group of the inertial bundle, $SO(1,3)$, is lifted to $Spin(1,3)$ owing to projection map $\pi_\Sp$.

\paragraph{gauge bundle:}
Although we quantise only $U(1)$ gauge theory with gravity in this report, we discuss a general $SU\!(\!N\!)$ gauge bundle here for completeness.
A principal $SU\!(\!N\!)$-gauge bundle is defined as a tuple such that:
\begin{align*}
\(\M{\otimes}V(\M),\pi^{~}_{\SU},\M,G^{~}_\cP\hspace{.1em}{\otimes}SU\!(\!N\!)\)\!.
\end{align*}
We introduce the gauge group in the inertial bundle and treat two fundamental fields, the gauge connection $\AAA^{~}_{SU}$ and the Higgs field $\bm{\phi}{\hspace{.1em}\in}V(\M)$.

A space of the $SU\!(\!N\!)$ Higgs is introduced as a section $\bm{\phi}{\hspace{.1em}\in\hspace{.1em}}\Gamma\left(\M,V,{SU\!(\!N\!)}\right)$ belonging to the fundamental representation of the $SU\!(\!N\!)$ symmetry.
The $SU\!(\!N\!)$ group operator $\Gsu$  acts on the section as
\begin{align*}
\Gsu:End(\Gamma^{~}_{\hspace{-.1em}\textsc{w}}):
\bm{\phi}\mapsto\Gsu\hspace{-.2em}\left(\bm{\phi}\right):=
\bm{g}_\SU^{~}\hspace{.1em}\bm{\phi},
\end{align*}
yielding the component representation as
\begin{align*}
\left[\Gsu\left(\bm{\phi}\right)\right]_I=
\left[\bm{g}_\SU^{~}\right]_{I}^{\hspace{.2em}J}\left[\bm{\phi}\right]_J^{~},
\end{align*}
where $\bm{g}_{SU}^{~}$ is a ($N\!\times\!N$)-unitary matrix.
The Higgs boson is the $SU\!(\!N\!)$-spinor and is $N$-independent scalar functions concerning the local Lorentz group.

A connection in the gauge bundle $\AAA^{~}_{\SU}:=\AAA_\SU^I\hspace{.2em}\tau^{~}_I\in\Omega^1(\TsM)\otimes\sss\uuu(\!N\!)$ is an $SU\!(\!N\!)$ Lie-algebra valued one-form object.
We define covariant differential $d^{~}_\SUO$ on $p$-form object $\aaa\in\Omega^p(\TsM)$ concerning $SU\!(\!N\!)$ and $SO(3,1)$ as
\begin{align*}
d^{~}_\SUO\hspace{.2em}\aaa\hspace{.1em}&:=
\bm{1}_\SU\hspace{.1em}d_\www\aaa-i\hspace{.1em}\cSU\hspace{.1em}
[\AAA^{~}_\SU,\aaa]_\wedge,
\intertext{where}
[\AAA^{~}_\SU,\aaa]_\wedge&:=\AAA^{~}_\SU\wedge\aaa-(-1)^{p}\hspace{.1em}\aaa\wedge\AAA^{~}_\SU,
\end{align*}
and $\cSU$ is a dimensionless coupling constant of the gauge interaction. 
Connection $\AAA^{~}_\SU$ belongs to an adjoint representation of the gauge group as
\begin{align*}
\Gsu\hspace{-.2em}\left(\AAA^{~}_\SU\right)&=
\bm{g}^{-1}_\SU\hspace{.2em}\AAA^{~}_\SU\hspace{.2em}\bm{g}_\SU^{~}
+i\hspace{.1em}{c^{-1}_\SU}\gdg,
\end{align*}
which ensures $SU\!(\!N\!)$ covariance of the covariant differential.
At the same time, $\AAA^{~}_\SU$ is a vector in $\TsM$ with the local  $SO(1,3)$-group.
Corresponding gauge curvature two-form $\FFF_\SU$ is defined through a structure equation such that:
\begin{align*}
\FFF_\SU=\FFF_\SU^I\hspace{.2em}\tau^{~}_I&:=d_\www\AAA^{~}_\SU
-i\hspace{.1em}{{\cSU}}\hspace{.1em}\AAA^{~}_\SU\wedge\AAA^{~}_\SU
=\left(d\AAA_\SU^I+\cG\hspace{.1em}\www\wedge\AAA_\SU^I
+\frac{{\cSU}}{2}f^I_{~JK}\hspace{.1em}\AAA_\SU^J\wedge\AAA_\SU^K
\right)\tau^{~}_I,
\end{align*}
where $f^I_{~JK}=f^{~}_{IJK}$.
We utilise the scaled spin connection; thus, the gravitational coupling constant is appearing in the Lagrangian, which provides the relative strength between gravitational and $SU\!(\!N\!)$ gauge forces.

The gauge connection and gauge curvature are represented using the trivial bases in $\TsM$, respectively, as
\begin{align*}
\AAA^{~}_\SU&=\AAA_\SU^I\hspace{.2em}\tau^{~}_I=:\Aa^I_{a}\eee^a\hspace{.2em}\tau^{~}_I,~~\textrm{and}~~
\FFF_\SU=\FFF_\SU^I\hspace{.2em}\tau^{~}_I=:\frac{1}{2}\f^I_{ab}\hspace{.1em}\eee^a\hspace{-.1em}\wedge\eee^b\hspace{.2em}\tau^{~}_I.
\end{align*}
In this expression, tensor coefficients of the gauge curvature is provided using those of the gauge connection such that: 
\begin{align*}
\f^I_{ab}&=\partial_a\Aa^I_b-\partial_b\Aa^I_a+
{{\cSU}}\hspace{.1em}f^I_{~JK}\Aa^J_a\Aa^K_b+\Aa^I_\bcdot\TT^\bcdot_{~ab},
\end{align*}
where $\TT$ is a tensor coefficient of the torsion form defined as  $\TTT^\bullet=:\TT^\bullet_{\hspace{.5em}\bcdots}\eee^\bcdot\wedge\eee^\bcdot/2$.
When the spacetime manifold is torsionless, the gauge curvature has the same representation as it in the flat spacetime.

The $Spin(1,3)$ covariant differential acts on the spinor as\!\cite{fre2012gravity, moore1996lectures,Nakajima:2015rhw}
\begin{align*}
d_\www:=d-i\frac{\cG}{2}\hspace{.1em}\eta_\bcdots\eta_\stars\www^{\bcdot\star}\frac{\sigma^{\bcdot\star}}{2},
\end{align*}
where $\sigma^{ab}/2:=i[\gamma^a,\gamma^b]/4$ is a generator of $Spin(1,3)$ and $\psi$ is a four-component Dirac spinor.
We define the Dirac operator as
\begin{align*}
\dSp\psi&:=\iota_{\gamma}d_\www:
\Gamma\left(\M,\Omega^0(\TsM)\right)
\overset{d_\www}{\longrightarrow}
\Gamma\left(\M,\Omega^1(\TsM)\right)
\overset{\iota_\gamma}{\longrightarrow}
\Gamma\left(\M,\Omega^0(\TsM)\right)\!,
\intertext{
which has an expression owing to the trivial basis as
}
\dSp\psi&=\left(\gamma^\bcdot\partial_\bcdot
-i\frac{\cG}{4}\gamma^\bcdot\omega_\bcdot^{~\stars}{\sigma_\stars}
\right)\psi,
\end{align*}
where $\iota_{\gamma}$ is a contraction with the Clifford algebra $\gamma^a$\!.

The connection and curvature of the spinor-gauge bundle are provided, respectively, as
\begin{align*}
\AAA^{ab}_\SG=\www^{ab}\otimes\bm{1}_\SU+[\bm{1}_\Sp]^{ab}\otimes\AAA^{~}_\SU,~~\textrm{and}~~
\FFF^{ab}_\SG=\RRR^{ab}_\Sp\otimes\bm{1}_\SU+[\bm{1}_\Sp]^{ab}\otimes\FFF_\SU,
\end{align*}
We introduce a pair of Dirac spinors such as $\bm{\psi}=(\psi^\nu,\psi^e)^T$.
The Dirac operator on $\bm{\psi}$ concerning the spinor-gauge bundle is provided as
\begin{align*}
\ds_{\SG}\bm{\psi}:=\(
\gamma^\bcdot\partial_\bcdot
-i\frac{\cG}{4}\gamma^\bcdot\omega_\bcdot^{~\stars}{\sigma_\stars}
-i{\cSU}\gamma^\bcdot\Aa^I_\bcdot \tau^{~}_I\)\bm{\psi}.
\end{align*}

\paragraph{spinor-gauge bundle:}
We construct the \YMU theory in the spinor-gauge bundle, which is a Whitney sum of spinor and gauge bundles given as
\begin{align*}
\left(\M{\otimes}\(
\VV^{\textsc{l}}_{\hlf}\oplus\VV^{\textsc{l}}_{\overline\hlf}
\){\otimes}\VV^{\textsc{e}}_{\hlf},\pi^{~}_{\Sp}\oplus\pi^{~}_{\SU},\M, G^{~}_\cP{\otimes}Spin(1,3)\otimes{\SU(2)}\right).
\end{align*}
The total space of the gauge bundle is lifted to the spin manifold owing to the bundle map $\pi^{~}_{\Sp}$.
A connection and a curvature are provided, respectively, as
\begin{align}
\AAA_\SG&=\AAA_\SU\otimes\bm{1}_\Sp+\bm{1}_\SU^{~}\otimes\AAA_\Sp,\quad\textrm{and}\quad
\FFF_\SG=\FFF_\SU^{~}\otimes\bm{1}_\Sp+\bm{1}_\SU^{~}\otimes\FFF_\Sp,\label{SGcc}
\end{align}
yielding the spinor-gauge covariant differential as
\begin{align}
d_{\SG}^{~}:=\(\bm{1}_\SU^{~}\otimes\bm{1}_\Sp\)d
-i\hspace{.1em}{c^{~}_\SU}({\AAA}_{\SU}\otimes\bm{1}_\Sp)
-i\hspace{.1em}\cG(\bm{1}_\SU^{~}\otimes{{\AAA}}_\Sp),\label{codiffSp1}
\end{align}

\section{Yang--Mills--Utiyama Lagrangian}
\subsection{Bare Lagrangian}\label{BLag}
We introduce the Lagrangian four-form in the spinor-gauge bundle: $\LLL^\bare:=\LL^\bare(\xi\in\M)\hspace{.1em}\vvv\in\Omega^4(\TsM)$, where $\LL^\bare(\xi)\in\Omega^0(\TsM)$ is a Lagrangian density.
Superscript ``$\bare$'' indicates that a given object is a \textit{bare} object before renormalisation.
We define the Lagrangian form as a null physical dimensional object; thus, the Lagrangian density has a $L^{-4}$ dimension.
A bear (unrenormalised) Lagrangian form of the the Yang--Mills--Utiyama theory consists of seven parts as follows:\begin{align*}
\LL_\text{\YMU}^\bare\hspace{.3em}:=&
 \LL_\GR^\bare+\LL_{\GR;\gfix}^\bare
+\LL_{\GR;gh}^\bare\hspace{.4em}
+\LL_{\SU}^\bare+\LL_{\SU;\gfix}^\bare
+\LL_{\SU;gh}^\bare\hspace{.4em}
+\LL_{\MT}^\bare,
\end{align*}
where
\begin{itemize}
\item Gravitational part:
\begin{itemize}
\item $\LL_\GR^\bare$ : a pure gravitational term, 
\item $\LL_{\GR;\gfix}^\bare$ : a $G^{~}_\cP$ gauge-fixing term,
\item $\LL_{\GR;gh}^\bare$ : a Fadeev--Poppov ghost term for the $G^{~}_\cP$ gauge group,
\end{itemize}
\item $SU\!(\!N\!)$ gauge part:
\begin{itemize}
\item  $\LL_{\SU}^\bare$: an $SU\!(\!N\!)$ gauge boson term,
\item $\LL_{\SU;\gfix}^\bare$ : an $SU\!(\!N\!)$ gauge-fixing term,
\item $\LL_{\SU;gh}^\bare$ : a Fadeev--Poppov ghost term for the $SU\!(\!N\!)$ gauge group,
\end{itemize}
\item Matter part:
\begin{itemize}
\item $\LL_{\MT}^\bare$ : a matter (fermion) field term.
\end{itemize}
\end{itemize}
The spin connection appears in $\LL_\GR^\bare$ through the structure equation and in $\LL_{\MT}^\bare$ through the local $SO(1,3)$-covariant differential.
We utilise the unscaled spin connection $\hat\omega$ in the \YMU Lagrangian.
On the other hand, we assume that the observable spin connection is the scaled one; thus, the scaled spin connection appears in the Feynman rule.

The a pure gravitational Lagrangian is given as\!\cite{Kurihara:2025tro}
\begin{align}
\LLL_\GR^\bare&:=\frac{1}{\hbar\kE}\Tr^{~}_{\cP}\hspace{-.2em}
\left[\|\FFF^{~}_\cP\|^2\right]\vvv=
\frac{1}{2\kE\hbar}\hat\RRR^{\bare\hspace{.1em}\bcdots}\!\wedge\SSS^\bare_{\hspace{.3em}\bcdots}=
\frac{1}{4{\kE\hbar}}\epsilon_{\bcdots\bcdots}
\hat\RRR^{\bare\hspace{.1em}\bcdots}\!\wedge\eee^{\bare\hspace{.1em}\bcdot}\wedge\eee^{\bare\hspace{.1em}\bcdot}\!,\label{Lgr}
\end{align}
which is nothing other than the Einstein--Hilbert Lagrangian.
We obtain the component representation of the gravitational Lagrangian in $\TsM$ concerning the scaled spin connection field as  
\begin{align}
\LLL_\GR^\bare&=
\frac{1}{4\kE\hbar}\epsilon_{\bcdots\bcdots}
\(\partial^{~}_{\hspace{-.1em}a}\hspace{.1em}\hat\omega_{\hspace{.4em}b}^{\bare\hspace{.2em}\bcdots}+
\hat\omega_{\hspace{.4em}a\hspace{.3em}\star}^{\bare\hspace{.2em}\bcdot}\hspace{.2em}
\hat\omega_{\hspace{.4em}b}^{\bare\hspace{.2em}\star\bcdot}\)
\eee^{\bare\hspace{.1em}a}\wedge\eee^{\bare\hspace{.1em}b}\wedge
\eee^{\bare\hspace{.1em}\bcdot}\wedge\eee^{\bare\hspace{.1em}\bcdot}\!.\label{Lgr2}
\end{align}
Other Lagrangian forms are provided using the standard bases as\footnote{See, e.g., Section 5.3 in Ref$.$\!\cite{fre2012gravity}}:
\begin{align}
\LLL_{\MT}^\bare&:=\bar{\psi}^\bare\(i\hspace{.1em}\gamma^\bcdot\partial_\bcdot
-\frac{1}{\hbar}{m^\bare}
-i\frac{1}{4}\gamma^\bcdot
\hat\omega_{\hspace{.4em}\bcdot}^{\bare\hspace{.1em}\stars}{\sigma_\stars}
+\cSUz\hspace{.2em}\gamma^\bcdot\Aa^{\bare I}_{\hspace{.7em}\bcdot}\hspace{.1em}\tau_I
\)\psi^\bare\vvv,\label{LMT}\\
\LLL_\SU^\bare&:=\Tr^{~}_{\SU}\hspace{-.2em}
\left[\FFF^{\bare}_\SU\wedge\hat{\FFF}^{\bare}_\SU\right]=
\frac{1}{4}
\sum_{I=1}^N
\f^{\bare I}_{\hspace{-.3em}\SU\hspace{.4em}\bcdots}\hspace{.2em}
\f^{\bare\hspace{.1em}I\bcdots}_{\hspace{-.3em}\SU} 
\hspace{.1em}\vvv,
\end{align}
where $\hat{\FFF}$ is the Hodge-dual of $\FFF$.
We omit subscript $\SU$ for simplicity, hereafter.
We note that the Lagrangian (\ref{LMT}) is a short-handed representation since the spin connection couples with the Dirac spinor but the $\SU(N)$ gauge connection couples with the $\SU(N)$ multiplet of the Dirac spinors. 
The exact form can be found in Ref$.$\!\cite{Kurihara:2025tro}.

Lagrangian densities are defined in $\TsM$.
We decompose the bare \YMU Lagrangian into free and interaction parts according to the standard perturbative quantisation method such as
\begin{subequations}
\begin{align}
&\left\{
\begin{array}{cl}
\LL_{\GR;free}^\bare&:=
\(\partial^{~}_{\hspace{-.1em}a}\hspace{.1em}\hat\omega_{\hspace{.4em}b}^{\bare\hspace{.1em}ab}-
\partial^{~}_{\hspace{-.1em}b}\hspace{.1em}\hat\omega_{\hspace{.4em}a}^{{\bare}\hspace{.1em}ab}\)/2\lp^2,\\
\LL_{\GR;int}^\bare&:=\eta_{\bcdots}
\(\hat\omega_{\hspace{.3em}a}^{{\bare}\hspace{.1em}a\bcdot}\hspace{.1em}\hat\omega_{\hspace{.3em}b}^{{\bare}\hspace{.1em}\bcdot b}-
\hat\omega_{\hspace{.3em}b}^{{\bare}\hspace{.1em}a\bcdot}\hspace{.1em}\hat\omega_{\hspace{.3em}a}^{{\bare}\hspace{.1em}\bcdot b}\)/2\lp^2,
\end{array}
\right.\label{LGrfreeint}\\
&\left\{
\begin{array}{cl}
\LL_{\MT;free}^\bare&:=
\bar{\psi}^\bare\(i\hspace{.1em}\gamma^\bcdot\partial^{~}_{\!\bcdot}
-\frac{1}{\hbar}{m_e^\bare}\)\psi^\bare,\\
\LL_{\MT;int}^\bare&:=
\bar{\psi}^\bare\(
-i\hspace{.1em}\gamma^\bcdot\hat\omega_\bcdot^{~\stars}\sigma_\stars/4
+\cSUz\gamma^\bcdot\Aa^{\bare I}_{\hspace{.7em}\bcdot}\hspace{.1em}\tau_I\)\psi^\bare,
\end{array}
\right.\label{LMTfreeint}\\
&\left\{
\begin{array}{cl}
\LL_{\SU;free}^\bare&:=
\sum_{I=1}^N\eta^\bcdots\eta^\stars
\(\partial^{~}_{\!\bcdot}\Aa^{\bare I}_{\hspace{.7em}\star}-\partial^{~}_\star\Aa^{\bare I}_{\hspace{.7em}\bcdot}\)
\(\partial^{~}_{\!\bcdot}\Aa^{\bare I}_{\hspace{.7em}\star}-\partial^{~}_\star\Aa^{\bare I}_{\hspace{.7em}\bcdot}\)
/4,\\
\LL_{\SU;int}^\bare&:=\cSUz\sum_{I=1}^N\eta^\bcdots\eta^\stars
\(\partial^{~}_{\!\bcdot}\Aa^{\bare I}_{\hspace{.7em}\star}-\partial^{~}_\star\Aa^{\bare I}_{\hspace{.7em}\bcdot}\)
\(f^I_{~JK}\Aa^{\bare J}_{\hspace{.7em}\bcdot}\Aa^{\bare k}_{\hspace{.7em}\star}\)/2\\
~&+\cSUz^2\sum_{I=1}^N\eta^\bcdots\eta^\stars
\(f^I_{~JK}\Aa^{\bare J}_{\hspace{.7em}\bcdot}\Aa^{\bare K}_{\hspace{.7em}\star}\)
\(f^I_{~LM}\Aa^{\bare L}_{\hspace{.7em}\bcdot}\Aa^{\bare M}_{\hspace{.7em}\star}\)/4.
\end{array}
\right.\label{LSUfreeint}
\end{align}
\end{subequations}
where $\psi^\bare$ is an $N$-spinors $\psi^\bare:=(\psi^\bare_i,\cdots,\psi^\bare_N)$. 
Trivial $\bm{1}_\SU$ is omitted in formulae.
Later after providing ghost Lagrangian, we decompose it into free and interaction parts, too.
We utilised identity
\begin{align*}
\epsilon_{\bcdots ab}\hspace{.2em}\eee^\bcdot\wedge\eee^\bcdot\wedge\eee^{c}\wedge\eee^{d}
=2\(\delta_a^{c}\hspace{.1em}\delta_b^{d}-\delta_a^{d}\hspace{.1em}\delta_b^{c}\)\vvv
\end{align*}
in (\ref{LGrfreeint}).

%
%
\subsection{BRST transformation}\label{BRST}
The standard procedure of canonical quantisation based on the BRST symmetry\!\cite{Becchi:1974md,Tyutin:1975qk} is well established.
In this study, we utilise a quantisation method proposed by Nakanishi\!\cite{Nakanishi01061966}, Kugo and Ojima\!\cite{kugo1979local,Kugo1978459}.
The Nakanishi--Kugo--Ojima method introduces the auxiliary field and the Fadeev--Popov ghost field and sets the BRST transformation on all fields that appear in theory.
This section provides a set of the BRST transformations for the Yang--Mills--Utiyama theory.

The BRST transformations for unrenormalised fields are common to those for renormalised fields; thus, we omit the suffix ``$\hspace{.1em}\bare$'' in the following.

\subsubsection{Yang--Mills theory}
In the standard quantum field theory textbooks, we can find the detailed procedure of quantising the Yang--Mills theory based on the BRST symmetry, e.g., Ref$.$\!\cite{peskin1995introduction,weinberg1996quantum}.
Even though that, this section summarises the BRST transformation for the Yang--Mils theory in the inertial manifold and provides a guide to constructing the BRST transformation of the gravitational fields.

We introduce classical auxiliary and Faddeev-Popov ghost/anti-ghost fields as follows:
\begin{itemize}
\item auxiliary field:\hspace{2.9em}$B^{I}(\xi)$,
\item ghost field:\hspace{4.5em}$\chi^{I}(\xi)$,
\item anti-ghost field:\hspace{2.4em}$\bar{\chi}^{I}(\xi)$,
\end{itemize}
where suffix $I$ concerns the $SU\!(\!N\!)$ gauge group.
Each number of auxiliary, ghost and anti-ghost fields is the same as that of the degree of freedom of the gauge group.
For the $SU\!(\!N\!)$ symmetry, it is $N^2-1$.
The connection field (gauge boson) belongs to the adjoint representation of the gauge group; thus, the number of the above fields is the same as that of gauge bosons. 
The ghost/anti-ghost fields are the Grassmann number following an anti-commutable product. 
This study denotes the BRST transformation concerning the $SU\!(\!N\!)$ gauge group as $\delBRST^\SU[\bullet]$.
The BRST transformations for classical fermions and $SU\!(\!N\!)$ gauge bosons are
\begin{align*}
\delBRST^\SU[\psi]&:=i\cSU\hspace{.1em}\chi^{I}\tau^{~}_I\psi,\\
\delBRST^\SU[\AAA_\SU^{I}]&:=d^{~}_\SUO\hspace{.1em}\chi^I
=\(\partial^{~}_{\!\bcdot}\chi^I+\cSU\hspace{.1em}f^I_{\hspace{.2em}JK}\Aa^J_\bcdot\chi^K\)\eee^\bcdot.
\end{align*}
We note that  $d_\www s(\xi)=ds(\xi)$ for a local scalar function $s(\xi)$.
The BRST transformation on the Grassmann numbers $X$ and  $Y$ fulfils the Leibniz rule such that
\begin{align*}
\delBRST^\SU[XY]=\delBRST^\SU[X]\hspace{.1em}Y+\epsilon_X X\hspace{.1em}\delBRST^\SU[Y],
\end{align*}
where the signature $\epsilon_X=-1$ for $X\in\{\chi^{I},\bar{\chi}^{I}\}$, and $\epsilon_X=+1$ otherwise.
We define the BRST transformations for auxiliary and ghost/anti-ghost fields as
\begin{align*}
\delBRST^\SU[B^I]&:=0,\\
\delBRST^\SU[\chi^I]&:=-\frac{1}{2}\cSU\hspace{.1em}f^I_{\hspace{.2em}JK}\hspace{.1em}\chi^J\chi^k,\\
\delBRST^\SU[\bar\chi^I]&:=B^I.
\end{align*}
The BRST transformation for vierbein and spin connection fields gives zero.
Simple calculations show the BRST transformation is nilpotent for any fields.

\subsubsection{General relativity}
The current author has developed the canonical quantisation of general relativity in the Heisenberg picture utilising the Nakanishi--Kugo--Ojima quantisation method\!\cite{doi:10.1140/epjp/s13360-021-01463-3}.
This section discusses the BRST transformation of gravitational fields in the interaction picture following the method in the preceding section.

We introduce the auxiliary field and Faddeev--Popov ghost/anti-ghost fields as follows:
\begin{itemize}
\item auxiliary field:\hspace{2.8em}$\beta_{\mu}^{~a}(x)$,
\item ghost fields:\hspace{4em}$\chi^{a}_{\hspace{.3em}b}(\xi)$ and $\chi^\mu(x)$,
\item anti-ghost field:\hspace{2.5em}$\bar{\chi}_{\mu}^{\hspace{.3em}a}(x)$.
\end{itemize}
Here, ghost fields are Grassmannian.
Ghost fields $\chi^{a}_{\hspace{.3em}b}(\xi)$ and $\chi^\mu(x)$ were introduced to preserve the unitarity of the scattering amplitude, corresponding to the local Lorentz and global coordinate transformations, respectively.
We assign a ghost number $+1$ for $\chi^{a}_{\hspace{.3em}b}$ and $-1$ for $\bar{\chi}_{\mu}^{\hspace{.3em}a}$. 
An anti-ghost field corresponding to the global ghost is unnecessary since it is decoupled from the physical system in the inertial space.

In this study, we denote the BRST transformation for gravitational fields as $\delBRST^\GR[\bullet]$, and require rules introduced by Nakanishi\!\cite{Nakanishi01071978} as follows: 
The BRST transformation of the coordinate vector in $\MM$ should obey the general linear transformation as follows:\begin{align}
\delBRST^\GR\left[x^\mu\right]&=\chi^\mu.\label{BRSTx}
\end{align}
In addition, we require the postulate given in \!\cite{Nakanishi01071978}, such as
\begin{align}
\delBRST^\GR\left[\partial^{~}_{\!\mu} X\right]
&=\partial^{~}_{\!\mu}\delBRST^\GR\left[X\right]
-\(\partial^{~}_{\!\mu}\delBRST^\GR\left[x^\nu\right]\)\partial^{~}_{\!\nu} X
=\partial^{~}_{\!\mu}\delBRST^\GR\left[X\right]
-\(\partial^{~}_{\!\mu}\chi^\nu\)\partial^{~}_{\!\nu} X,\label{brstdelX}
\end{align}
where $X$ is any field defined in $\TMM$; thus, the BRST transformation acts on a one-form object in $\TsMM$ as
\begin{align*}
\delBRST^\GR\left[dx^\mu\right]&=\left(\partial^{~}_{\!\nu}\delBRST^\GR\left[x^\mu\right]\right)dx^\nu
=d\left(\delBRST^\GR\left[x^\mu\right]\right)
=~d\chi^\mu.
\end{align*}
Consequently, the BRST transformation and external derivative are commute with each other, i.e., 
\begin{align*}
\left[\delBRST^\GR,d\right]\bullet&=\delBRST^\GR\left[d\bullet\right]-
d\left(\delBRST^\GR\left[\bullet\right]\right)=~0,
\end{align*}

We define the BRST transformations of auxiliary and ghost/anti-ghost fields by reference to those of the Yang--Mills theory as
\begin{subequations}
\begin{align}
\delBRST^\GR\left[\beta_{\mu}^{\hspace{.3em}a}\right]&=\delBRST^\GR\left[\chi^\mu\right]~=~0,\label{BRSTchi1}\\
\delBRST^\GR\left[\chi^{a}_{\hspace{.3em}b}\right]\hspace{.3em}&=
\chi^{a}_{\hspace{.3em}\bcdot}\hspace{.1em}\chi^{\bcdot}_{\hspace{.3em}b},\label{BRSTchi2}\\
\delBRST^\GR\left[\bar{\chi}_{\mu}^{~a}\right]&=\lp\hspace{.1em}\beta_{\mu}^{~a}.\label{BRSTchi3}
\end{align}
\end{subequations}
Here, we have set the BRST transformations to preserve the physical dimension of each field.
The BRST transformation of the position vector given in (\ref{BRSTx}) provides the global ghost field with a Length dimension.
The auxiliary field has the same physical dimension as the spin connection field; it is $L^{-1}$ dimension.
The ghost field must be a null dimension object owing to (\ref{BRSTchi2}).
We also set anti-ghost fields as a null physical dimension; thus, we introduce the Planck length in (\ref{BRSTchi3}) to adjust a physical dimension.

The BRST transformation satisfies the following Leibniz rule:
\begin{align}
\delBRST^\GR\left[XY\right]&=\delBRST^\GR\left[X\right]Y+\epsilon_XX\delBRST^\GR\left[Y\right],\label{Leib}
\end{align}
where the signature $\epsilon_X=-1$ for $X\in\{\chi^\mu,\chi_a^{\hspace{.3em}b},\bar{\chi}_{\mu}^{\hspace{.3em}a}\}$, and $\epsilon_X=+1$ otherwise.
The BRST transformations of vierbein and spin connection-forms are defined as
\begin{subequations}
\begin{align}
\delBRST^\GR\left[\eee^a\right]&:=
\chi^{a}_{~\bcdot}\hspace{.1em}\eee^\bcdot,\label{brsteee}\\
\delBRST^\GR\left[\hat\www^{ab}\right]&:=
d\chi^{ab}
-\(\hat\www^{a}_{~\bcdot}\hspace{.1em}\chi^{\bcdot b}
+\hat\www^{b}_{\hspace{.3em}\bcdot}\hspace{.1em}\chi^{a \bcdot}\)\!,\label{brstwww}
\end{align}
\end{subequations}
where $\chi^{ab}:=\chi^{a}_{\hspace{.3em}\bcdot}\eta^{\bcdot b}$.
The local ghost field is anti-symmetric owing to the above definition.
They induce the BRST transformations of vierbein and spin connection fields as
\begin{subequations}
\begin{align}
\delBRST^\GR\left[\E^a_\mu\right]&=
-\E^a_\nu\hspace{.1em}\partial^{~}_{\!\mu}\chi^{\nu}
+\chi^{a}_{\hspace{.3em}\bcdot}\hspace{.1em}\E^{\bcdot}_\mu,\\
\delBRST^\GR\left[\hat\omega^{\hspace{.3em}ab}_{\mu}\right]&=
\E_\mu^\bcdot\hspace{.1em}\partial^{~}_{\!\bcdot}\chi^{a b}+\(\partial^{~}_{\!\mu}\chi^\nu\)\hat\omega^{\hspace{.3em}ab}_{\nu}
-\(
\hat\omega_{\mu\hspace{.3em}\bcdot}^{~a}\hspace{.2em}\chi^{\bcdot b}+
\hat\omega_{\mu\hspace{.3em}\bcdot}^{\hspace{.3em}b}\hspace{.2em}\chi^{a\bcdot}_{~}\)\!.\label{brstwabc}
\intertext{The BRST transformation of the vierbein inverse is obtained after simple calculations as}
\delBRST^\GR\left[\E_a^\mu\right]&=
\E_a^\nu\hspace{.1em}\partial^{~}_{\!\nu}\chi^{\mu}
-\chi^{\bcdot}_{\hspace{.3em}a}\hspace{.1em}\E_{\bcdot}^{\mu},
\end{align}
where $\delBRST^\GR\left[\E^a_\mu\hspace{.1em}\E_a^\nu\right]=\delBRST^\GR\left[\delta_\mu^\nu\right]=0$ is used.
\end{subequations}

We introduce one-form objects of auxiliary and ghost/anti-ghost fields as
\begin{subequations}
\begin{align}
\bbb^{a}&:=\(\lp\hspace{.1em}\beta_{\mu}^{~a}-\bar\chi^{~a}_{\nu}\hspace{.1em}\partial^{~}_{\!\mu}\chi^\nu\)dx^\mu,\label{bfGR}\\
\ccc^{a}&:=\chi^{a}_{\hspace{.3em}\bcdot}\hspace{.1em}\eee^\bcdot,\label{cfGR}\\
\bar{\ccc}^{a}&:=\bar{\chi}_{\mu}^{~a}\hspace{.1em}dx^\mu=\bar\chi^{~a}_{\bcdot}\hspace{.1em}\eee^\bcdot,\label{cbfGR}
\end{align}
where $\bar\chi_{b}^{~a}:=\bar\chi_\mu^{~a}\hspace{.1em}\E^\mu_{b}$.
\end{subequations}
They have a physical dimension as $\left[\bbb^{a}\right]=\left[\ccc^{a}\right]=\left[\bar{\ccc}^{a}\right]=L$, and their BRST transformations are
\begin{align}
\delBRST^\GR\left[\bbb^{a}\right]=\delBRST^\GR\left[\ccc^a\right]=0~~&\textrm{and}~~
\delBRST^\GR\left[\bar{\ccc}^{a}\right]=\bbb_{~}^a.\label{Dbccb}
\end{align}
Auxiliary form has representations in the inertial space as
\begin{align*}
\text{(\ref{bfGR})}=\(\lp\beta_{\bcdot}^{\hspace{.3em}a}-\bar\chi^{\hspace{.3em}a}_{\star}\hspace{.1em}
\partial^{~}_{\!\bcdot}\chi^\star\)\eee^\bcdot\in\TsM,
~~&\text{where}~~
\beta_{b}^{\hspace{.3em}a}:=\beta_{\mu}^{\hspace{.3em}a}\hspace{.1em}\E^\mu_{b},~~
\chi^a:=\E^a_\mu\hspace{.1em}\chi^\mu.
\end{align*}
We represent the auxiliary form in $\TsM$ as $\bbb^{a}=B_{\hspace{.2em}\bcdot}^{a}\eee^\bcdot$ using components defined as
\begin{align}
B_{\hspace{.2em}b}^{a}&:=\(\lp\beta_{b}^{\hspace{.3em}a}-
\bar\chi^{\hspace{.3em}a}_{\bcdot}\hspace{.1em}
\partial^{~}_b\chi^\bcdot\)\!.\label{Bab}
\end{align}
Since we define the Lagrangian form in $\TsM$, Romanised forms appear in the Lagrangian.
We provide the BRST transformation of (\ref{brstdelX}) and (\ref{brstwabc}) in the inertial space for future convenience:
\begin{subequations}
\begin{align}
\delBRST^\GR\left[\partial^{~}_aX(\xi)\right]
&=\partial^{~}_a\(\delBRST^\GR\left[X\right](\xi)\)
-\chi^\bcdot_{\hspace{.3em}a}\(\partial^{~}_{\!\bcdot} X(\xi)\)\!,\label{DdX}\\
\delBRST^\GR\left[\hat\omega_c^{\hspace{.3em}ab}\right]
&=\partial^{~}_c\chi^{ab}-\(
 \hat\omega_{c\hspace{.3em}\bcdot}^{\hspace{.3em}a}\hspace{.1em}\chi^{\bcdot b}
+\hat\omega_{c\hspace{.3em}\bcdot}^{\hspace{.3em}b}\hspace{.1em}\chi^{a\bcdot}
+\hat\omega_{\bcdot}^{\hspace{.3em}ab}\hspace{.1em}\chi^{\bcdot}_{\hspace{.3em}c}
\)\!.\label{Dwabc}
\end{align}
\end{subequations}
We note that the BRST transformation on the object defined in the inertial space does not include the global ghost field, as shown above.
Consequently, the global ghost field does not appear in the ghost Lagrangian defined in the inertial space.

Direct calculations show that the BRST transformations are nilpotent for all forms and fields.
Consequently, the classical Lagrangian $\LL_\GR$ is BRST-invariant.
 In this study, we simplified definitions of auxiliary and ghost/anti-ghost fields from those in Ref$.$\!\cite{doi:10.1140/epjp/s13360-021-01463-3}.
Although calculations are almost identical to those in Ref$.$\!\cite{doi:10.1140/epjp/s13360-021-01463-3}, we provide proof of nilpotency for all necessary forms in \textbf{Appendix~\ref{app1}} for completeness.

%
%
\subsection{Ghost and gauge-fixing Lagrangian}\label{Lghost}
This section introduces gauge-fixing and ghost Lagrangians to the bare quantum Lagrangian according to the standard prescription:
Prepare the necessary number of functions $F[\Aa, B, \chi, \bar\chi]$, which includes the same number of ghost and anti-ghost fields. 
Then, add a term 
\begin{align*}
\LL_{\bullet;\gfix}+\LL_{\bullet;gh}:=\delBRST^\bullet[\bar{\chi}\hspace{.1em}F]
\end{align*}
to the Lagrangian.
As a result, the total Lagrangian is BRST-invariant, and after fixing all gauge degrees, the Lagrangian is still keeping the BRST symmetry.
We omit suffix ``$\hspace{.1em}\bare$'' suffix for the bare fields in this section, too.

\subsubsection{Yang--Mills theory}
For the $SU\!(\!N\!)$ gauge part, we introduce the scalar field as
\begin{align*}
H^{I}&:=\eta^\bcdots\partial^{~}_{\!\bcdot}\Aa^{I}_{\hspace{.4em}\bcdot}+
\frac{1}{2}\xi_AB^{I},
\intertext{which provides the gauge-fixing and ghost Lagrangians, such that}
\LL_{\SU;\gfix}+\LL_{\SU;gh}&:=\sum_{I=1}^N\delBRST^\SU[\bar{\chi}^{I}H^{I}].
\end{align*}
We decompose it into
\begin{subequations}
\begin{align}
\LL_{\SU;\gfix}&:=\sum_{I=1}^N\delBRST^\SU[\bar{\chi}^{I}]H^{I}=\sum_{I=1}^N\(
B^{I}\eta^\bcdots\partial^{~}_{\!\bcdot}\Aa^{I}_{\hspace{.4em}\bcdot}
+\frac{\xi_A}{2}\hspace{.1em}B^{I}\hspace{-.1em}B^{I}\)\!,\label{Lsugf1}\\
\LL_{\SU;gh}&:=\sum_{I=1}^N\bar{\chi}^{I}\delBRST^\SU[H^{I}]
=\bar\chi^{I}\eta^\bcdots\partial^{~}_{\!\bcdot}\(
\partial^{~}_{\!\bcdot}\chi^{I}+\cSU\hspace{.1em}f^I_{\hspace{.2em}JK}\Aa^{J}_{\hspace{.4em}\bcdot}\chi^{K}\)\!.\label{LsuFP}
\end{align}
We can write the gauge-fixing Lagrangian as
\begin{align}
\text{(\ref{Lsugf1})}&=\sum_{I=1}^N\frac{1}{2\xi_A}\left\{
\(C^{I}\)^2-\(\eta^\bcdots\partial^{~}_{\!\bcdot}\Aa^{I}_{\hspace{.4em}\bcdot}\)^2
\right\},\label{Lsugf}
\intertext{where}
C^{I}&:=\eta^\bcdots\partial^{~}_{\!\bcdot}\Aa^{I}_{\hspace{.4em}\bcdot}+\xi_AB^{I}\!.\notag
\end{align}
\end{subequations}
Therefore, scalar field $C^{I}$ (and thus $B^{I}$, too) is decoupled from the other parts, and the standard gauge-fixing term 
\begin{align}
\LL_{\SU;\gfix}&:=-\frac{1}{2\xi^{~}_{\hspace{-.1em}A}}\sum_{I=1}^N
\left(\eta^{\bcdots}\partial^{~}_{\bcdot}\Aa^{I}_{\hspace{.4em}\bcdot}\right)^2\!,\label{Acovg}
\end{align}
and interaction vertices among gauge bosons and ghosts remain in the Lagrangian.
The gauge-fixing condition provides one constraint on four components of $\Aa^a$\!.
Together with an on-shell (mass-less) condition, two dynamical degrees (transverse polarisation) remain on a $SU\!(\!N\!)$ gauge boson. 

Owing to definitions, we can confirm that the Lagrangian is BRST-invariant:
\begin{align*}
\delBRST^\SU\left[\LL_{\SU}+\LL_{\SU;\gfix}
+\LL_{\SU;gh}
+\LL_{\MT}\right]=0.
\end{align*}

\subsubsection{General relativity}
This section provides gauge-fixing and ghost Lagrangians for a gravity part according to the same prescription used for the Yang--Mills theory.
We introduce a one-form object
\begin{align*}
\HHH^{a}&:=
\ddd\hspace{-.1em}\www^{a}
+\frac{1}{2\lpt}\xi_{\omega}\hspace{.1em}\bbb^{a}\!,
~\text{with}~~
\ddd\hspace{-.1em}\www^{a}:=\(\partial^{~}_{\!\bcdot}\hat\omega^{\hspace{.3em}\star\bcdot}_{\star}\)\eee^a\!.
\end{align*}
We note that $\ddd\hspace{-.1em}\www^{a}$ has a length-inverse dimension and $\bbb^{a}$ has length dimension; thus, $\HHH^{a}$ has a length-inverse dimension. 
We define gauge-fixing and ghost Lagrangians using this one-form object as
\begin{align*}
\LLL_{\GR;\gfix}+\LLL_{\GR;gh}&:=\frac{1}{\lpt}\delBRST^\GR\left[
\bar\ccc^{ \bcdot}\hspace{-.2em}\wedge\HHH^{ \bcdot}
\hspace{-.2em}\wedge\SSS_{\bcdots}
\right],
\end{align*}
where the Planck length appears to adjust the Lagrangian forms to a null-dimension.
Owing to the definition and nilpotent of each field, this is BRST invariant. 
We decompose it into the gauge-fixing and ghost parts, such that:
\begin{subequations}
\begin{align}
\lpt\hspace{.1em}\LLL_{\GR;\gfix}&:=\delBRST^\GR\left[\bar\ccc^{\bcdot}\right]\wedge\HHH^{\bcdot}
\hspace{-.2em}\wedge\SSS_{\bcdots}
\overset{\text{(\ref{Dbccb})}}{=}\bbb^{\bcdot}\hspace{-.2em}\wedge\HHH^\bcdot\wedge\SSS_{\bcdots}
,\label{Lgrgf}\\
\lpt\hspace{.1em}\LLL_{\GR;gh}&:=-\bar\ccc^{\bcdot}\wedge
\delBRST^\GR\left[\HHH^{\bcdot}\hspace{-.2em}\wedge\SSS_{\bcdots}\right]
\label{LgrFP}
\end{align}
\end{subequations}

\paragraph{gauge-fixing Lagrangian:}p=
First, we fix the gauge-fixing Lagrangian.
We introduce a new one-form object 
\begin{align}
\CCC^{a}&:=\ddd\hspace{-.1em}\www^{a}+\frac{\xi_\omega}{\lpt}\hspace{.1em}\bbb^{a} \label{defCCC}
\end{align}
and eliminate $\bbb$ from $\HHH$ as
\begin{align}
\text{\ref{defCCC})}&\implies\bbb^{a} =\frac{\lpt}{\xi_\omega}\(\CCC^{a}-\ddd\hspace{-.1em}\www^{a}\)\implies
\HHH^a=\frac{1}{2}\(\CCC^a+\ddd\hspace{-.1em}\www^{a}\)\!,
\end{align}
yielding 
\begin{align}
\text{(\ref{Lgrgf})}\rightarrow
\LLL_{\GR;\gfix}&=\frac{1}{2\xi_\omega}\left\{
\CCC^{\bcdot}\wedge\CCC^{\bcdot}\wedge\SSS_{\bcdots}-
\(\hspace{.1em}\partial^{~}_{\!\bcdot}\hat\omega_\star^{~\star\bcdot}\)^2\vvv
\right\}\!.\label{Lgrgf2}
\end{align}
The first term of (\ref{Lgrgf2}) has representation in the global space like
\begin{align*}
\CCC^{\bcdot}\wedge\CCC^{\bcdot}\wedge\SSS_{\bcdots}
=C^{\bcdot}_{\hspace{.2em}\star}\hspace{.1em}C^{\times}_{\hspace{.2em}\ast}
\(\delta^\star_\bcdot\hspace{.1em}\delta^\ast_\times-\delta^\ast_\bcdot\hspace{.1em}\delta^\star_\times\)
\vvv,
\end{align*}
where $C^{a}_{\hspace{.2em}b}$ is a coefficient function defined through $\CCC^{a}=:C^{a}_{\hspace{.2em}\bcdot}\hspace{.1em}\eee^{\bcdot}$.
Thus, $\CCC^{a}$ is isolated from the physical system and cab be ignored.

The spin connection $\hat\omega_{c}^{\hspace{.3em}ab}$ has $24$ components due to anti-symme concerning the superscripts. 
In addition, the torsionless condition provides twelve constraints on the spin connection field and implies a relation on the Riemann curvature such that:
\begin{align}
R_{ab;cd}&=R_{cd;ab},~~\text{where}~~
R_{ab;cd}:=R^\bcdots_{\hspace{.7em}cd}\hspace{.1em}\eta^{~}_{\bcdot a}\hspace{.1em}\eta_{\bcdot b}.\label{Rabcdcdab}
\end{align}
The Riemann curvature fulfiling (\ref{Rabcdcdab}) can be constructed using a half of the spin connection fields of
\begin{subequations}
\begin{align}
\hat\omega_{\dot{a}}^{\hspace{.4em}b}&:=\hat\omega_{\hat a}^{\hspace{.4em}\hat{a}b}.\label{wpolvec}
\end{align}
We do not apply the Einstein convention of repeated suffixes for those with a hat. 
When we take the summation, we denote it as
\begin{align}
\hat\omega_{\dot{\Sigma}}^{\hspace{.4em}b}&:=\sum_{\hat a}\hat\omega_{\hat a}^{\hspace{.4em}\hat{a}b}.\label{wpolvec2}
\end{align}
\end{subequations}
There are twelve independent components in (\ref{wpolvec}) due to four constraints of $\hat{a}\neq b$.
We set the remaining twelve components to 
\begin{align}
\hat\omega_{a}^{\hspace{.4em}bc}&=0~~\text{for}~~a\neq b\land a\neq c,\label{wconst}
\end{align}
without loss of generality (See section 2.7 of Ref.\cite{Kurihara:2025tro}.)
Therefore, we redefine the gauge-fixing Lagrangian as
\begin{subequations}
\begin{align}
\LLL_{\GR;\gfix}&:=-\frac{1}{2\xi_\omega}
\(\partial^{~}_{\!\bcdot}\hat\omega^{\hspace{.3em}\bcdot}_{\dot{\Sigma}}\)^2\vvv,
\label{Wgauge}
\intertext{which corresponds to set the covariant gauge-fixing condition as}
\partial^{~}_{\!\bcdot}\hat\omega^{\hspace{.3em}\bcdot}_{\dot{\Sigma}}&=0.\label{Wgauge2}
\end{align}
\end{subequations}
We note that the dimensional factor, $\lp$, is encapsulated in $\CCC$ and disappears in the gauge-fixing Lagrangian, as in (\ref{Wgauge}).
The equation (\ref{Wgauge2}) provides four constraints on $\hat\omega_{\dot{a}}^{\hspace{.4em}b}$.
In addition, there are six on-shell conditions, which leaves two physical degrees of freedom in the theory.
The graviton is the quantized spin connection field in our theory, and it is a massless spin-two particle.
The little group concerning the graviton in four-dimensional spacetime is $SO(2)$, which has two degrees of freedom and is consistent with the above discussion.

\paragraph{ghost Lagrangian:}
Next, we construct the ghost Lagrangian, which provides an interaction term between the ghosts and the physical fields.
We need a further simplification for the ghost/anti-ghost fields.
The role of the Faddeev-Popov ghost is to preserve the Ward identity and to ensure the unitarity of the gravitational scattering amplitudes.
Quantum theory requires as many ghost fields as the dimension of the gauge group, which is six for the Lorentz group.
However, all six components of the graviton are not independent, but only two independent components, as discussed above.
So we need one scalar field each for ghost and anti-ghost fields.
Owing to the BRST transformation of the spin connection field, the local ghost is anti-symmetric concerning its two suffixes.

We parameterize the local ghost field as
\begin{subequations}
\begin{align}
\eta_{a\bcdot}\hspace{.1em}\chi^\bcdot_{\hspace{.3em}b}(\xi)&=\chi_{ab}(\xi):=\chi(\xi)\hspace{.2em}c_ac_b,
\end{align}
where $c_a:=(c_0,c_1,c_2,c_3)$ is a Grassmannian vector.
On the other hand, we set the anti-ghost fields a symmetric tensor:
\begin{align}
\bar\chi^{\hspace{.3em}a}_{b}(\xi)=\bar\chi(\xi)\hspace{.2em}\delta^{a}_{b}.\label{scalarchi}
\end{align}
Both $\chi(\xi)$ and $\bar\chi(\xi)$ are Grassmannian scalar fields.
The auxiliary form also has the same degree of freedom and the same convention as that of the anti-ghost field, such that
\begin{align}
\text{(\ref{DdX})}&=
B_{\hspace{.2em}b}^{a}(\xi)=B(\xi)\hspace{.2em}\delta^{a}_{b}.\label{scalarB}
\end{align}
\end{subequations}
In the following calculations, we keep all components of the spin connection field without requiring the torsionless condition.

The ghost Lagrangian splits into two parts owing to the Leibniz rule of the BRST transformation, such that:
\begin{align}
-\text{(\ref{LgrFP})}=~&
\bar\ccc^{\bcdot}\wedge
\delBRST^\GR\left[\HHH^{\bcdot}\right]\wedge\SSS_{\bcdots}
+\bar\ccc^{\bcdot}\wedge\HHH^{\bcdot}\hspace{-.2em}\wedge
\delBRST^\GR\left[\SSS_{\bcdots}\right]\!,\notag\\\overset{\text{(\ref{Dbccb})}}{=}&
\bar\ccc^{\bcdot}\wedge
\delBRST^\GR\left[\ddd\hspace{-.1em}\www^{a}\right]\wedge\SSS_{\bcdots}
+\bar\ccc^{\bcdot}\wedge\ddd\hspace{-.1em}\www^{\bcdot}\wedge\delBRST^\GR\left[\SSS_{\bcdots}\right]
+\frac{1}{2\lpt}\xi_{\omega}\hspace{.1em}\bar\ccc^{\bcdot}\wedge\bbb^{\bcdot}\wedge
\delBRST^\GR\left[\SSS_{\bcdots}\right]\!.\label{LgrFP2}
\end{align}
Simple calculations provide the BRST transformation of the surface form and $\HHH^a$, respectively,  as
\begin{subequations}
\begin{align}
\delBRST^\GR\left[\SSS_{ab}\right]
&=\frac{1}{2}\epsilon_{ab\bcdots}\delBRST^\GR\left[\eee^{\bcdot}\wedge\eee^{\bcdot}\right]
\overset{\text{(\ref{brsteee})}}{=}\epsilon_{ab\bcdot\star}
\chi\hspace{.1em}c_{~}^\bcdot c^{~}_\times\hspace{.1em}\eee^\times\hspace{-.2em}\wedge\eee^\star\!,\label{DSab}\\
\delBRST^\GR\left[\HHH^a\right]&=
\delBRST^\GR\left[\(\partial^{~}_{\!\bcdot}\hat\omega^{\hspace{.4em}\bcdot}_{\dot\Sigma}\)\eee^a\right]\overset{\text{(\ref{DdX})}}{=}
\(\partial^{~}_{\!\bcdot}\delBRST^\GR\left[\hat\omega_{\dot\Sigma}^{\hspace{.4em}\bcdot}\right]
-\chi\hspace{.1em}c_{~}^\star c^{~}_\bcdot\hspace{.1em}\partial^{~}_{\!\star} \hat\omega_{\dot\Sigma}^{\hspace{.4em}\bcdot}\)\eee^a-
\(\partial^{~}_{\!\bcdot}\hat\omega^{\hspace{.3em}\star\bcdot}_{\star}\)\delBRST^\GR[\eee^a],
\notag\\&\hspace{-1.2em}\overset{\text{(\ref{Dwabc})}}{=}
\left\{\(\partial^{~}_{\!\bcdot}\partial^{~}_{\!\star}\chi\){c^\star}{c^\bcdot}
-\(\partial^{~}_{\!\bcdot}\hat\omega_{\dot\Sigma\hspace{.1em}\star}^{~}\chi\){c^\star}{c^\bcdot}
-\(\partial^{~}_{\!\bcdot}\hat\omega_{\star\hspace{.2em}\times}^{\hspace{.3em}\bcdot}\hspace{.1em}\chi\){c^\star}{c^\times}
-\(\partial^{~}_{\!\bcdot}\hat\omega_{\star}^{\hspace{.3em}\times\bcdot}\hspace{.1em}\chi\){c^\star}{c_\times}
-\(\partial^{~}_{\!\bcdot} \hat\omega_{\dot\Sigma}^{\hspace{.4em}\star}\)\chi\hspace{.1em}c_{~}^\bcdot c^{~}_\star
\right\}\eee^a
\notag\\&\hspace{1.5em}-\(\partial^{~}_{\!\bcdot}\hat\omega_{\dot\Sigma}^{\hspace{.4em}\bcdot}\)\chi{c^a}{c_\star}\eee^\star,
\notag\\&=\left\{\(\partial^{~}_{\!\bcdot}\partial^{~}_{\!\star}\chi\){c^\star}{c^\bcdot}
-\hat\omega_{\dot\Sigma\hspace{.1em}\star}^{~}\(\partial^{~}_{\!\bcdot}\chi\){c^\star}{c^\bcdot}\right\}\eee^a
-\(\partial^{~}_{\!\bcdot}\hat\omega_{\dot\Sigma}^{\hspace{.4em}\bcdot}\)\chi{c^a}{c_\star}\eee^\star,\notag\\
\xrightarrow{\text{(\ref{Wgauge2})}}
\delBRST^\GR\left[\HHH^a\right]&=\left\{\(\partial^{~}_{\!\bcdot}\partial^{~}_{\!\star}\chi\){c^\star}{c^\bcdot}
-\hat\omega_{\dot\Sigma\hspace{.1em}\star}^{~}\(\partial^{~}_{\!\bcdot}\chi\){c^\star}{c^\bcdot}\right\}\eee.\label{DHa}
\end{align}
\end{subequations}
Since $c^\bullet$ is a Grassmann number, the last two terms in the r.h.s in (\ref{LgrFP2}) disappears like
\begin{align*}
\bar{\ccc}^{\bcdot}\wedge\bbb^{\bcdot}\wedge\delBRST^\GR\left[\SSS_{\bcdots}\right]&=
\epsilon_{\bcdots\bcdots}\hspace{.1em}\bar{\chi}\hspace{.1em}B\chi\hspace{.1em}c^\bcdot c_\star\hspace{.1em}
\eee^\bcdot\wedge\eee^\bcdot\wedge\eee^\star\wedge\eee^\bcdot
=3!\hspace{.1em}\bar{\chi}\hspace{.1em}B\chi\(\delta^\star_\bcdot\hspace{.1em}c^\bcdot c_\star\)\vvv=0,
\intertext{and}
\bar\ccc^{\bcdot}\wedge\ddd\hspace{-.1em}\www^{\bcdot}\hspace{-.2em}\wedge
\delBRST^\GR\left[\SSS_{\bcdots}\right]&=3!\hspace{.2em}
\bar{\chi}\(\partial^{~}_{\!\bcdot}\hat\omega_{\dot\Sigma}^{\hspace{.4em}\bcdot}\)
\chi\(\delta^\star_\bcdot\hspace{.1em}c^\bcdot c_\star\)\vvv=0.
\end{align*}
The first term in (\ref{LgrFP2}) is calculated using (\ref{DHa}) as
\begin{align*}
\bar\ccc^{\bcdot}\wedge\delBRST^\GR\left[\ddd\hspace{-.1em}\www^{\bcdot}\right]\wedge\SSS_{\bcdots}&=
\(\bar{\chi}\(\partial^{~}_{\!\bcdot}\partial^{~}_\star\chi\){c^\star}{c^\bcdot}-
\bar{\chi}\hspace{.1em}\hat\omega_{\star\hspace{.2em}\bcdot}^{\hspace{.3em}\star}
\(\partial^{~}_\times\chi\){c^\bcdot}{c^\times}\)\vvv.
\end{align*}
As a result, we obtain the ghost Lagrangian such that:
\begin{align}
\LLL_{\GR;gh}&=-\frac{1}{\lpt}\bar\ccc^{\bcdot}\wedge\delBRST^\GR\left[\ddd\hspace{-.1em}\www^{\bcdot}\right]\wedge\SSS_{\bcdots}
=\frac{1}{\lpt}\(\(\partial^{~}_{\!\bcdot}\bar{\chi}\)\(\partial^{~}_\star\chi\)+
\bar{\chi}\hspace{.1em}\hat\omega_{\dot\Sigma\hspace{.1em}\bcdot}^{~}\(\partial^{~}_{\!\star}\chi\)\){c^\bcdot}{c^\star}
\hspace{.1em}\vvv\label{Lgrgh}.
\end{align}
 
 Although we keep all components of the spin connection fields in the above calculations, the result includes only components appearing in (\ref{wpolvec}); thus, our parameterisations of ghost field, (\ref{scalarchi}), is consistent with  that of the connection fields.
Consequently, the interaction terms consist of scalar ghost/anti-ghost fields and polarisation vectors $\hat\omega_{\dot{a}}^{\hspace{.3em}b}$.
 
The nilpotent of gravitational fields ensures the BRST invariance of the gravitational Lagrangian, such that:
\begin{align*}
\delBRST^\GR\left[\LLL_{\GR}+\LLL_{\GR;\gfix}+\LLL_{\GR;gh}\right]=0.
\end{align*}
When we set $\hat\omega\rightarrow\cG\omega$, we can obtain the BRST transformations for scaled fields.

%
%
\section{\QGED Feynman rules}\label{FeynmanRules}
This section presents a formulation of Feynman rules concerning the Yang--Mills--Utiyama Lagrangian.
For curved spacetime, a momentum space, in which Feynman rules are defined, is not trivial.
In this study, we utilise the definition of the momentum space introduced in Ref$.$\!\cite{Kurihara:2022green} and denote it by $\TstM$.
We follow a phase-convention of the Feynman rules in Ref$.$\!\cite{10.1143/PTPS.73.1}. 
We utilise the scaled fields in the Feynman rule for the spin connection since we assumed it must be observable.
Feynman rules for bare fields have a common shape to those for renormalised fields. Therefore, continuing from the previous section, this section also omits the suffix ``$\hspace{.1em}\bare$''.

In the following, we will focus exclusively on the $U(1)$ gauge as the gauge group and a single spinor field (electron field).
Consequently, the resulting theory is quantum electromagnetic dynamics (QED) with the gravitational force, which is referred to as \textit{Quantum GraviElectro Dynamecs} (\QGED). 
We set the fine structure constants involved in the electromagnetic and gravitational interactions as dimensionless objects,
\begin{align}
\alpha&=\frac{\cSU^2}{4\pi}\implies\cSU=\frac{e}{\hbar^\hlf},~~\text{and}~~
\aGR=\frac{\cG^2}{4\pi}.\label{alphae2}
\end{align}
The perturbation expansion is performed with respect to a power of these fine structure constants.

The gravitational fine-structure constant $\aGR$ appears in the Lagrangian through the local $\SO(1,3)$-invariant differential initially in the \QGED\!, despite no such constant in the standard method. 
It is an indispensable tool for quantising general relativity, utilising the established method of the quantum Yang-Mills theory.
This study maximally relies on the standard theory of particle physics and follows its established procedure for introducing the gravitational interaction into the theory.
Even at the classical level, the gravitational coupling constant determines the relative strength of the gravitational force compared to the other fundamental forces, namely the electroweak and strong forces.
There is no reason not to appear a coupling constant only for the local $\SO(1,3)$-invariant differential, despite its appearance in the $\SU(2)\otimes U\!(1)$- and $\SU(3)$-invariant differentials.

The perturbative expansion is carried out concerning the fine structure constant in the standard procedure of the standard theory.
The lowest order solutions of the equation of motion provide the propagators of the \textit{free} particles and the vertices of interactions among them.
For the vierbein field,  an equation of motion is provided by the torsion equation.
The spacetime torsion is provided owing to the fermion field as
\begin{align}
0=
\frac{\delta\LL_\text{\YMU}}{\delta\www^{ab}}=
\frac{\delta(\LL_{\GR;int}+\LL_{\MT;int})}{\delta\www^{ab}}=
\frac{1}{2}\epsilon_{ab\bcdots}\hspace{.1em}\TTT_{\hspace{-.1em}\textsc{f}_{\hspace{-.1em}\textsc{m}}}^{\hspace{.3em}\bcdot}
\wedge\eee^\bcdot_{~}+
\cG\(\bpsi^{\dagger_{\hspace{-.1em}\epsilon}}_{~}\gamma_{~}^\bcdot{\sigma}^{~}_{ab}\hspace{.2em}\bpsi\)\VVV_\bcdot.\label{torsionful}
\end{align}
Thus, spacetime is torsionless at the lowest order of perturbation.
Furthermore, the second term on the right-hand side of the equation is non-zero only when $\bpsi\neq0$.
In a three-dimensional Cauchy surface where any two points are spacelike separated, we can expect spacetime to have torsion at only a countable number of points.
It is reasonable to assume that the spacetime is torsionless almost everywhere, even at all orders of the perturbation expansion.

We put the gauge-fixing term into the Lagrangian to quantise the gauge theory.
After the gauge fixing, the BRST symmetry remaining in the Lagrangian ensures the unitarity of the scattering amplitude, which is independent of the gauge parameter included in the Feynman rule.
This also applies to the $\SO(1,3)$ gauge invariance of the quantum \YMU theory.
First, we restrict a spacetime to the local inertial space where the local $\SO(1,3)$ symmetry remains, choosing a specific coordinate to break the global $GL(4)$ symmetry by setting a vanishing Levi-Civita tensor. 
The full $\SO(1,3)$ symmetry still has too many degrees of freedom to quantise the theory.
After putting the $\SO(1,3)$ gauge-fixing term in the Lagrangian, the spin connection has two dynamic degrees of freedom while keeping the gauge parameter independence. 
As a result, the spin connection quantum can be interpreted as the massless spin-two graviton.
%
%
\subsection{Green's function for vierbein and spin connection fields}\label{greensfunctionWE}
This section provides Green's function for spin connection and vierbein fields in the configuration space based on Ref$.$\!\cite{Kurihara:2022green}.
The \YMU theory treats the gravitational fields parallel to the internal gauge fields as much as possible.
Though the internal gauge field Lagrangian is in second-order differential formalism for the field strength, the Einstein--Hilbert Lagrangian is in first-order one. 
The second-order formalism is desirable to quantise the gravitational part of the quantum \YMU theory.

We have introduced the co-Poincar\'{e} symmetry and gave the gravitational Lagrangian in the second formalism using the co-Poinca\'{e} curvature as (\ref{Lgr}). 
The variational operation on the gravitational action with respect to the translation operator provides\footnote{See section 5.6.3 in Ref$.$\!\cite{fre2012gravity} and section 2.1 in Ref$.$\!\cite{Kurihara_2020}.}
\begin{subequations}
\begin{align}
\text{(\ref{Lgr})}\rightarrow{\frac{\delta \I_{\hspace{-.1em}\GR}}{\delta P^a}}=\int\frac{\delta\LLL_\GR^{~}}{\delta P^a}\overset{\text{(\ref{cPLieA2})}}{=}
\frac{1}{2\kE\hbar}\epsilon_{a\bcdot\bcdots}\int\hat\RRR^{\bcdots}\!\wedge\TTT_{~}^{\bcdot}\overset{\text{(\ref{torsionful})}}{=}0.\label{delPLgr1}
\end{align}
We are further rewriting the formula as
\begin{align}
\text{(\ref{delPLgr1})}=
\frac{1}{2\kE\hbar}\epsilon_{a\bcdot\bcdots}\hat\RRR^{\bcdots}\!\wedge{d}_{\hat{\www}}\eee_{~}^{\bcdot}
\overset{\text{(\ref{BianchiRST})}}{=}
\frac{1}{2\kE\hbar}\epsilon_{a\bcdot\bcdots}{d}_{\hat{\www}}
\(\hat\RRR^{\bcdots}\!\wedge\eee_{~}^{\bcdot}\)=0,\label{delPLgr2}
\end{align}
\end{subequations}
owing to the Bianchi identity.
We construct propagators of the gravitational bosons using the second-order formalism of (\ref{delPLgr1}) and (\ref{delPLgr2}).  
We note that the last equality in (\ref{delPLgr2}) holds in both the local and global cotangent bundle due to the soldering relation (\ref{LaplaceGw}) under the torsionless assumption.

\paragraph{spin connection:}
As mentioned for the spin connection in \textbf{section \ref{BLag}}, the observed field is the scaled spin connection.
Thus, we give the Feynman rule for the scaled one here.
We consider the Einstein equation as an equation of motion for the spin connection,
The Einstein equation is a first-order partial differential equation concerning the spin connection. 
We start from (\ref{delPLgr2}), then set the coupling constant to zero to obtain a free-field equation as
\begin{subequations}
\begin{align}
\text{(\ref{delPLgr2})}&\implies
d_{\hat\www}\(\epsilon_{a\bcdots\bcdot}\hat\RRR^\bcdots\wedge\eee^\bcdot\)=0,\notag\\
&\xrightarrow{\hat\omega\rightarrow\cG\omega}
\partial^{~}_{\!\bcdot}\!\({R}^{a\bcdot}-\frac{1}{2}\eta^{a\bcdot}{R}\)+\cG(\cdots)=
\partial_{~}^\bcdot\Ri^{a\star}_{\hspace{.7em}\bcdot\star}
-\frac{1}{2}\partial_{~}^a\Ri^{\bcdot\star}_{\hspace{.7em}\bcdot\star}+\cG(\cdots)
=0,\label{EinsteinEq2}
\end{align}
where $\cG(\cdots)$ is terms proportional to $\cG$.
We used the anti-symmetry of $\Ri$ concerning both the subscripts and superscripts in the above. 
Owing to the structure equation, we obtain an equation for the classical spin connection field as\!\cite{Kurihara:2022green}
\begin{align}
\text{(\ref{EinsteinEq2})}&\xrightarrow{\text{\ref{RRR}}}
\(\partial_{~}^\bcdot\partial^{~}_{\!\bcdot}\omega_{\star}^{~a\star}-
\partial_{~}^\bcdot\partial^{~}_\star\omega_{\bcdot}^{~a\star}-
\partial_{~}^a\partial^{~}_\star\omega_\bcdot^{~\star\bcdot}\)+
\cG(\cdots)=0\underset{\cG=0}{\xrightarrow{\text{(\ref{Wgauge2})}}}
\(\partial_{~}^\bcdot\partial^{~}_{\!\bcdot}\omega_{\star}^{~\star a}\)\(\xi\)=0.\label{deltaomega}
\end{align}
\end{subequations}
The fundamental solution of (\ref{deltaomega}) provides Green's function of the spin connection as
\begin{align*}
{G^{~}_{\omega}}(\xi):=\frac{\omega_{\hspace{1em}\dot{a}}^{(\lambda)\hspace{.4em}b}}
{-(\xi^0)^2+(\xi^1)^2+(\xi^2)^2+(\xi^3)^2-i\delta},
\end{align*}
where $\omega_{\hspace{1em}\dot{a}}^{(\lambda)\hspace{.4em}b}$ is polarisation vectors of the spin connection field.
Infinitesimal constant $\delta>0$ is a parameter to determine the analyticity of Green's function.

\paragraph{vierbein:}
The torsion two-form has a tensor representation as in (\ref{torsionFMST2}) with \ref{torsionFM}.
The torsionless equation, which is obtained as an Euler-Lagrange equation concerning the Lagrangian (\ref{Lgr}), is
\begin{align}
\TT^a_{\hspace{.6em}bc}&=\frac{1}{2}\E^a_\mu
\(\partial^{~}_c\E_b^\mu-\partial^{~}_b\E_c^\mu\)+
\frac{1}{2}\cG\(\omega_{b\hspace{.3em}c}^{\hspace{.3em}a}-\omega_{c\hspace{.3em}b}^{\hspace{.3em}a}\)=0.\label{torsionlesscp}
\end{align}
The divergence concerning the superscript of the torsion form, $\partial^{~}_{\!\bcdot}\TT^\bcdot_{\hspace{.6em}bc}$, and the gauge fixing conditionof the spin connection  (\ref{Wgauge2}) provides the constraint in the global space such as\!\cite{Kurihara:2022green} 
\begin{subequations}
\begin{align}
\partial^{~}_{\!\mu}\E^\mu_a&=0,\label{dE0}
\end{align}
which is consistent with the de\hspace{.15em}Donder \textit{gauge}\footnote{Strictly speaking, it is not the gauge fixing condition since a vierbein is not a gauge boson but a section.}: $\partial^{~}_{\!\mu}(\sqrt{-g}g^{\mu\nu})=0$ and induces a constraint in the inertial space as
\begin{align}
(\ref{dE0})\implies&0=\E_\nu^a\partial^{~}_{\!\mu}\E^\mu_a=\partial^{~}_{\!\mu}\(\E_\nu^a\hspace{.1em}\E^\mu_a\)-\E^\mu_a\partial^{~}_{\!\mu}\E_\nu^a=
-\partial^{~}_{a}\E_\nu^a,\label{dE0In}
\intertext{owing to}
&\partial^{~}_{\!\mu}\(\E_\nu^a\hspace{.1em}\E^\mu_a\)=\partial^{~}_{\!\mu}\delta^\mu_\nu=0.\notag
\end{align}
\end{subequations} 

On the other hand, the divergence with respect to the subscript of the torsionless equation provides an equation of motion for a free vierbein field in the local and global spacetime manifold.
The Bianchi identity can be written as
\begin{align}
d^{~}_\www\TTT^a=\(d^{~}_\www\hspace{.1em}d^{~}_\www\)\eee^a=0\overset{\text{(\ref{LaplacewG})}}{\iff}
\(d^{~}_\Gamma\hspace{.1em}d^{~}_\Gamma\)\eee^a=0,
\end{align}
where we have used notations and the soldering relation given in \textbf{Appendix \ref{Soldering}}.
Thus, we obtain that
\begin{align}
 g_{~}^{\mu\nu}\partial^{~}_{\mu}\partial^{~}_{\nu}\hspace{.1em}\E^{a}_{\rho}(x)\simeq0\simeq
\eta_{~}^{\bcdots}\partial^{~}_{\bcdot}\partial^{~}_{\bcdot}\hspace{.1em}\E^{a}_{\rho}(x(\xi)),\label{deltaE}
\end{align}
where $\bullet\!\simeq\!\bullet$ means equality as free fields with $\cG\!=\!0$. 
The fundamental solution of equation (\ref{deltaE}) is
\begin{align}
{G^a_{\E}}(\xi):=\frac{\E_\mu^{(a)}(\xi)}
{-(\xi^0)^2+(\xi^1)^2+(\xi^2)^2+(\xi^3)^2-i\delta}\(\simeq
\frac{\E_\mu^{(a)}(x)}
{-g_{\mu\nu}x^\mu x^\nu-i\delta}\)\!,
\label{GE}
\end{align}
where ${\E}_{\mu}^{(a)}$ is polarisation vectors of the vierbein field.
Polarisation vectors are components of a spin-one vector defined in $\TsMM$ such that
\begin{align*}
{\eee}^{(a)}:={\E}^{(a)}_\mu\hspace{-.1em}(x)\hspace{.2em}dx^\mu,
\end{align*}
where $a=1,2$ shows two independent polarisation vectors.
We interpret the subscript as a vector component concerning the global spacetime manifold, and the superscript as an inner degree of freedom corresponding to two polarisation vectors.
If we interpret a four-fold field $\E^a_\mu$ as a component of the global vector defined in $\TsMM$ with respect to the subscript $\mu$, the local $\SO(1,3)$ symmetry is an inner degree of freedom with respect to the superscript ``$a$'', which is interpreted as a degree of polarisation of a four-fold field.
We can construct a spin-two polarisation vector of the metric tensor by coupling the angular momenta\!\cite{Kurihara:2022green}.
We note that the free spin connection field propagates in the inertial manifold; on the other hand, the free vierbein field does in the global spacetime manifold.
We interpret that the long-range phenomena, such as gravitational wave propagation, are owned by the vierbein field.

%
%
\subsection{External field}
The \QGED Lagrangian includes four external physical fields: the vierbein field, the spin connection field, the photon and the electron.
Additionally, the \QGED has ghost/anti-ghost fields.
Among them, the photon and the electron are defined purely in inertial space.
The spin connection and the local ghost/anti-ghost fields are also objects defined in inertial space according to the romanisation of the subscript.
The spin connection field propagates as a wave in the inertial bundle according to equation (\ref{deltaomega}).
On the other hand, the vierbein field $\E_\mu^a(x\in\MM)$ is an object intrinsically defined in the global spacetime and propagating in it according to the wave equation (\ref{deltaE}).
The vierbein field has a dual role in the \QGED: it is a dynamic field that propagates in global spacetime, and it serves as a transformation function that converts a Greek index into a Roman index.
An electron and a photon do not couple directly to the vierbein field via the interaction Lagrangian.
This study considers the vierbein field as an asymptotic state in the asymptotically flat spacetime.

\begin{figure}[t]
 \begin{center} 
   \includegraphics[width=9cm]{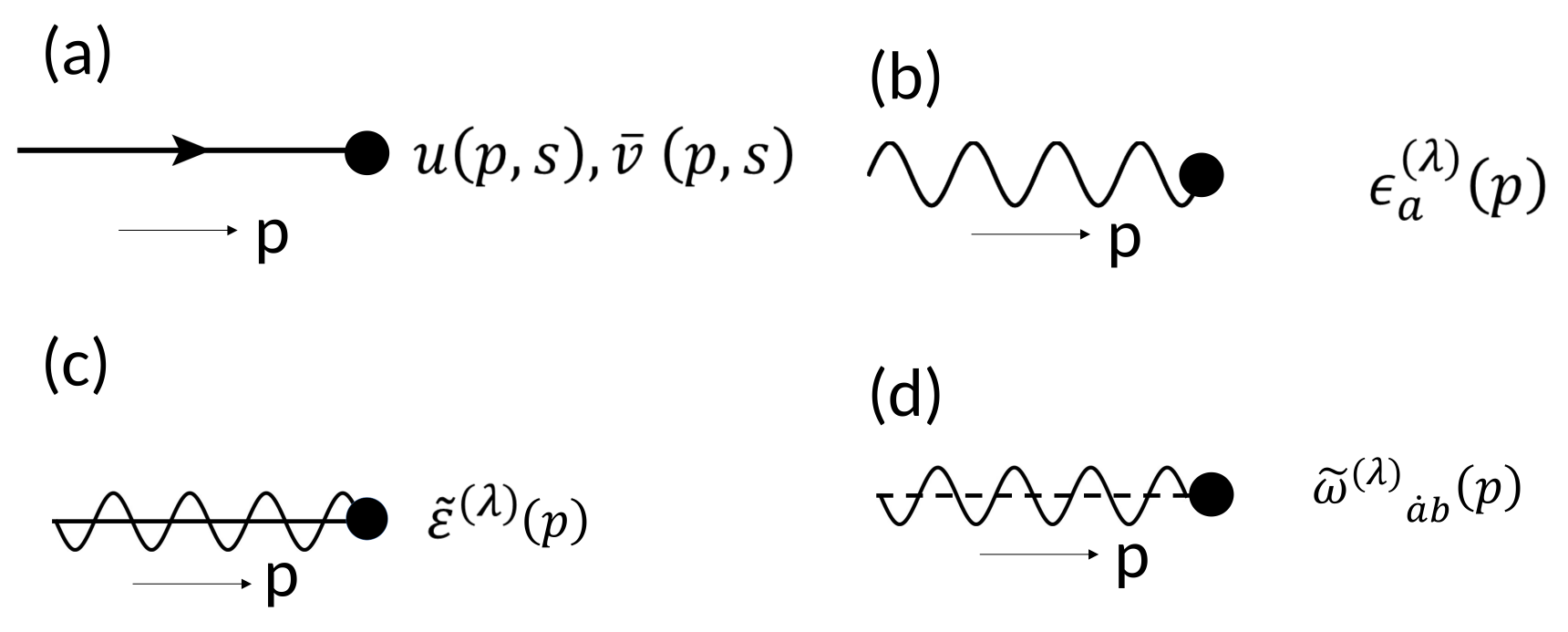}
 \caption{\label{fig1}\small
Figures provide pictorial representations of the external QEDG fields.
A solid circle shows the interaction vertex.
We take the momentum of the external field in the incoming direction.} 
 \end{center} 
\end{figure}

\paragraph{Spinor (Figure \ref{fig1}-(a)):}
According to the standard perturbation method, we introduce electron and positron wave functions as follows:
Spinor fields of electron and positron, $u(p,s)$ and  $v(p,s)$, are positive  and negative frequency parts of the plane-wave solution of the Dirac equation with four momentum $p_{~}^a=(p^{~}_0,\vp)\in V^1(\TtM)$  and spin $s$ as follows:
\begin{align}
\psi(\vxi\hspace{.2em})=:\frac{1}{(2\pi\hbar)^{3}}\int\frac{d^3\vp}{\sqrt{2E_p}}
\sum_{s=1,2}\(
a^{~}_e(\vp,s)u(\vp,s)e^{-i\vxi\cdot\vp/\hbar}+
b^{\dagger}_e(\vp,s)v(\vp,s)e^{+i\vxi\cdot\vp/\hbar}
\)\!,\label{spnorm}
\end{align}
where $E_p:=\sqrt{\vp\cdot\vp+m^2}$ is an on-shell energy, and $a^{~}_e(p,s)$ and $b^{\dagger}_e(p,s)$ are an electron's creation and annihilation operators, respectively.
The canonical commutation relation of the creation/annihilation operators is
\begin{align}
\left\{a^{~}_e(\vp,s),a^{\dagger}_e(\vp\hspace{.1em}',s')\right\}=
\left\{b^{~}_e(\vp,s),b^{\dagger}_e(\vp\hspace{.1em}',s')\right\}=(2\pi\hbar)^3
\delta^3(\vp-\vp\hspace{.1em}')\delta_{ss'}\!,\label{ccrpsi}
\end{align}
and otherwise zero.
Thus, the physical dimension of electron's creation and annihilation operators are 
\begin{align*}
\text{(\ref{ccrpsi})}\implies
[a^{~}_e]_\text{pd}=[b^{~}_e]_\text{pd}=L^{3/2}\!.
\end{align*}
The creation operator creates the electron spinor in the momentum space.
The time-independent spinor $u$ has the physical dimension $E^{\hlf}$\!, and the integration measure has $L^{-3}E^{-\hlf}$\!.
As a consequence, the operator (\ref{spnorm}) consistently produces a quantum with the physical dimension $L^{-2/3}$\!.

Spinors are normalised as 
\begin{align*}
\bar{u}(\vp,s)\hspace{.1em}\gamma_0\hspace{.1em}u(\vp,s')=2E_p\hspace{.1em}\delta_{ss'},~&\textrm{and}~~
\bar{v}(\vp,s)\hspace{.1em}\gamma_0\hspace{.1em}v(\vp,s')=2E_p\hspace{.1em}\delta_{ss'}\!.
\end{align*}
The polarisation sums of an electron and a positron provides
\begin{align}
\sum_su(\vp,s)\bar{u}(\vp,s)=\sp+m,~&\textrm{and}~~
\sum_sv(\vp,s)\bar{v}(\vp,s)=\sp-m.\label{fnorm}
\end{align}

\paragraph{Photon (Figure \ref{fig1}-(b)):}
The Fourier transformation of transversely polarised $U(1)$ gauge field $\Aa^{T}_a(\vxi)$ provides a polarisation vector of a physical photon, such that: 
\begin{align}
\Aa^{T}_a(\vxi\hspace{.2em})=:\frac{1}{(2\pi\hbar)^{3}}
\int\frac{d^3\vk}{\sqrt{2k_0}}
\sum_{\lambda=1,2}\epsilon_a^{(\lambda)}
\({a}^{~}_\gamma(\vk,\lambda)e^{-i\vxi\cdot\vk/\hbar}+
{a}_\gamma^\dagger(\vk,\lambda)e^{+i\vxi\cdot\vk/\hbar}
\)\!,\label{Acreate}
\end{align}
where $k_0$ is an on-shell energy such that $k_0^2-\vk\cdot\vk=0$.  
The creation and annihilation operators of a photon with momentum $k$ and polarisation $\lambda$ are defined as
\begin{align*}
\hat{a}^{~}_\gamma(\vk,\lambda):={a}^{~}_\gamma(\vk,\lambda)/\hbar^{\hlf}
~~\text{and}~~ 
\hat{a}_\gamma^\dagger(\vk,\lambda):={a}_\gamma^\dagger(\vk,\lambda)/\hbar^{\hlf}\!.
\end{align*}
Owing to the canonical quantisation conditions, we obtain their physical dimensions as
\begin{align}
\left[\hat{a}^{~}_\gamma(\vk,\lambda),\hat{a}_\gamma^\dagger(\vk\hspace{.1em}',\lambda')\right]&=
(2\pi\hbar)^3\delta^3\hspace{-.2em}\(\vk-\vk\hspace{.1em}'\)\delta_{\lambda'\lambda}\implies
[{a}^{~}_\gamma]_\text{pd}=[a^{\dagger}_\gamma]_\text{pd}=L^2E^\hlf\!.\label{CQe}
\end{align}
Polarisation vector $\epsilon_a^{(\lambda)}$ is defined to have null physical dimension.
Thus, after the multiplication of $\hbar$ on both sides of (\ref{Acreate}), the creation operator creates a photon with the physical dimension $E$ together with the integration measure.

A photon polarisation vector is normalized as
\begin{align}
\sum_{\lambda=1,2}\epsilon_a^{(\lambda)}(k)\epsilon_b^{(\lambda)}(k)&=
-\eta_{ab}+\(1-\xi_A\)\frac{k_ak_b}{\eta_\bcdots k^\bcdot k^\bcdot},\label{Anorm}
\end{align}
in the covariant gauge (\ref{Acovg}).

\paragraph{Vierbein (Figure \ref{fig1}-(c)):}
We consider an asymptotic state of the vierbein field in the asymptotically flat spacetime as $g_{\mu\nu}\xrightarrow{\vx\rightarrow\infty}\textup{diag}\(1,-1,-1,-1\)$. 
The Fourier transformation of the vierbein field of the fundamental solution (\ref{GE}) provides a polarisation vector in the momentum space, such that:
\begin{align}
\lp^{-1}\E_\mu^{(\lambda)}(x)=:\frac{1}{(2\pi\hbar)^{3}}
\int\frac{d^3\vp}{\sqrt{2p_0}}\hspace{.2em}
\tilde{\E}_\mu^{(\lambda)} \(
\Pmass{a}_{\E}^{~}(p,\lambda)e^{-i\vx\cdot\vp/\hbar}+
\Pmass{a}_{\E}^\dagger(p,\lambda)e^{+i\vx\cdot\vp/\hbar}
\)\!,\label{FourierE}
\end{align}
where $p_0$ is an on-shell energy such that $p_0^2-\vp\cdot\vp=0$.  
We note that the vacuum concerning the vierbein field is not a flat metric in general, and the Fourier transformation and the momentum space in the curved spacetime are not trivial\!\cite{Kurihara:2022green}.
The vierbein fields appear in the integral kernel of the Fourier transformation as
\begin{align*}
x\cdot p:=g_{\mu\nu}(x)\hspace{.1em}x^\mu p^\nu=
\eta_\bcdots\hspace{.1em}\E^\bcdot_\mu(x)\E^\bcdot_\nu(x)x^\mu p^\nu\!.
\end{align*}
Thus, in curved space time, the integration (\ref{FourierE}) is not simply the Fourier transformation: the integration kernel contains the vierbein itself.
We consider the Fourier transformation only in the asymptotic frame, which is assumed to be flat.

The equation of motion of the free vierbein field in the flat spacetime, shown in (\ref{deltaE}),  is the same as that of a photon; thus, the Fock space of a free (asymptotic) vierbein operator is the same as that of a photon.
We define the creation and the annihilation operators as
\begin{align*}
\hat{a}^{~}_\E(p,\lambda):=\kE^{\hspace{-.1em}\hlf}{a}^{~}_\E(p,\lambda)
~~\text{and}~~ 
\hat{a}_\gamma^\dagger(p,\lambda):=\kE^{\hspace{-.1em}\hlf}{a}_\gamma^\dagger(p,\lambda).
\end{align*}
The canonical quantisation condition requires an equal-time commutation relation on creation and annihilation operators of the physical vierbein fields, such as\!\cite{nakanishi1990covariant}
\begin{align}
\left[\hat{a}^{~}_\E(p,\lambda),\hat{a}_\E^\dagger(p',\lambda')\right]&=(2\pi\hbar)^3
\delta^3\hspace{-.2em}\(\vp-\vp\hspace{.1em}'\)
\delta_{\lambda'\lambda}\implies
[{a}^{~}_\E]_\text{pd}=L_{~}^2E^{-\hlf}\!.\label{CRaEaE}
\end{align}
If we set a polarisation vector $\tilde{\E}_\mu^{(\lambda)}$ to an object with a null physical dimension, the creation operator together with the integration measure creates a vierbein-quantum with the energy dimension.
Thus, the operator ``$\lp^{-1}\E_\mu^{(\lambda)}(x)$'', which creates a vierbein quantum with the inverse length dimension in configuration space, provides the operator ``$\Pmass{a}_{\E}^{~}(p,\lambda)$'', which creates a vierbein quantum with the energy dimension (together with the integration measure) in momentum space, after the Fourier transformation.

We introduce four polarisation vectors of the asymptotic vierbein field propagating along, e.g., the $x^1$-axis, such as
\begin{align*}
&{\tilde{\E}}_{\mu}^{(\lambda)}:=\frac{1}{\sqrt{2}}
\left\{
\begin{array}{lc}
(1,\hspace{.5em}i,~0,\hspace{.8em}0), & (\lambda=0),\\
(i,\hspace{.5em}1,~0,\hspace{.8em}0), & (\lambda=1),\\
(0,~0,~1,\hspace{.8em}1), & (\lambda=2),\\
(0,~0,~1,-1), & (\lambda=3).
\end{array}
\right.
\end{align*}
with a circular polarisation.
The vierbein field is constructed from polarisation vectors, such as 
\begin{align*}
\E_\mu^a:=&\tilde{\E}_\mu^{(\lambda=a)},
\intertext{which provides a flat metric tensor}
g_{\mu\nu}\hspace{.3em}=&\eta^{~}_\bcdots\E_\mu^\bcdot\E_\nu^\bcdot=
\textup{diag}\(1,-1,-1,-1\)\!.
\end{align*}
This relation implies the normalisation of the polarisation vector as
\begin{align}
\sum_{\lambda,\lambda'}\eta^{~}_{\lambda'\!\lambda}\tilde{\E}_\mu^{(\lambda)}(p)\hspace{.1em}
\tilde{\E}_\nu^{(\lambda')}(p)&=g_{\mu\nu}.\label{EEnorm}
\end{align}
Each vierbein polarisation vector represents the spin-one state, and a symmetric product (\ref{EEnorm}) concerning two Lorentz indexes constructs a spin-two state of the quantised metric tensor.

Two polarisation vectors, 
\begin{align*}
\ppp^{(\lambda=\bm{+})}:={\tilde{\E}}_{\mu}^{(2)}dp^\mu=\frac{1}{\sqrt{2}}\(dp^2+dp^3\)
~&\textrm{and }~~
\ppp^{(\lambda=\bm{\times})}:={\tilde{\E}}_{\mu}^{(3)}dk^\mu=\frac{1}{\sqrt{2}}\(dp^2-dp^3\)\!,
\end{align*}
have a dynamic degree and are physically observable in the quantum field theory, similar to the polarisation vector of the transversely polarised photon in the QED.
We use representations $\lambda=2=:\bm{+}$ and $\lambda=3=:\bm{\times}$ according to the standard convention.
Two polarisation vectors,  
\begin{align*}
\tilde{\E}^{(s)}_\mu:=\frac{1}{\sqrt{2}}\(\tilde{\E}^{(0)}_\mu-i\tilde{\E}^{(1)}_\mu\)\!,~&\textrm{and}~~
\tilde{\E}^{(l)}_\mu:=\frac{1}{\sqrt{2}i}\(\tilde{\E}^{(0)}_\mu+i\tilde{\E}^{(1)}_\mu\)\!,
\end{align*}
correspond to a scalar vierbein and a longitudinally polarised vierbein, respectively.
Polarisation vectors satisfy the covariant constraint for the on-shell momentum $\tilde{p}_\mu=(\tilde{p},0,0,\tilde{p})$ as
\begin{align*}
g^{\mu\nu}\tilde{p}_\mu\hspace{.1em}\tilde{\E}^{(\lambda)}_\nu=0
\end{align*}
for $\lambda\in\{\bm{+},\bm{\times},s\}$.

\paragraph{Spin connection (Figure \ref{fig1}-(d)):}
The discussion is entirely parallel to that of the photon field.
The Fourier transformation of the spin connection field yields a polarisation vector in the momentum space such that:
\begin{align*}
\omega^{(\lambda)\hspace{.3em}b}_{\hspace{1.em}\dot{a}}(\xi)=:\frac{1}{(2\pi\hbar)^{3}}
\int\frac{d^3\vp}{\sqrt{2p_0}}
\sum_{\lambda=1,2}\tilde\omega^{(\lambda)\hspace{.3em}b}_{\hspace{1.em}\dot{a}}(\vp)\(
a_{\omega}^{~}(\vp,\lambda)e^{-i\vxi\cdot \vp/\hbar}+
a_{\omega}^\dagger(\vp,\lambda)e^{+i\vxi\cdot \vp/\hbar}
\)\!.
\end{align*}
In contrast to the Vierbein field defined in the global manifold, the spin connection field is described in the local inertial manifold after Romanising the subscripts. Its Fourier transform is, therefore, well-defined in inertial space.
The equation of motion for the free spin connection field shown in (\ref{deltaomega}) is the same as that of a photon; thus, the Fock space of a free vierbein operator is the same as that of a photon. 
We set a canonical quantisation condition of an equal-time commutation relation on creation and annihilation operators of physical spin connection fields as
\begin{align*}
\left.\left[\hat{a}^{~}_\omega(\vp,\lambda),\hat{a}_\omega^\dagger(\vp',\lambda')\right]\right|_{p^{~}_0={p^{~}_0}'}&=
(2\pi\hbar)^3\hspace{.1em}\delta^3\hspace{-.2em}\(\vp-{\vp'}\)\delta_{\lambda'\lambda}
%
%
\implies
[{a}^{~}_\omega]_\text{pd}=[a^{\dagger}_\omega]_\text{pd}=L^2E^\hlf\!,
~\text{where}~\hat{a}^{~}_\omega:={a}^{~}_\omega/\hbar^{\hlf}\!.
\end{align*}
Consequently, a polarisation vector of the spin connection field in the momentum space has null physical dimension.

We can construct polarisation vectors of the spin-two field for the spin connection in the momentum space.
For a spin connection momentum $p^{~}_a:=(p^{~}_0,p\sin{\vartheta}\cos{\varphi},p\sin{\vartheta}\sin{\varphi},p\cos{\vartheta})$, we introduce three linearly polarised  vectors, such that:
\begin{align*}
\tilde{\varepsilon}_{a}^{(\lambda=1)}&:=
\(0,\cos{\vartheta}\cos{\varphi},\cos{\vartheta}\sin{\varphi},-\sin{\theta}\)\!,\\
\tilde{\varepsilon}_{a}^{(\lambda=2)}&:=\(0,-\sin{\varphi},\cos{\varphi},0\)\!,\\
\tilde{\varepsilon}_{a}^{(\lambda=3)}&:=
\frac{1}{\sqrt{p_0^2-p^2}}\(p,p_0\sin{\vartheta}\cos{\varphi},p_0\sin{\vartheta}\sin{\varphi},p_0\cos{\vartheta}\)\!.
\end{align*}
The polarisations $\lambda=1,2$ are the transverse components of the physical spin connection, and $\lambda=3$ corresponds longitudinal one appearing in the virtual state in the propagator.
The circularly polarised vectors are constructed using them as
\begin{subequations}
\begin{align}
\tilde{\varepsilon}_{a}^{(+)}&:=\frac{1}{\sqrt{2}}\(\tilde{\varepsilon}_{1}^{(1)}+i\tilde{\varepsilon}_{a}^{(2)}\)\label{wlambda1},\\
\tilde{\varepsilon}_{a}^{(-)}&:=\frac{1}{\sqrt{2}}\(\tilde{\varepsilon}_{1}^{(1)}-i\tilde{\varepsilon}_{a}^{(2)}\)\label{wlambda2},\\
\tilde{\varepsilon}_{a}^{(l)}&:=\tilde{\varepsilon}_{1}^{(3)}\label{wlambda3}.
\end{align}
\end{subequations}
They satisfy the covariant gauge-fixing conditions of $\eta^\bcdots p_\bcdot\tilde{\varepsilon}_\bcdot^{(\lambda)}=0$.

We can construct polarisation vectors of the spin connection fields owing to spin-one polarisation vectors (\ref{wlambda1})$\sim$(\ref{wlambda3}) as
\begin{subequations}
\begin{align}
\tilde{\omega}^{(0)}_{\hspace{1em}\dot{a}b}&:=
\tilde{\varepsilon}_{a}^{(+)}\tilde{\varepsilon}_{b}^{(-)}-
\tilde{\varepsilon}_{a}^{(-)}\tilde{\varepsilon}_{b}^{(+)},\label{wlambda21}\\
\tilde{\omega}^{(+1)}_{\hspace{1em}\dot{a}b}&:=
\tilde{\varepsilon}_{a}^{(+)}\tilde{\varepsilon}_{b}^{(l)}-
\tilde{\varepsilon}_{a}^{(l)}\tilde{\varepsilon}_{b}^{(+)},\label{wlambda22}\\
\tilde{\omega}^{(-1)}_{\hspace{1em}\dot{a}b}&:=
\tilde{\varepsilon}_{a}^{(-)}\tilde{\varepsilon}_{b}^{(l)}-
\tilde{\varepsilon}_{a}^{(l)}\tilde{\varepsilon}_{b}^{(-)},\label{wlambda23}
\end{align}
where a value of $\dot{a}$ on the spin connection equals to that of $a$ on the polarisation vector.
\end{subequations}
Here, we utilise a convention
\begin{align*}
\tilde{\omega}^{(\lambda)}_{\hspace{1em}\dot{a}b}:=\tilde{\omega}^{(\lambda)\hspace{.3em}\hat{a}\bcdot}_{\hspace{1em}\hat{a}}\eta^{~}_{\bcdot b}\big|_{\hat{a}\rightarrow\dot{a}}
~&\text{and}~~
\tilde{\omega}^{(\lambda)}_{\hspace{1em}a\dot{b}}:=\tilde{\omega}^{(\lambda)\hspace{.3em}\bcdot\hat{b}}_{\hspace{1em}\hat{b}}\eta^{~}_{\bcdot a}\big|_{\hat{b}\rightarrow\dot{b}}
\end{align*}
to distinguish two different polarisation vectors concerning two subscripts. 
They are anti-symmetric concerning two suffixes and satisfy the covariant gauge condition as
\begin{align*}
\tilde{\omega}^{(\lambda)}_{\hspace{1em}\dot{a}b}=-\tilde{\omega}^{(\lambda)}_{\hspace{1em}b\dot{a}}~&\text{and}~~
\eta^\bcdots p^{~}_\bcdot\hspace{.1em}\tilde{\omega}^{(\lambda)}_{\hspace{1em}\dot{a}\bcdot}=0.
\end{align*}
Polarisation vectors $\tilde{\omega}^{(+1)}_{\hspace{1em}\dot{a}b}$, $\tilde{\omega}^{(0)}_{\hspace{1em}\dot{a}b}$ and $\tilde{\omega}^{(-1)}_{\hspace{1em}\dot{a}b}$ provide the spin-two polarisation tensors with a helicity state $-1$, $0$ and $+1$.

Polarisation vectors (\ref{wlambda21})$\sim$(\ref{wlambda23}) provide the polarisation sum such that
\begin{align}
\sum_{\lambda=-1,0,1}
\tilde{\omega}^{(\lambda)}_{\hspace{1em}\dot{a}b}\hspace{.2em}
\tilde{\omega}^{(\lambda)}_{\hspace{1em}\dot{c}d}&=
\(-\eta_{ac}^{~}+\(1-\xi^{~}_\omega\)\frac{p^{~}_a p^{~}_c}
{\eta_{~}^\bcdots\hspace{.1em} p^{~}_\bcdot p^{~}_\bcdot}\)
\(-\eta_{bd}^{~}+\(1-\xi^{~}_\omega\)\frac{p^{~}_b p^{~}_d}
{\eta_{~}^\bcdots\hspace{.1em} p^{~}_\bcdot p^{~}_\bcdot}\)\notag\\
&-
\(-\eta_{ad}^{~}+\(1-\xi^{~}_\omega\)\frac{p^{~}_a p^{~}_d}
{\eta_{~}^\bcdots\hspace{.1em} p^{~}_\bcdot p^{~}_\bcdot}\)
\(-\eta_{bc}^{~}+\(1-\xi^{~}_\omega\)\frac{p^{~}_b p^{~}_c}
{\eta_{~}^\bcdots\hspace{.1em} p^{~}_\bcdot p^{~}_\bcdot}\)\!,
\label{Wnorm}
\end{align}
with the covariant gauge-fixing (\ref{Wgauge}).

\paragraph{Ghosts (no figure):}
Although the ghost does not appear as an external field, we give here the canonical quantisation condition for the ghost and the anti-ghost field.
We define generalised four-momenta concerning ghost/anti-ghost fields as
\begin{align*}
\text{(\ref{Lgrgh})}\rightarrow\pi^a_\chi:=
\frac{\delta\LL_{\GR;gh}}{\delta\(\partial^{~}_a\chi\)}=\frac{1}{\lpt}
c^{\bcdot}\(\partial^{~}_{\!\bcdot}\bar{\chi}\)c^a
~~&\text{and}~~
\pi^a_{\bar\chi}:=\frac{\delta\LL_{\GR;gh}}{\delta\(\partial^{~}_a\bar\chi\)}=\frac{1}{\lpt}\(
-c^{\bcdot}\(\partial^{~}_{\!\bcdot}{\chi}\)c^a
+\cG\hspace{.1em}\bar\chi\hspace{.1em}\omega_{\dot\sigma\bcdot}\hspace{.1em}c^\bcdot c^a\).
\end{align*}
The Fourier transformation of ghost/anti-ghost fields provides those in the momentum space, such that:
\begin{align}
\lp^{-1}\hspace{.1em}\chi(\xi)=:\frac{1}{(2\pi\hbar)^{3}}
\int\frac{d^3\vp}{\sqrt{2p_0}}\hspace{.2em}
\tilde{\chi}(p)\(
{a}_{\chi}^{~}(p)e^{-i\vxi\cdot\vp/\hbar}+
{a}_{\chi}^\dagger(p)e^{+i\vxi\cdot\vp/\hbar}
\)~~\text{and}~~(\chi\rightarrow\bar\chi,~\tilde\chi\rightarrow\tilde{\bar\chi}).
\end{align}
Commutation relations for creation and annihilation operators of ghost/anti-ghost fields are
\begin{align*}
\left\{a^{~}_\chi(p),a_\chi^\dagger(p')\right\}&=
\left\{a^{~}_{\bar\chi}(p),a_{\bar\chi}^\dagger(p')\right\}=
(2\pi\hbar)^3
\delta^3\hspace{-.2em}\(\vp-\vp\hspace{.1em}'\)\!.
\end{align*}
Consequently, ghost/anti-ghost fields in the momentum space have a null physical dimension.
\vskip 5mm\noindent
Fields appearing in the \QGED and their physical dimensions are summarised in Table \ref{table3}.
\begin{table}[b]
\begin{center}
\caption{\label{table3}\small
Fields in the \QGED and its physical dimension in the Lagrangian and Feynman rules.}
\vskip 2mm
\begin{tabular}{ll||ll|ll}	
\multicolumn{2}{c||}{Field} & configurations space &dim. & momentum space &dim.\\
\hline
\multirow{2}{*}{section}
&electron& \hspace{3em}$\psi(\xi)$ & $L^{-3/2}$& \hspace{3em}$u(p,s)$ & $E^{\hlf}$\\
&vierbein  &\hspace{2.3em}$\lp^{-1}\E^a_\mu(x)$ &$L^{-1}$& \hspace{2.5em}$\tilde{\E}^{(\lambda)}_\mu\hspace{-.1em}(p)$&$\textit{1}$ \\
\hline
\multirow{2}{*}{connection}
&photon& \hspace{3em}$\Aa_a(\xi)$ & $L^{-1}$ & \hspace{3em}$\epsilon^{(\lambda)}_a\hspace{-.1em}(p)$ & {\it 1}\\
&spin connection & \hspace{3em}$\omega_c^{~ab}(\xi)$ & $L^{-1}$ &\hspace{3em}$\tilde{\omega}^{(\lambda)}_{\hspace{.7em}ab}\hspace{-.1em}(p)$  &{\it 1}\\
\hline
\multicolumn{2}{c||}{ghost/anti-ghost}&\hspace{2.0em}$\lp^{-1}\chi(\xi)$/$\bar\chi(\xi)$&$L^{-1}$&\hspace{2.5em}$\tilde\chi(p)$/$\tilde{\bar\chi}(p)$&$\textit{1}$
\end{tabular}
\end{center}
\end{table}

%
%
\subsection{Vertices}
\begin{figure}[bt]
\begin{center}
    \includegraphics[width=9cm]{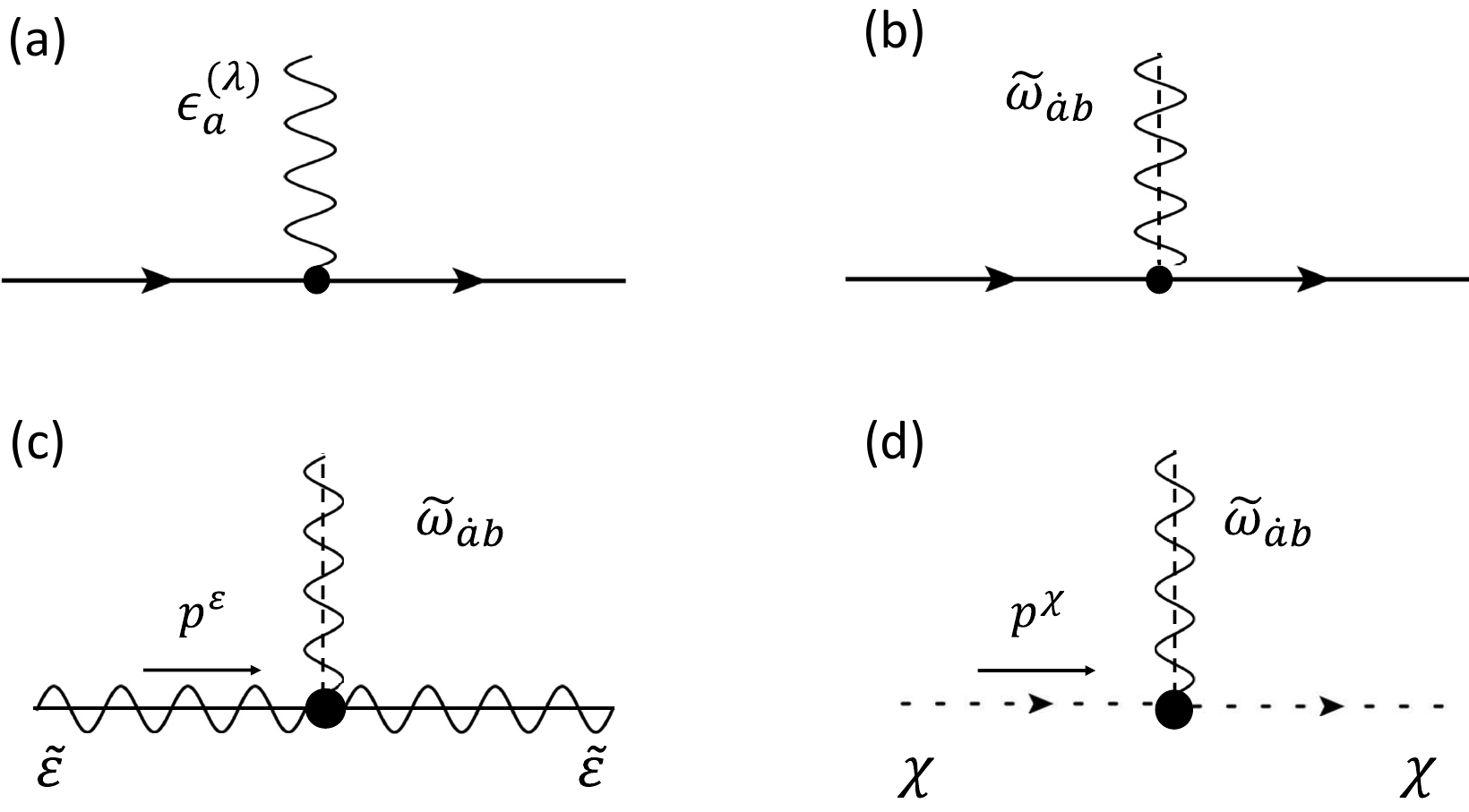}
 \caption{\label{fig2}\small
Figures provide pictorial representations of the QEDG vertices for (a) electron-photon vertex, (b) electron-spin connection vertex, (c) spin connection-vierbein vertex, and (c) ghost-spin connection vertex.
Vertex Feynman rules are given in (\ref{v1})$\sim$(\ref{v4}).
}
\end{center}
\end{figure}

\paragraph{(spinor)-(photon) (Figure \ref{fig2}-(a)):}
Although the (spinor)-(photon) vertex Feynman rule is found in standard textbooks of quantum field theory, we provide it here to illustrate our convention between the interaction Lagrange and the Feynman rule.
The vertex rule provides a multiplicable factor after truncation of the wave functions and the creation/annihilation operators.

An electromagnetic interaction vertex truncated the electron fields is provided as
\begin{align}
\text{(\ref{LMTfreeint})}\big|^{~}_{\omega\rightarrow0}\rightarrow
{V}_{\epsilon\psi}&:=\frac{e}{\hbar^\hlf}\hspace{.1em}\gamma^\bcdot\Aa^{~}_{\bcdot}\!,\label{Vepsi0}
\end{align}
where we omit $\SU(\!N\!)$ generator since it is $U(1)$ gauge.
Here, $\hbar^\hlf$ appease owing to (\ref{alphae2}).
After the Fourier transformation, we obtain it in the momentum space as 
\begin{align*}
\text{(\ref{Vepsi0})}\xrightarrow{\text{F.T.}}
\widetilde{V}_{\epsilon\psi}&=e\hspace{.1em}\gamma^\bcdot\epsilon^{(\lambda)}_{\hspace{.3em}\bcdot}\hat{a}^{~}_\gamma(\vk,\lambda).
\end{align*}
Thus, we obtain the vertex rule by truncating the polarisation vector and the creation operator as
\begin{align*}
\textrm{(photon)--(electron)$^2$}&=e\hspace{.1em}\gamma^a\!.
\end{align*}
The truncated polarisation vector from the vertex is moved to the numerator of the photon propagator.

\paragraph{(electron)-(spin connection) (Figure \ref{fig2}-(b)):}
A gravitational interaction vertex of the electron truncated the electron fields is provided as
\begin{align}
\text{(\ref{LMTfreeint})}\big|^{~}_{\Aa\rightarrow0}\rightarrow
\widehat{V}_{\hat\omega\psi}&:=-i\frac{\cG}{4}\gamma^\bcdot\omega_\bcdot^{~\stars}{\sigma_\stars}
=\frac{\cG}{2}\gamma^\bcdot\omega_\bcdot^{~\stars}\gamma_\star\gamma_\star
=\frac{\cG}{2}\gamma_{~}^\bcdot\omega_{\dot{\Sigma}}^{\hspace{.5em}\bcdot}\hspace{.1em}\eta^{~}_\bcdots,
\label{Vwpsi0}
\end{align}
where we use 
$\eta^{\hat{a}\bcdot}\gamma_\bcdot^{~}\hspace{.1em}\gamma_{\hat{a}}^{~}=\delta^{\hat{a}}_{\hat{a}}$ for $\hat{a}\in\{0,1,2,3\}$; thus,
\begin{align*}
\sum_{a,b,c}\gamma^a_{~}\hspace{.1em}\gamma_b^{~}\hspace{.1em}\gamma_c^{~}\hspace{.1em}\omega_{a}^{\hspace{.3em}bc}~\overset{\text{(\ref{wconst})} }{=}~
\sum_{a,b,c}\delta^a_b\hspace{.1em}\gamma_c^{~}\hspace{.1em}\omega_{a}^{\hspace{.3em}bc}=
\gamma_\bcdot^{~}\hspace{.1em}\omega_{\dot{\Sigma}}^{\hspace{.5em}\bcdot}.
\end{align*}
We note that the scaled spin connection field and its in the momentum space as
\begin{align}
\cG\omega=\hat\omega \xrightarrow{\text{F.T.}}
\cG\tilde\omega=\tilde{\hat\omega}.
\end{align}
Consequently, we obtain the vertex rule as
\begin{align*}
\text{(spin connection)-(electron)}^2&=\cG\gamma_{~}^a\!.
\end{align*}
The spin connection-form is the one-form object in $\TsM$ and the rank-two tensor in $\TM$.
Therefore, $\omega_{\dot{a}}^{\hspace{.4em}b}$ is the Lorentz tensor in $\TM$, and  $\gamma_{~}^\bcdot\tilde\omega_{\dot\Sigma\bcdot}$ is the Clifford algebra values Lorentz scalar.

\paragraph{(vierbein)-(spin connection) (Figure \ref{fig2}-(c)):}
The pure gravitational Lagrangian density given in (\ref{LGrfreeint}) provides the interaction vertex among vierbein and spin connection fields in the configuration space with the unscaled field as
\begin{align}
\text{(\ref{LGrfreeint})}\rightarrow
{\hat{V}}_{\omega\E}&=
-\frac{1}{2\lp^2}\eta_\bcdots^{~}
\(\delta^{a}_{c}\hspace{.1em}\delta^{b}_{d}-\delta^{b}_{c}\hspace{.1em}\delta^{a}_{d}\)
{\hat\omega}_{a}^{\hspace{.4em}c\bcdot}\hspace{.1em}{\hat\omega}_{b}^{\hspace{.4em}d\bcdot}
=\frac{1}{2\lp^2}\eta_\bcdots^{~}\hspace{.1em}{\hat\omega}_{\star}^{\hspace{.4em}\times\bcdot}\hspace{.1em}{\hat\omega}_{\times}^{\hspace{.4em}\star\bcdot}
-\frac{1}{2\lp^2}\eta_\bcdots^{~}\hspace{.1em}\hat{\omega}_{\dot\Sigma}^{\hspace{.4em}\bcdot}\hspace{.1em}{\hat\omega}_{\dot\Sigma}^{\hspace{.4em}\bcdot}.\label{Lintddww}
\end{align}
Given the vierbein field, we can solve the torsionless equation (\ref{torsionlesscp}) algebraically using the unscaled fields as
\begin{align}
\text{(\ref{torsionlesscp})}&\xrightarrow{\cG\omega\rightarrow\hat\omega}
\hat\omega_{b\hspace{.3em}c}^{\hspace{.3em}a}-\hat\omega_{c\hspace{.3em}b}^{\hspace{.3em}a}
=-\E^a_\mu
\(\partial^{~}_c\E_b^\mu-\partial^{~}_b\E_c^\mu\)\!.\label{W2E}
\end{align}
The first term in the l.h.s. of  (\ref{Lintddww}) can be written as
\begin{align}
\eta^{~}_{\bcdots}\hspace{.2em}{\hat\omega}_{\star}^{\hspace{.4em}\times\bcdot}\hspace{.1em}{\hat\omega}_{\times}^{\hspace{.4em}\star\bcdot}&\overset{\text{(\ref{wconst})}}{=}
\eta^{~}_{\bcdots}\hspace{.2em}\eta_{~}^\stars\eta_{~}^\timess
\hat\omega_{\star\hspace{.2em}\times}^{\hspace{.3em}\bcdot}
\(\hat\omega_{\star\hspace{.2em}\times}^{\hspace{.3em}\bcdot}-\hat\omega_{\times\hspace{.2em}\star}^{\hspace{.3em}\bcdot}\)
\overset{\text{(\ref{W2E})}}{=}
-\eta^{~}_{\bcdots}\hspace{.2em}\eta_{~}^\stars\eta_{~}^\timess
\hat\omega_{\star\hspace{.2em}\times}^{\hspace{.3em}\bcdot}\hspace{.2em}
\E^\bcdot_\mu\hspace{-.1em}\(\partial^{~}_\star\E_\times^\mu-\partial^{~}_{\hspace{-.1em}\times}\E_\star^\mu\),\notag\\
&\overset{\text{(\ref{wconst})}}{=}
-\hat\omega_{\dot{\Sigma}}^{\hspace{.4em}\times}\hspace{.2em}
\E^\bcdot_\mu\hspace{-.1em}\(\partial^{~}_{\!\bcdot}\E_\times^\mu-\partial^{~}_{\hspace{-.1em}\times}\E_\bcdot^\mu\)
\overset{\text{(\ref{dE0})}}{=}
\hat\omega_{\dot{\Sigma}}^{\hspace{.4em}\times}\hspace{.1em}
\E^\bcdot_\mu\hspace{.1em}\partial^{~}_{\hspace{-.1em}\times}\E_\bcdot^\mu.
\end{align}
We note that 
\begin{align*}
\E^\bcdot_\mu\hspace{.1em}\partial^{~}_{\!\bcdot}\E_a^\mu=\partial^{~}_{\!\bcdot}\(\E^\bcdot_\mu\hspace{.1em}\E_a^\mu\)-
\E_a^\mu\partial^{~}_{\!\bcdot}\E^\bcdot_\mu
\overset{\text{(\ref{dE0In})}}{=}0.
\end{align*}
For the second term in the l.h.s. of  (\ref{Lintddww}), we take summation for both sides of (\ref{W2E}) keeping $a=c$, yielding
\begin{align*}
\sum_{a=c}\text{(\ref{W2E})}&\rightarrow
\(\hat\omega_{a\hspace{.3em}\bcdot}^{\hspace{.3em}\bcdot}-\hat\omega_{\bcdot\hspace{.3em}a}^{\hspace{.3em}\bcdot}\)
=-\E^\bcdot_\mu
\(\partial^{~}_{\!\bcdot}\E_a^\mu-\partial^{~}_a\E_\bcdot^\mu\)\!
\implies
\hat\omega^{\hspace{.5em}\bcdot}_{\dot\Sigma}\hat\omega^{\hspace{.5em}\bcdot}_{\dot\Sigma}\hspace{.1em}\eta_{\bcdots}=
-\hat\omega^{\hspace{.5em}\star}_{\dot\Sigma}\E^\bcdot_\mu\partial^{~}_\star\E_\bcdot^\mu.
\end{align*}
Thus, we obtain the interaction vertex for the scaled spin connection as
\begin{align*}
\text{(\ref{Lintddww})}&=
\frac{1}{\lp^2}\hspace{.1em}
{\hat\omega}^{\hspace{.6em}\bcdot}_{\dot\Sigma}
\({\E^\star_\mu}\hspace{.1em}\partial^{~}_{\!\bcdot}{\E_\star^\mu}\)
\xrightarrow{\hat\omega\rightarrow\cG\omega}
\tilde{V}_{\omega\E}=\frac{\cGR}{\lp^2}\hspace{.1em}
\omega^{\hspace{.6em}\bcdot}_{\dot\Sigma}
\({\E^\star_\mu}\hspace{.1em}\partial^{~}_{\!\bcdot}{\E_\star^\mu}\)\!.
\end{align*}
The Fourier transform (\ref{FourierE}) provides the creation operator together with the vierbein polarisation vector, and it converts the differential operator into the momentum operator $\hat{p}^a:=i\hbar\hspace{.1em}\eta^{a\bcdot}\partial_\bcdot$.
We define the creation operator to create an energy quantum as
\begin{align*}
\Pmass\E_\mu^{a}(x)&=\frac{\hbar}{\lp}\E_\mu^{a}(x)\xrightarrow{\text{F.T.}}
\tilde{\E}_\mu^{a}
\(\hbar\hspace{.1em}{a}_{\E}^{~}\!\)(p,\lambda)
~~\text{and}~~
i\hbar\(\partial^{~}_{\hspace{-.15em}a}\hspace{-.1em}\Pmass\E_b^\mu\)(x)\xrightarrow{\text{F.T.}}
\eta^{~}_{a\bcdot}\hspace{.1em}p^\bcdot\eta^{~}_{b\star}\hspace{.1em}
\tilde{\E}_\mu^\star{\(\hbar\hspace{.1em}{a}_{\E}^{~}\!\)(p,\lambda)},
\end{align*}
where $p^a$ is an incoming momentum of the vierbein field. 
We note that $\hbar\hspace{.1em}\lp^{-1}=\Pmass$ in our units.
The operator $\(\hbar\hspace{.1em}{a}_{\E}^{~}\!\)(p,\lambda)$ creates an energy-quantum together with the integration measure concerning the Fourier transformation.
As a result, we obtain the (vierbein)-(spin connection) vertex rule by truncating the creation operator and the polarisation vector as
\begin{align*}
\text{(spin connection)--(vierbein)}^2={\cG}\hspace{.1em} p_{~}^a\!.
\end{align*}

\paragraph{(ghost)-(spin connection) (Figure \ref{fig2}-(d)):}
The Feynman rule for (spin connection)-(ghost) vertex is obtained from the interaction Lagrangian in (\ref {Lgrgh}) using the unscaled spin connection  as
\begin{align*}
\hat{V}_{\bar{\chi}\omega\chi}&=
\frac{1}{\lpt}
\bar{\chi}\hspace{.1em}\hat\omega_{\dot\Sigma\hspace{.1em}\bcdot}^{~}\(\partial^{~}_{\!\star}\chi\){c^\bcdot}{c^\star}
\xrightarrow{\hat\omega\rightarrow\cG\omega}
{V}_{\bar{\chi}\omega\chi}=
\frac{\cG}{\lpt}
\bar{\chi}\hspace{.1em}\omega_{\dot\Sigma\hspace{.1em}\bcdot}^{~}\(\partial^{~}_{\!\star}\chi\){c^\bcdot}{c^\star},
\end{align*}
for the ghost vertex owing to the Lagrangian (\ref{Lgrgh}).
In the momentum space, we set the creation operator to create an energy-quantum as the same as the vierbein fields such that:
\begin{align*}
\Pmass\chi(\xi)&=\frac{\hbar}{\lp}\chi\xrightarrow{\text{F.T.}}
\tilde{\chi}(\xi)
\(\hbar\hspace{.1em}{a}_{\chi}^{~}\!\)(p).
\end{align*}
As a result, we obtain the (ghost)-(spin connection) vertex rule by truncating the creation operator and the polarisation vector as
\begin{align*}
\text{(spin connection)--(ghost)--(anti-ghost)}={\cG}\hspace{.1em} p_{~}^a\!,
\end{align*}
where $p_{~}^a$ is a incoming momentum of the ghost.
\vskip 5mm\noindent
Here, we summarise the \QGED interaction vertices shown in Figure \ref{fig2} as follows: 
\begin{subequations}
\begin{align}
\bullet~&\textrm{(photon)--(electron)$^2$}& \textrm{(Figure \ref{fig2}-(a)):} & &e\hspace{.1em}\gamma_{~}^a&\label{v1}\\
\bullet~&\textrm{(spin connection)--(electron)$^2$}& \textrm{(Figure \ref{fig2}-(b)):}& &\cG\gamma_{~}^a&\label{v2}\\
\bullet~&\textrm{(spin connection)--(vierbein)$^2$}& \textrm{(Figure \ref{fig2}-(c)):}&
&\cG\hspace{.1em} p_{\hspace{-.1em}\E}^{\hspace{.1em}b}
&\label{v3}\\
\bullet~&\textrm{(spin connection)--(ghost)$^2$}& \textrm{(Figure \ref{fig2}-(d)):}& 
&\cG\hspace{.1em} p_{\hspace{-.1em}\chi}^{\hspace{.1em}a}
&\label{v4}
\end{align}
where $p^{\E}_{\bullet}$ and $p^{\chi}_{\bullet}$ are incoming momenta of the vierbein and ghost fields, respectively.
\end{subequations}

%
%
\subsection{Propagators}
A quantum propagator is the vacuum-to-vacuum transition amplitude of the time-ordered product of quantum fields consisting of a sum of creation and annihilation operators, such that:
\begin{align}
D_{\hspace{-.1em}F}(\xi-\eta)&:=i\langle0|T\phi(\xi)\phi^\dagger(\eta)|0\rangle=
\frac{i}{(2\pi\hbar)^{4}}\int{dp^4}
\frac{e^{-ip\cdot(\xi-\eta)/\hbar}}{-\eta_\bcdots p^\bcdot p^\bcdot-i\delta}\xrightarrow{\text{F.T.}}D^\phi(p),\label{Dfxy}
\end{align}
where $\phi(\xi)$ shows a general quantum field at $\xi\in\TsM$.
Additional degrees of freedom, e.g., a Lorentz index of vector bosons or a spinor index of electrons, are omitted here.  
A suffix ``$F$'' of $D_{\hspace{-.1em}F}$ stands for the \textit{Feynman propagator} and is omitted hereafter.
The Feynman propagator naturally appears in some extended Riemannian metric (the amphometric\!\cite{Kurihara:2022green,Kurihara:2025tro}).
The classical counterpart of a quantum propagator is Green's function for the equation of motion for the free gauge field, which is a wave equation; thus, their Fock space of creation and annihilation operators is the same as that of the QED gauge field. 

The Fourier transformation of (\ref{Dfxy}) provides a quantum propagator in the momentum space such that:
\begin{align}
D^\phi(p)&=i\int_{\TsM}\vvv\hspace{.2em}e^{-ip\cdot\xi/\hbar}
\langle0|T^*\phi(\xi)\phi^\dagger(0)|0\rangle=
\frac{i}{-\eta_\bcdots p^\bcdot p^\bcdot-i\delta},\label{Dphi}
\end{align}
where $p$ is a momentum of the field.
We define the Feynman rule for a particle propagator as
\begin{align}
\text(\ref{Dphi})&\implies
\frac{\sum\tilde{\phi}(p,\dots)\tilde{\phi}^\dagger(p,\dots)}{-\eta_\bcdots p^\bcdot p^\bcdot-i\delta},
\label{Dphi2}
\end{align}
where $\tilde{\phi}^{(\dagger)}(p,\cdots)$ is a creation (annihilation) operator in the momentum space with quantum numbers depending on the particle species, which is the truncated one from the vertices at both ends of the propagator.
The summation takes place for all possible quantum numbers.

We obtain propagators for \QGED fields owing to  (\ref{Dphi}) as follows:
\begin{itemize}
\begin{subequations}
\item Photon propagator:
\begin{align}
D^{\Aa}_{\hspace{.2em}ab}(p)&=
\frac{\sum_{\lambda}\epsilon_a^{(\lambda)}(p)\epsilon_b^{(\lambda)}(p)}
{-\eta_\bcdots p^\bcdot p^\bcdot-i\delta}
\end{align}
\item Electron propagator: 
\begin{align}
S_{\psi}(p)=
\frac{\sum_{s}u(p,s)\bar{u}(p,s)}
{-\eta_\bcdots p^\bcdot p^\bcdot+m^2-i\delta}
\end{align}
\item Vierbein propagator:
\begin{align}
D^\E_{\hspace{.1em}\mu\nu}(p)=
\frac{\sum_{\lambda}\tilde{\E}_\mu^{(\lambda)}(p)\tilde{\E}_\nu^{(\lambda)}(p)}
{-\eta_\bcdots p^\bcdot p^\bcdot-i\delta}\label{DE}
\end{align}
\item Spin-connection propagator:
\begin{align}
D_{\hspace{.2em}\dot{a}b;\dot{c}d}^{\omega}(p)=
\frac{\sum_{\lambda}
\tilde{\omega}^{(\lambda)}_{\hspace{1em}\dot{a}b}\hspace{.2em}
\tilde{\omega}^{(\lambda)}_{\hspace{1em}\dot{c}d}}
{-\eta_\bcdots p^\bcdot p^\bcdot-i\delta}\label{Dchi}
\end{align}
\item Ghost propagator:
\begin{align}
D_{\chi}(p)=
\frac{1}
{-\eta_\bcdots p^\bcdot p^\bcdot-i\delta}
\end{align}
\end{subequations}
\end{itemize}
Polarisation summations for the \QGED fields are given in  (\ref{fnorm}), (\ref{Anorm}), (\ref{EEnorm}) and (\ref{Wnorm}).
The vierbein propagator (\ref{DE}) is defined in the momentum space corresponding to the (flat) asymptotic spacetime.

\begin{figure}[tb]
 \begin{center} 
   \includegraphics[width=8.5cm]{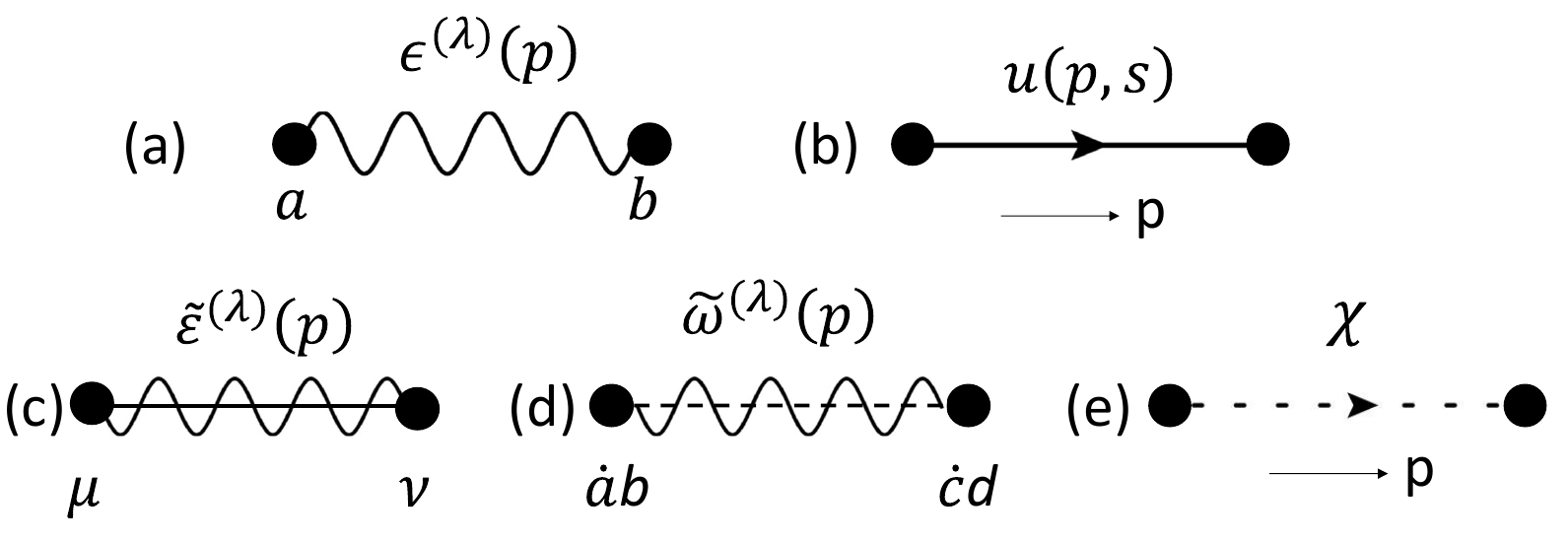}
 \caption{\label{fig3}\small
Figures provide pictorial representations of the QEDG propagators.
A solid circle shows the interaction vertex.
Equation numbers show the propagator function and filed normalisation of the corresponding field.} 
 \end{center} 
\end{figure}

%
%
\subsection{Integration measure and S-matrix}
An integration measure for each loop momentum $l_j$ and outgoing external momentum $k_j$ are, respectively, taken to be\begin{align}
\prod_j\(-1\)^{\sigma_j}\(\mu_R^2\)^{\varepsilon^{~}_{\hspace{-.1em}U\hspace{-.1em}V}}\frac{d^Dl_j}{i(2\pi\hbar)^D } ~~&{\rm and}~~
\prod_j\frac{d^3k_j}{(2\pi\hbar)^3}\frac{1}{2k^0_j},\label{intmeas}
\end{align}
where $k^0_j$ is the zeroth component of the $j$'th outgoing momentum $k_j$.
In the dimensional regularisation method, we need to put an arbitrary parameter with energy dimension, $\mu_R^{~}$, to adjust the physical dimension of the loop contribution.
We set $\sigma_j=1$, if a $j$'th loop particle is an electron or a ghost; otherwise, $\sigma_j=0$. 
Owing to the analytic continuation (the Wick rotation) concerning the zeroth component $ dl_j^0$, an imaginary unit in the loop integration is eliminated.
The $S$-matrix is defined owing to the interaction Lagrangian as
\begin{align*}
S:=T^* \exp{\left[i\int d^4\xi\LL_{int}\(\xi\)\right]},
\end{align*}
and its perturbative expansion is
\begin{align*}
S=1+\sum_{N=1}^{\infty}\frac{i^N}{N!}\int d^4\xi_1\cdots\int d^4\xi_N
T^*\left[\LL_{int}\(\xi_1\)\cdots\LL_{int}\(\xi_N\)\right].
\end{align*}
The scattering matrix, namely $T$-matrix, is defined owing to the $S$-matrix as
\begin{align}
S=1+i T.\label{TMatrix}
\end{align}
Matrix elements $S_{fi}:=\langle f|S|i\rangle$ is expressed as
\begin{align*}
S_{fi}=\delta_{fi}+i(2\pi\hbar)^4\delta(p^{~}_f-p^{~}_i)\sum_{pol}\big| T_{fi}\big|^2,
\end{align*}
where $|i\rangle$ and $|f\rangle$ are, respectively,  initial and final states, and $p^{~}_i$ and $p^{~}_f$ are their total momentum.
Our convention of the Feynman rules, including the gravitational bosons, gives a factor of the scattering matrix in total, such as 
\begin{align*}
i_{~}^N&=i\times i_{~}^{N+L-1}\times i_{~}^{-L},
\end{align*}
where $L$ is the number of loops.
Here, the second factor is absorbed into the propagators and the last factor into loop integrals; thus, the first $i$ gives a correct factor of $iT$.

\subsection{Spin connection, vierbein and torsion}\label{torsion}
Before we move on to the concrete calculations of loop amplitudes, this section will examine the relationship between the spin connection and vierbein, as well as a subtle point related to them. 

The Einstein--Hilbert gravitational Lagrangian comprises two fundamental fields: a vierbein and a spin connection.
We treat the spin connection defined in the inertial space using the local variable as
\begin{align}
\omega_c^{\hspace{.3em}ab}(\xi)&=\E_c^\mu\(x(\xi)\)\omega_\mu^{\hspace{.3em}ab}(\xi)\implies
\www=\omega_\bcdot^{\hspace{.3em}\stars}(\xi)(\partial_\star\times\partial_\star)\eee^\bcdot
\in V^2(\TM)\otimes\Omega^1(\TsM)\otimes\SO(1,3).\label{EWform}
\end{align}
We treat the vierbein in (\ref{EWform}) as a map from a global form to a local one, rather than a dynamic field.

The equations of motion for the vierbein and the spin connection are, respectively, the torsion equation and the Einstein equation.
They have the same degree of freedom and are not independent of each other.
In classical physics, they consist of a set of simultaneous differential equations.
The torsion equation is obtained as the Euler--Lagrange equation with respect to the vierbein form to be (\ref{torsionful}).
The spin coupling term appears only in the matter interaction Lagrangian, apart from the gravitational Lagrangian; thus, the spacetime torsion is created only by an electron in \QGED.
Moreover, as mentioned in \textbf{section \ref{FeynmanRules}}, we can assume that the spacetime is torsionless almost everywhere in the three-dimensional space-like Cauchy surface.

On the other hand, the spacetime curvature, equivalently the spin connection, is created by a stress-energy tensor obtained by varying the Lagrangian with respect to the vierbein form; thus, primarily the Yang--Mills gauge fields concern the spacetime curvature.
The matter Lagrangian has the vierbein only in the volume form, which plays as the integration measure and does not have a dynamic degree of freedom\footnote{See, e.g., section 5.6 in Ref.\cite{fre2012gravity}.} in the inertial frame.

\begin{figure}[t]
 \begin{center} 
   \includegraphics[width=8cm]{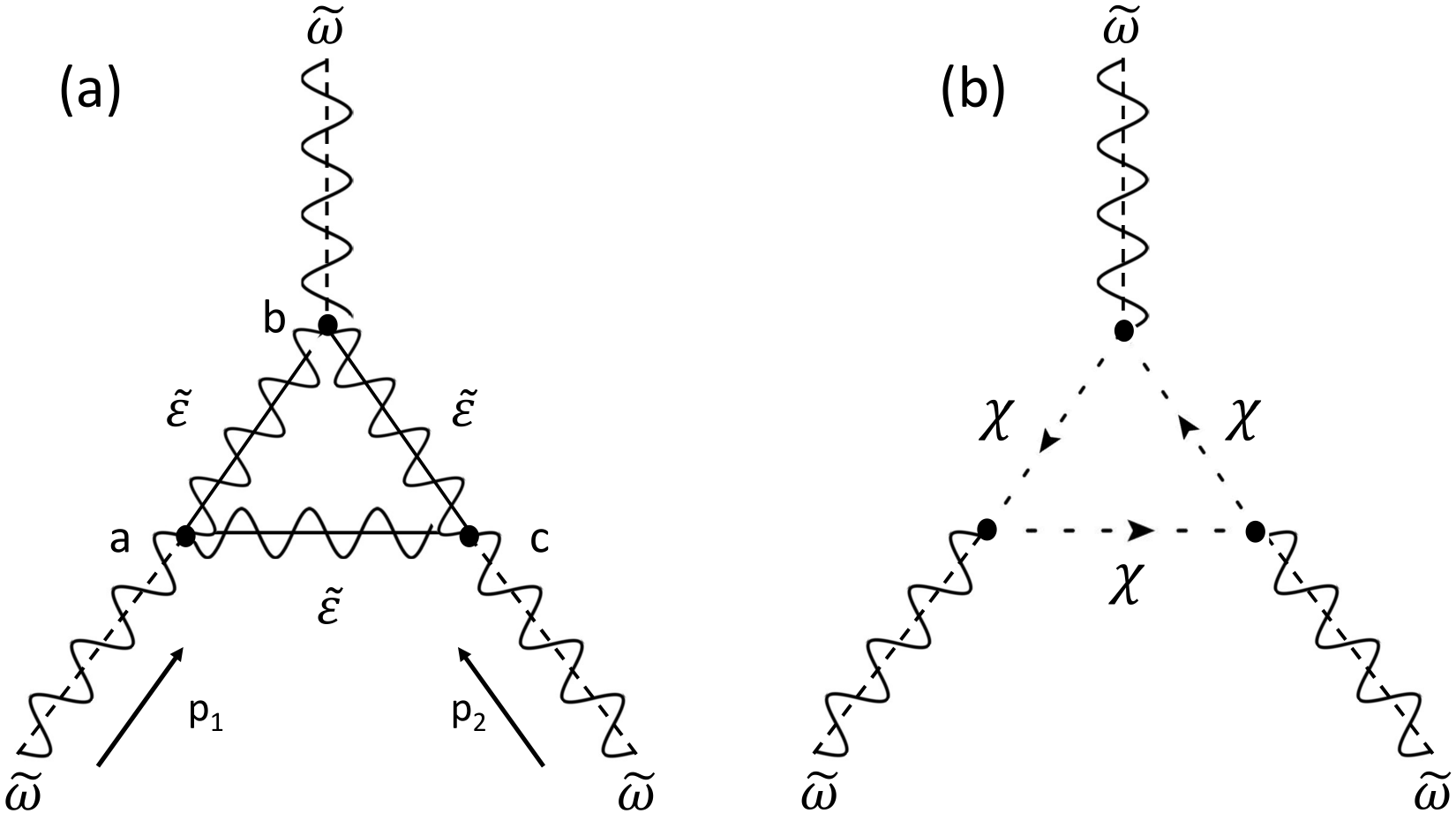}
 \caption{\label{anomlVtx}\small
Spin-connection self-interaction Feynman-diagrams induced by the vierbein loop (a) and the ghost loop (b). 
Cross diagrams, momentum $p_1$ connected to $c$ and  momentum $p_2$ connected to $a$, also exist. 
} 
 \end{center} 
\end{figure}

In classical physics, the torsionless condition is ensured during the simultaneous solving process, as sources of curvature and torsion are separated from each other.
At the quantum level, we need to verify that no anomaly exists to violate the torsionless condition after quantisation.
A quadratic term concerning the spin connection in the Einstein--Hilbert Lagrangian induces an unexpected mass term for the spin connection in the standard procedure.
We have introduced a trick to avoid the spin connection having a mass by using the torsionless equation, as in (\ref{W2E}).
Consequently, a vertex shown in Figure \ref{anomlVtx}  is a possible source of quantum anomaly.
Here, we investigate the absence of anomaly in Figure \ref{anomlVtx}.

Vertex functions are provided as
\begin{align}
\text{Figure \ref{anomlVtx}-(a)}&=:\Gamma_{(\E)}^{abc}(p_1,p_2)=
\cGR^{\hspace{-.2em}3}\(\mu_R^2\)^{\varepsilon^{~}_{\hspace{-.1em}U\hspace{-.1em}V}}\int\frac{d^Dk}{i(2\pi\hbar)^D }
\frac{N^{abc}(p_1,p_2)}
{k^2(k+p_1)^2(k-p_2)^2}+(p1\leftrightarrow p_2),\label{Fig4a}\\
\text{Figure \ref{anomlVtx}-(b)}&=:\Gamma_{(\chi)}^{abc}(p_1,p_2)
=-\cGR^{\hspace{-.2em}3}\(\mu_R^2\)^{\varepsilon^{~}_{\hspace{-.1em}U\hspace{-.1em}V}}\int\frac{d^Dk}{i(2\pi\hbar)^D }
\frac{k^a(k^b+p_{1}^b)(k^c-p_{2}^c)}{k^2(k+p_1)^2(k-p_2)^2}+(p1\leftrightarrow p_2),\label{Fig4b}
\intertext{where}
N^{abc}(p_1,p_2)&:=\frac{1}{D}\(\eta^{ab}k.(k+p_1)(k-p_2)^c+\eta^{bc}k^a(k+p_1).(k-p_2)+\eta^{ac}k.(k-p_2)(k+p_1)^a\)\!,\notag
\end{align}
using Feynman rules.
Here, we set $p_1$ and $p_2$ to on-shell.
These amplitudes conserve a current as 
\begin{align*}
\Gamma_{(\E)}^{abc}(p,-p)=\Gamma_{(\chi)}^{abc}(p,-p)=0.
\end{align*}
The QGED does not have a spin connection self-coupling at a tree level.
Thus, when vertices (\ref{Fig4a})+ (\ref{Fig4b}) have divergence, there is no counter vertex to absorb infinity. 
In reality, each of the two vertices has a divergent integral; however, they are cancelled out after summing them up, such as
\begin{align*}
\Gamma_{(\E)}^{abc}(p_1,p_2)\approx-\Gamma_{(\chi)}^{abc}(p_1,p_2)\approx-\cGR^{\hspace{-.2em}3}C_{UV}\frac{1}{16\pi^2}
\frac{1}{8}\((p_1+p_2)^a\eta^{bc}+2(p_1+p_2)^c\eta^{ab}\)\!,
\end{align*}
where we show only a divergent part.

In conclusion, the torsionless condition embedded in the (spin connection)-(vierbein)-(vierbein) vertex introduced through (\ref{W2E}) does not lead to a quantum anomaly.

\section{Renormalisation of the \QGED}
This section demonstrates a renormalisation of the \QGED at one loop order due to the Feynman rules provided in the previous section.
In the following calculations, we exploit the Feynman gauge for both the photon and spin connection fields, such that $\xi^{~}_{\hspace{-.1em}A}=\xi_{\hspace{-.1em}\omega}=1$.

%
%
\subsection{Power counting}
Before starting the one-loop renormalisation calculations for the \QGED, we apply power counting to the theory to confirm its degree of divergence.
For details on power counting, see, e.g., Ref$.$\!\cite{kaku1993quantum}.
For each vertex in the interaction Lagrangian, we can assess a degree of ultraviolet divergence.
In the interaction Lagrangian, each field has a physical dimension, in the $D$-dimensional momentum space, as
\begin{align*}
\begin{array}{cclcl}
&&&&D=4\\
\left[\text{fermion}\right]_\text{pd}&=&E^{(D-1)/2}&\implies&\rho^{~}_f=3/2\\
\left[\text{boson}\right]_\text{pd}&=&E^{(D-2)/2}&\implies&\rho^{~}_b=1\\
\left[\text{derivative}\right]_\text{pd}&=&E^{1}&\implies&\rho^{~}_\partial=1
\end{array},
\end{align*}
and induces a UV-divergence (a divergence at the high energy limit) degree as shown in the right formula above.
Here, a vierbein and an (anti-)ghost belong to the ``boson'' since their high-energy behaviour is the same as that of a photon and a spin connection.
In addition, each vertex has the $D$-dimensional Dirac $\delta$-function owing to the energy-momentum conservation, which gives $[\delta^{\!D}_{~}\!(p)]_\text{pd}=E^{-D}$ physical dimensions.
Thus, we introduce a degree of divergence for the given vertex in the four-dimensional spacetime as
\begin{align}
\rho(V)&:=\sum_{a\in\{f,b,\partial\}}\rho^{~}_a N_a-D
=\frac{3}{2}N_f+N_b+N_\partial-4,\label{rhoV}
\end{align}
where $N_f$, $N_b$ and $N_\partial$ are the number of electrons, gauge bosons (photon and spin connection) and derivative couplings in the vertex, respectively.
If the degree is greater than zero, it induces the UV-divergence, and the theory is not renormalisable.
On the other hand, if the degree is equal to zero, it induces a logarithmic divergence, and the theory may be renormalisable.
The divergence degree of each vertex in Figure \ref{fig2} in four-dimensional spacetime is given as
\begin{align*}
\rho\({V}_{\epsilon\psi}\)&=\rho\({V}_{\omega\psi}\)=\frac{3}{2}\times2+1+0-4=0,~~~
\rho\({{V}}_{\omega\E}\)=\rho\({V}_{\bar{\chi}\omega\chi}\)=\frac{3}{2}\times0+3+1-4=0.
\end{align*}
This means that all the \QGED vertices do not have a potential divergence.

We can assign the superficial degree of divergence, $d(\Gamma)$, to the Feynman diagram $\Gamma$ employing the momentum dimension.
When some Feynman diagram has $d(\Gamma)\ge0$, it gives ultraviolet divergent integrals.
We denote the number of loops, external/internal lines and vertices of the given Feynman diagram $\Gamma$ as shown in Table \ref{table33}.
\begin{table}[bt]
\begin{center}
\caption{\label{table33}\small
The number of ingredients of the Feynman diagrams.}
\vskip 2mm
\begin{tabular}{lcc}
\multicolumn{2}{c}{Type}& number of objects in the diagram $\Gamma$\\
\hline
\multirow{4}{*}{External lines} & electron& $E_\psi$\\
 &photon& $E_\gamma$\\
 &vierbein& $E_\E$\\
 &spin connection& $E_\omega$\\
\hline
\multirow{4}{*}{Internal lines}& electron& $I_\psi$\\
 &photon& $I_\gamma$\\
 &vierbein& $I_\E$\\
 &spin connection& $I_\omega$\\
 &ghost/anti-ghost& $I_\chi$\\
\hline
 \multirow{3}{*}{Vertex} &(electron)$^2$-(photon)& $V_\gamma$\\
 &(electron)$^2$-(spin connection)& $V_\omega$\\
 &(vierbein)$^2$-(spin connection)$^{*)}$& $V_\E$\\
 &(ghost)-(anti-ghost)-(spin connection)$^{*)}$& $V_\chi$\\
\end{tabular}
\end{center}
\begin{center}$*)$ derivative coupling\end{center}
\end{table}
In the four-dimensional spacetime, the superficial degree of divergence is provided as
\begin{align}
d(\Gamma)=2\sum_{a\in\{\gamma,\omega,\E,\chi\}}I_a +
I_\psi+\sum_{a\in\{\E,\chi\}}V^{~}_{\hspace{-.1em}a}-4\(\sum_{a\in\{\gamma,\omega,\E,\chi\}} V^{~}_{\hspace{-.1em}a}-1\).
\end{align}
We can express the superficial degree of divergence by the number of external particles only, using identities such as
\begin{align*}
V^{~}_{\hspace{-.1em}a}&=2I_a+E_a~~\text{for}~~a\in\{\gamma,\omega,\E\},~~~
2V_\gamma+2V_\omega=2I_\psi+E_\psi,~~
V_\chi=I_\chi.
\end{align*}
Consequently, we obtain the superficial degree of divergence in four-dimensional spacetime, such that:
\begin{align*}
d(\Gamma)=4-E_\gamma-E_\E-E_\omega-\frac{3}{2}E_\psi+
\sum_{a\in\{\gamma,\omega,\E,\chi\}}V^{~}_{\hspace{-.1em}a}\hspace{.1em}V\hspace{-.2em}\(\rho^{~}_a\).
\end{align*}
Therefore, the UV-divergent diagrams must have external particles as 
\begin{align}
E_\gamma+E_\E+E_\omega+\frac{3}{2}E_\psi<4.\label{numOCT}
\end{align}
As a result, only a finite number of diagrams will give an ultraviolet divergence, which can be eliminated by the finite number of counter terms.
This analysis suggests that our theory could be renormalised in all orders of perturbative expansion\footnote{
To confirm the renormalisability of all orders, we need to verify a concrete demonstration of higher-order calculations, which is beyond the scope of this article.}.

At first glance, this conclusion appears to contradict the known result that quantum general relativity is \textit{not} renormalisable.
We show in \textbf{Appendix \ref{appC}} that this is not the case.
%
%
\subsection{One-loop renormalisation}\label{OLR}
This section provides renormalisation constants and counter terms of the \QGED at a one-loop level.
Although the QED's renormalisation procedure is established and well known, we repeat it as a guide for renormalising the gravitational part.
We exploit the on-shell renormalisation conditions such that
\begin{enumerate}
\item the pole position of propagators should be located at physical mass,
\item residues of propagators at the pole should be unity,
\end{enumerate}
and the charge renormalisation utilised in Refs.\!\cite{10.1143/PTPS.73.1, 10.1143/PTPS.100.1,BELANGER2006117}.
This report refers to charge renormalisation as the coupling constant renormalisation.
Owing to the second condition of the on-shell condition, we can omit self-energy correction diagrams on the external line.

Throughout this study, we exploit dimensional regularisation to regulate the ultraviolet divergence and put fictitious masses into massless bosons to treat the infrared divergence.

\subsubsection{Renormlisation constants}
For a QED part, renormalisation constants $Y^{~}_{\hspace{-.2em}\Aa}$, $Z^{~}_\psi$, $Z^{~}_{\hspace{-.1em}\Aa}$ and $\delta{m^{\Aa}_{e}}$ are introduced to the Lagrangian, and for the \QGED, additional renormalisation constants,  $Y^{~}_{\hspace{-.2em}\omega}$, $Z^{~}_{\hspace{-.1em}\omega}$, $Z^{~}_{\hspace{-.1em}\E}$ and $\delta{m^{\omega}_{e}}$ appears.
Bare objects in the Lagrangian are replaced as 
\begin{align*}
\psi^\bare&=
{Z^{~}_{\psi\Aa}}^{\hspace{-.2em}\hlf}\hspace{.1em}
{Z^{~}_{\psi\omega}}^{\hspace{-.2em}\hlf}\hspace{.1em}\psi,\\
\Aa^\bare_{\hspace{.7em}a}\hspace{.1em}&=
{Z^{~}_{\hspace{-.1em}\Aa}}^{\hlf}\hspace{.2em}\Aa^{~}_a,~~~
\omega^{{\bare}ab}_{\hspace{.7em}c}\hspace{.1em}=\hspace{.2em}
{Z^{~}_{\hspace{-.1em}\omega}}^{\hlf}\hspace{.2em}\omega^{\hspace{.3em}ab}_{c},~~~
\E^{{\bare}a}_{\hspace{.8em}\mu}\hspace{.1em}=\hspace{.2em}
{Z^{~}_{\hspace{-.1em}\E}}^{\hlf}\hspace{.2em}\E^a_\mu\\
e^\bare&=Y^{~}_{\hspace{-.2em}\Aa}\hspace{.2em}e,~~~\cGz=Y^{~}_{\hspace{-.2em}\omega}\cG,\\
m_e^\bare&=m_e+\(\delta{m^{\Aa}_{e}}+\delta{m^{\omega}_{e}}\)\!.
\end{align*}
In addition to that, there are several mixed corrections at $\mathcal{O}(e\hspace{.1em}\cGR)$.
Then, we add the counter-term Lagrangian to the renormalised one. 
%

\subsubsection{Photon and spin connection renormalisation}\label{PSR}
\begin{figure}[]
 \begin{center} 
   \includegraphics[width=7cm]{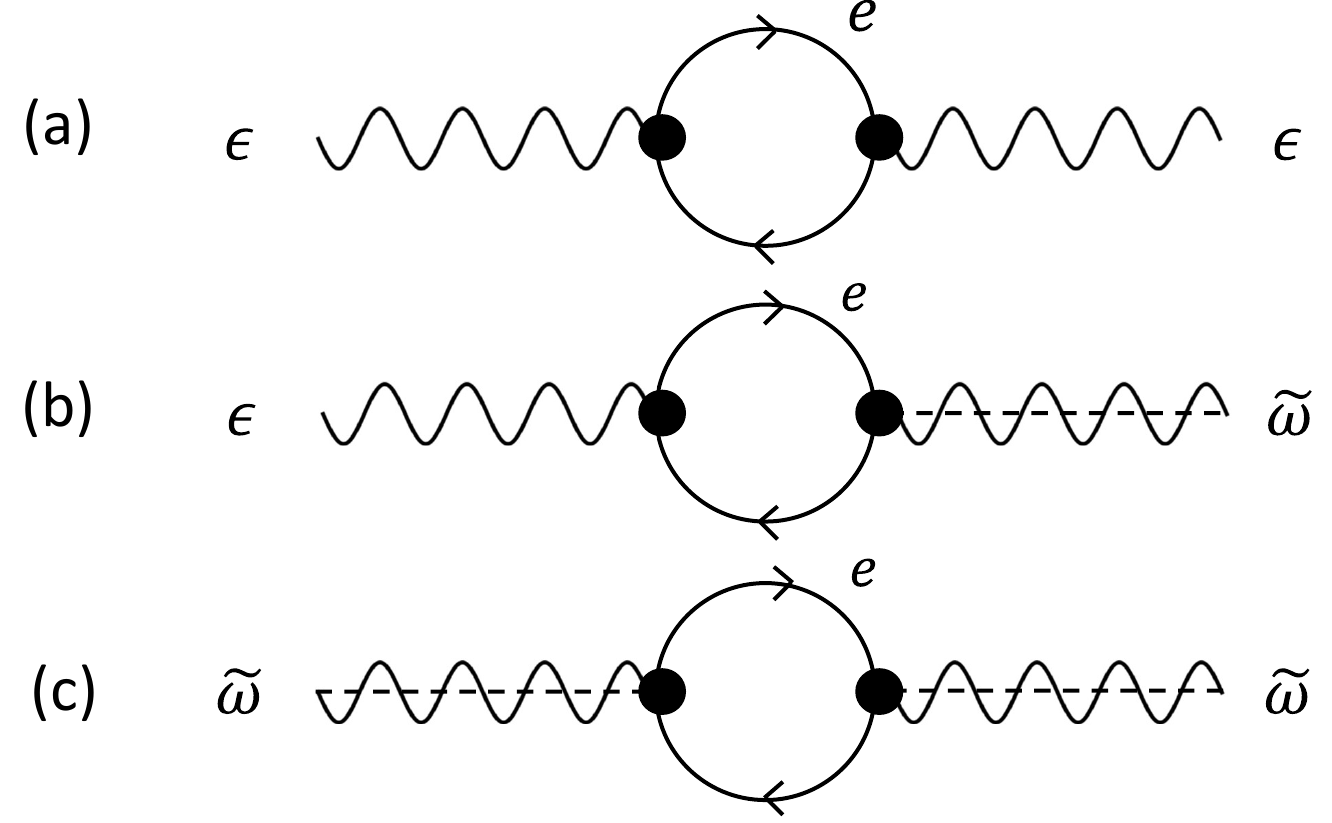}
 \caption{\label{figvacpol}\small
Photon and spin connection vacuum-polarisation diagrams with an electron loop.} 
 \end{center} 
\end{figure}
\begin{figure}[t]
 \begin{center} 
   \includegraphics[width=6cm]{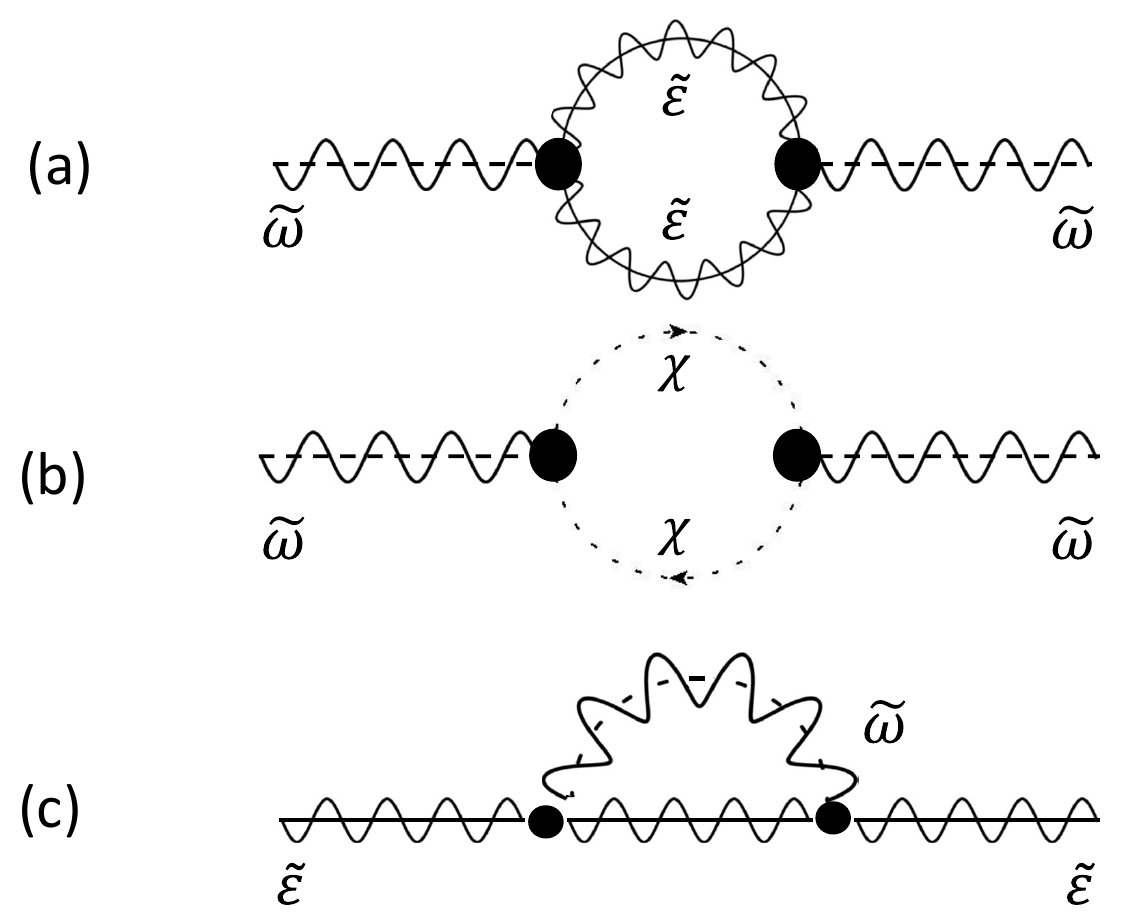}
 \caption{\label{figvacpol2}\small
spin connection and vierbein vacuum-polarisation diagrams with a boson loop.} 
 \end{center} 
\end{figure}

First, we provide a gauge boson renormalisation in the \QGED.
Vacuum polarisation diagrams in the \QGED are depicted in Figures \ref{figvacpol} and \ref{figvacpol2}; the former shows the electron-loop diagrams, and the latter shows boson-loop diagrams.
We calculate vacuum polarisation at $\mathcal{O}(e^2)$, $\mathcal{O}(\cGR^{\hspace{-.3em}2})$ and $\mathcal{O}(e\hspace{.1em}\cGR)$.

Owing to the Lorentz invariance of the photon wave function, the vacuum polarisation diagram of a photon shown in Figure \ref{figvacpol}-(a) can be expressed, in general,  as
\begin{subequations}
\begin{align}
\text{Figure \ref{figvacpol}-(a)}&=:\Pi^{(\gamma{e}\gamma)}_{ab}(q^2)=\(\eta^{~}_{ab}-\frac{q_a\hspace{.1em}q_b}{q^2}\)a^{~}_{\gamma{e}\gamma}(q^2)
+\frac{q_a\hspace{.1em}q_b}{q^2}b^{~}_{\gamma{e}\gamma}(q^2),\label{PIgeg}
\end{align}
where $q_a$ is a photon momentum.
After standard calculations, we obtain the results as
\begin{align}
a^{~}_{\gamma{e}\gamma}(q^2)&=-\frac{\alpha}{4\pi}\frac{4}{3}q^2
\(C_{UV}-\log{\frac{m_e^2}{\mu^2_R}}\)\!,\label{PIgeg2}\\
b^{~}_{\gamma{e}\gamma}(q^2)&=0,\label{PIgeg3}
\end{align}
\end{subequations}
where $C_{UV}:=1/\varepsilon^{~}_{UV}-\gamma^{~}_E+\log{4\pi}$.
Direct calculations provide that the above vacuum polarisation fulfils a current conservation, such that
\begin{align*}
\eta^\bcdots q_\bcdot\Pi^{(\gamma{e}\gamma)}_{a\bcdot}(q^2)&=0.
\end{align*}
We obtain the renormalisation constant for a photon wave function as
\begin{align}
{Z^{~}_{\hspace{-.1em}\Aa}}^{\hlf}&=
1-\frac{1}{2}\left.\frac{\partial}{\partial q^2}a^{~}_{\gamma{e}\gamma}(q^2)\right|_{q^2=0}=
1-\frac{1}{2}\frac{\alpha}{4\pi}
\frac{4}{3}\(C_{UV}-\log{\frac{m_e^2}{\mu^2_R}}\)\!.\label{ZAA}
\end{align}
 
Vertex Feynman rules for a (photon)-(electron) and a (spin connection)-(electron) are the same as each other except the coupling constant, as shown in (\ref{v1}) and (\ref{v2}); thus, vacuum polarisations, including the spin connection fields, are easily provided from that for a photon as follows:
\begin{align*}
\text{Figure \ref{figvacpol}-(b)}&=:\Pi^{(\gamma{e}\omega)}_{ab}(q^2)=\(\eta^{~}_{ab}-\frac{q_a\hspace{.1em}q_b}{q^2}\)a^{~}_{\gamma{e}\omega}(q^2)
+\frac{q_a\hspace{.1em}q_b}{q^2}b^{~}_{\gamma{e}\omega}(q^2),\\
\text{Figure \ref{figvacpol}-(c)}&=:\Pi^{(\omega{e}\omega)}_{ab}(q^2)=\(\eta^{~}_{ab}-\frac{q_a\hspace{.1em}q_b}{q^2}\)a^{~}_{\omega{e}\omega}(q^2)
+\frac{q_a\hspace{.1em}q_b}{q^2}b^{~}_{\omega{e}\omega}(q^2),
\intertext{where}
a^{~}_{\gamma{e}\omega}(q^2)&=
-\frac{\(\alpha\aGR\)^{1/2}}{4\pi}
\frac{4}{3}q^2\(C_{UV}-\log{\frac{m_e^2}{\mu^2_R}}\)\!,\\
a^{~}_{\omega{e}\omega}(q^2)&=
-\frac{\aGR}{4\pi}
\frac{4}{3}q^2\(C_{UV}-\log{\frac{m_e^2}{\mu^2_R}}\)\!,\\
b^{~}_{\gamma{e}\omega}(q^2)&=b^{~}_{\omega{e}\omega}(q^2)=0.
\end{align*}
Again, they fulfil the current conservation by themselves. 

The spin connection has two other vacuum polarisation diagrams; a vierbein loop shown in Figure \ref{figvacpol2}-(a) and a ghost loop shown in Figure \ref{figvacpol2}-(b). 
The calculations of both diagrams are very similar owing to their derivative couplings.
A difference comes from scalar or vector propagators in loops.
Calculations owing to the Feynman rules provide the results as
\begin{align*}
\text{Figure \ref{figvacpol2}-(a)}=:\Pi^{(\omega\E\omega)}_{ab}(q^2)&=
\frac{1}{2}\cGR^{\hspace{-.2em}2}\(\mu_R^2\)^{\varepsilon^{~}_{\hspace{-.1em}U\hspace{-.1em}V}}\int\frac{d^Dk}{i(2\pi\hbar)^D }
\frac{\eta^\bcdots k_\bcdot(k_\bcdot+q_\bcdot)}{k^2(k-q)^2}\eta^{~}_{ab},\\
&=\(\eta^{~}_{ab}-\frac{q_a\hspace{.1em}q_b}{q^2}\)a^{~}_{\omega\E\omega}(q^2)
+\frac{q_a\hspace{.1em}q_b}{q^2}b^{~}_{\omega\E\omega}(q^2),\\
\text{Figure \ref{figvacpol2}-(b)}=:\Pi^{(\omega\chi\omega)}_{ab}(q^2)
&=-\cGR^{\hspace{-.2em}2}\(\mu_R^2\)^{\varepsilon^{~}_{\hspace{-.1em}U\hspace{-.1em}V}}\int\frac{d^Dk}{i(2\pi\hbar)^D }
\frac{k_a(k_b+q_b)}{k^2(k-q)^2},\\
&=\(\eta^{~}_{ab}-\frac{q_a\hspace{.1em}q_b}{q^2}\)a^{~}_{\omega\chi\omega}(q^2)
+\frac{q_a\hspace{.1em}q_b}{q^2}b^{~}_{\omega\chi\omega}(q^2).
\end{align*}
A factor of one-half in the vierbein loop is due to the symmetric factor of two identical particles in the loop.\begin{align*}
&a^{~}_{\omega\E\omega}(q^2)+
a^{~}_{\omega\chi\omega}(q^2)=-\frac{\aGR}{4\pi}q^2\hspace{.2em}
\frac{1}{6}\(C_{UV}+\frac{5}{3}-\log{\frac{-q^2}{\mu^2_R}}\)\!,\\
&b^{~}_{\omega\E\omega}(q^2)=-b^{~}_{\omega\chi\omega}(q^2)=-\frac{\aGR}{4\pi}q^2\hspace{.2em}
\frac{1}{4}\(C_{UV}+2-\log{\frac{-q^2}{\mu^2_R}}\)
\implies b^{~}_{\omega\E\omega}(q^2)+b^{~}_{\omega\chi\omega}(q^2)=0.
\end{align*}
Thus, the vacuum polarisation of boson loops fulfils current conservation non-trivially after summing up two contributions.
This result encourages us to continue renormalisation calculations further.

At last in current section, we provide a vacuum polarisation of a vierbein field depicted in Figure \ref{figvacpol2}-(c).
After calculations following the Feynman rules, we obtain that
\begin{align*}
\text{Figure \ref{figvacpol2}-(c)}=:\Pi^{(\E\omega\E)}_{ab}(q^2)&=
\cGR^{\hspace{-.2em}2}\(\mu_R^2\)^{\varepsilon^{~}_{\hspace{-.1em}U\hspace{-.1em}V}}\int\frac{d^Dk}{i(2\pi\hbar)^D }
\frac{\eta_{~}^{\bcdots}q_\bcdot(k_\bcdot+q_\bcdot)\eta^{~}_{ab}-q_a(q_b+k_b)}{k^2(k-q)^2},\\
&=\(\eta^{~}_{ab}-\frac{q_a\hspace{.1em}q_b}{q^2}\)a^{~}_{\E\omega\E}(q^2)
+\frac{q_a\hspace{.1em}q_b}{q^2}b^{~}_{\E\omega\E}(q^2),
\end{align*}
where
\begin{align*}
a^{~}_{\E\omega\E}(q^2)&=-\frac{\aGR}{4\pi}\frac{1}{2}q^2
\(C_{UV}+2-\log{\frac{-q^2}{\mu^2_R}}\)\!,\\
b^{~}_{\E\omega\E}(q^2)&=0,
\end{align*}
which fulfils the current conservation by itself. 

We note that all two-point functions appearing above have the form
\begin{align*}
\Pi^{~}_{ab}(q^2)=\(\eta^{~}_{ab}-\frac{q_a\hspace{.1em}q_b}{q^2}\)a(q^2)
+\frac{q_a\hspace{.1em}q_b}{q^2}b(q^2),
\end{align*}
and the $b(q^2)$-term disappears after all.
It ensures the Ward--Takahashi identity in the QGED, including the gravitational sector.

We introduce total $\Nf$ fermions, in which $\Ncf$ fermions are electromagnetically charged, in the \QGED and summarise the renormalisation constant for bosonic fields:
\begin{subequations}
\begin{align}
{Z^{~}_{\hspace{-.1em}\Aa}}^{\hlf}&=
1-\frac{1}{2}\left.\frac{\partial}{\partial q^2}\sum_{i=1}^\Ncf a^{~}_{\gamma{e}\gamma}(m^{~}_i;q^2)\right|_{q^2=0}=
1-\frac{1}{2}\frac{\alpha}{4\pi}
\frac{4}{3}\sum_{i=1}^\Ncf\(C_{UV}-\log{\frac{m_i^2}{\mu^2_R}}\)\!,\label{ZAA2}\\
{Z^{~}_{\hspace{-.1em}\omega}}^{\hspace{-.2em}\hlf}&=
1-\frac{1}{2}\left.\frac{\partial}{\partial q^2}\(
\sum_{i=1}^\Nf\hspace{.1em}a^{~}_{\omega{e}\omega}\(q^2;m^{~}_i\)+
a^{~}_{\omega\chi\omega}(q^2)+
a^{~}_{\omega\E\omega}(q^2)
\)\right|_{q_{~}^2=-\mu_{I\hspace{-.2em}R}^2}\notag
\\&=
1-\frac{1}{2}\frac{\aGR}{4\pi}\left\{
\frac{4}{3}\sum_{i=1}^\Nf\(C_{UV}-\log{\frac{m_i^2}{\mu^2_R}}\)
-\frac{1}{6}\( C_{UV}+\frac{3}{2}+\log{\frac{\mu^2_R}{\mu^{2}_{\hspace{-.1em}I\hspace{-.2em}R}}}\)\right\}\!,\label{ZWW}
\\
{Z^{~}_{\Aa\hspace{-.1em}\omega}}^{\hspace{-.2em}\hlf}&=
1-\frac{1}{2}\left.\frac{\partial}{\partial q^2}\sum_{i=1}^\Ncf a^{~}_{\gamma{e}\omega}(m^{~}_i;q^2)
\right|_{q_{~}^2=0}=
1-\frac{1}{2}\frac{\(\alpha\aGR\)^{1/2}}{4\pi}
\frac{4}{3}\sum_{i=1}^\Ncf\(C_{UV}-\log{\frac{m_i^2}{\mu^2_R}}\)\!,\label{ZAW}\\
{Z^{~}_{\hspace{-.1em}\E}}^{\hspace{-.2em}\hlf}&=1-\frac{1}{2}\left.\frac{\partial}{\partial q^2}
a^{~}_{\E\omega\E}(q^2)\right|_{q_{~}^2=-\mu_{I\hspace{-.2em}R}^2}=1-\frac{1}{2}\frac{\aGR}{4\pi}
\frac{1}{2}\(C_{UV}+1+\log{\frac{\mu^2_R}{\mu^{2}_{\hspace{-.1em}I\hspace{-.2em}R}}}\)\!,\label{ZEE}
\end{align}
where $m^{~}_i$ is a mass of $i$'th fermion appearing in the \QGED and $\mu^{~}_{\hspace{-.1em}I\hspace{-.2em}R}$ is an artificial energy scale to regularise an infrared divergence, which is eliminated by adding real spin connection radiation diagrams.
We align fermions such that the first $\Ncf$ charged fermions are followed by $(\Nf-\Ncf)$ neutral ones. 
\end{subequations}

\subsubsection{Electron field and mass renormalisation}
\begin{figure}[t]
 \begin{center} 
   \includegraphics[width=6.5cm]{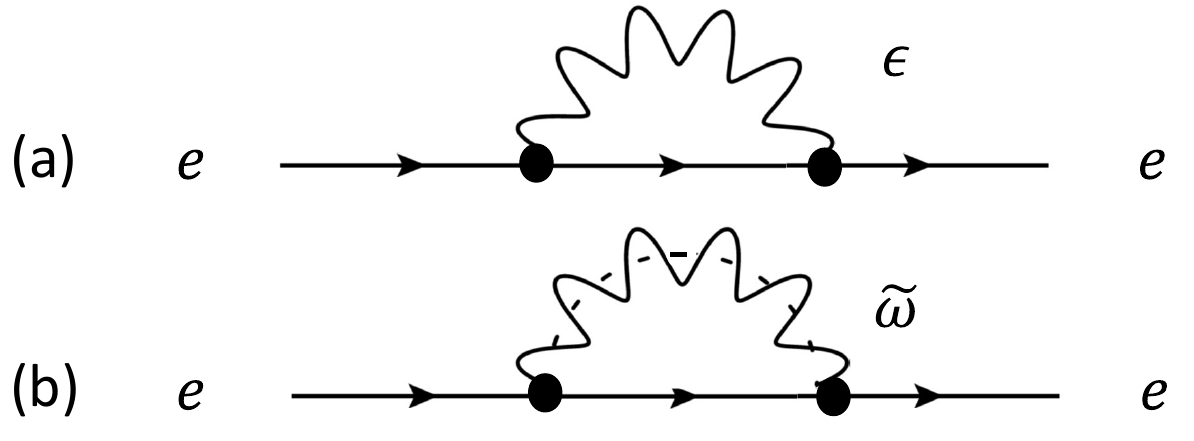}
 \caption{\label{fig4}\small
Electron self-energy diagrams
} 
 \end{center} 
\end{figure}
Electron self-energy diagrams in Figure \ref{fig4} provide electron field and mass renormalisation constants.
We denote the Feynman parameter integration of the two-point function $F_n$ as
\begin{align*}
F_n(\mu_1,\mu_2,\mu_3):=\int dx\hspace{.2em}x^n 
\log{\(\mu_1\hspace{.1em}(1-x)+\mu_2\hspace{.1em}x-\mu_3\hspace{.1em}x(1-x)\)}.
\end{align*}
A QED part given by Figure \ref{fig4}-(a) with electron momentum $p_a$ is provided as
\begin{align*}
&\text{Figure \ref{fig4}-(a)}=:\Sigma^{~}_{\Aa}(\sp)=\Sigma^{1}_\Aa(p^2)\bm{1}+
\Sigma^{\gamma}_\Aa(p^2)\!\sp,
\intertext{with}
\Sigma^{1}_\Aa(p^2)&=-\frac{\alpha}{4\pi}\hspace{.2em}4m^{~}_e
\(C_{UV}-\frac{1}{2}-F_0(m_e^2,\lambda^2,p^2)\)\rightarrow
\Sigma^{1}_\Aa(m_e^2)=
-\frac{\alpha}{4\pi}\hspace{.2em}4m^{~}_e\(C_{UV}+\frac{3}{2}-\log{\frac{m_e^2}{\mu_R^2}}\)\!,\\
\Sigma^{\gamma}_\Aa(p^2)&=\hspace{.7em}\frac{\alpha}{4\pi}
\(C_{UV}-1-2F_1(m_e^2,\lambda^2,p^2)\)\hspace{2.4em}\rightarrow
\Sigma^{\gamma}_\Aa(m_e^2)=\hspace{.8em}
\frac{\alpha}{4\pi}\(C_{UV}+2-\log{\frac{m_e^2}{\mu_R^2}}\)\!,
\end{align*}
where $\slash{\hspace{-.13em}p}:=\eta^\bcdots\gamma_\bcdot p_\bcdot$ and $\bm{1}$ is a unit spinor.
We first put the fictitious photon mass $\lambda$ to regularise an infrared divergence in the loop integration.
Then, we require the on-shell condition for the electron.
The right-most functions have no infrared divergence at $p^2_{~}=m^2_e$; thus, the fictitious photon mass is set to zero in the rightmost expressions.

Owing to these results, we can obtain a derivative of these functions as
\begin{align*}
\partial\Sigma^{1}_\Aa(m_e^2)&:=\left.\frac{\partial}{\partial\hspace{-.3em}\sp}
\Sigma^{1}_\Aa(p^2)\right|_{p^2=m_e^2}=\hspace{.8em}
\frac{\alpha}{4\pi}\frac{1}{m_e^{~}}4\(2+1\log{\frac{\lambda^2}{m_e^{2}}}\)\!,\\
\partial\Sigma^{\gamma}_\Aa(m_e^2)&:=\left.\frac{\partial}{\partial\hspace{-.3em}\sp}\Sigma^{\gamma}_\Aa(p^2)\right|_{p^2=m_e^2}=
-\frac{\alpha}{4\pi}\frac{1}{m_e^2}2\(3+1\log{\frac{\lambda^2}{m_e^{2}}}\)\!.\\
\end{align*}
Adding real-photon radiation diagrams will eliminate the fictitious photon mass from scattering amplitudes.
Owing to the above results, we obtain the QED renormalisation constants of the electron field and mass as
\begin{subequations}
\begin{align}
{Z^{~}_{\psi\Aa}}^{\hlf}&=1-\frac{1}{2}\(
\Sigma^{\gamma}_\Aa\(m_e^2\)+m_e^{~}\(m_e^{~}\hspace{.1em}\partial\Sigma^{\gamma}_\Aa\(m_e^2\)+
\partial\Sigma^{1}_\Aa\(m_e^2\)\)
\)\!,\notag\\
&=1-\frac{1}{2}\frac{\alpha}{4\pi}\(
C_{UV}+4-\log{\frac{m_e^{2}}{\mu_R^2}}+2\log{\frac{\lambda^2}{m_e^{2}}}\label{ZpsiA}
\)
\intertext{and}
\delta{m^{\Aa}_{e}}&=m_e^{~}\hspace{.1em}\Sigma^{\gamma}_\Aa\(m_e^2\)+\Sigma^{1}_\Aa\(m_e^2\)
=-\frac{\alpha}{4\pi}\hspace{.1em}3{m_e^{~}}\!\(C_{UV}+\frac{4}{3}-\log{\frac{m_e^{2}}{\mu_R^2}}\)\!.\label{dmeA}
\end{align}
\end{subequations}
Similarly, we obtain the electron self-energy in the \QGED as
\begin{align*}
\text{Figure \ref{fig4}-(b)}=:\Sigma^{~}_{\omega}(\sp)&=
-\cGR^{\hspace{-.2em}2}\(\mu_R^2\)^{\varepsilon^{~}_{\hspace{-.1em}U\hspace{-.1em}V}}\int\frac{d^Dk}{i(2\pi\hbar)^D }
\frac{\gamma_\bcdot(\sp+\slash{k}+m_e)\gamma_\bcdot}{k^2\((k-p)^2-m_e^2\)}
\(\eta^\bcdots\eta^\stars-\eta^{\star\bcdot}\eta^{\star\bcdot}\)\eta_\stars,\\
&=
\Sigma^{1}_\omega(p^2)\bm{1}+
\Sigma^{\gamma}_\omega(p^2)\sp.
\end{align*}
After similar calculations with the QED one, we obtain that
\begin{align*}
\Sigma^{1}_\omega(p^2)&=-\frac{\aGR}{4\pi}\hspace{.1em}12m^{~}_e\!\(
C_{UV}-\frac{1}{2}-F_0\(m^2_e,\lambda_\omega^2,p^2\)
\)\rightarrow
\Sigma^{1}_\omega(m_e^2)=
-\frac{\alpha}{4\pi}\hspace{.1em}12m^{~}_e\(C_{UV}+\frac{3}{2}-\log{\frac{m_e^2}{\mu_R^2}}\)\!,\\
\Sigma^{\gamma}_\omega(p^2)&=\hspace{.7em}\frac{\aGR}{4\pi}\hspace{.2em}3
\(C_{UV}-1-2F_1\(m^2_e,\lambda_\omega^2,p^2\)\)
\hspace{1.8em}\rightarrow
\Sigma^{\gamma}_\omega(m_e^2)=\hspace{.7em}
\frac{\aGR}{4\pi}\hspace{.2em}3\(C_{UV}+2-\log{\frac{m_e^2}{\mu_R^2}}\)\!,
\intertext{and}
\partial\Sigma^{1}_\omega(m_e^2)&:=
\left.\frac{\partial}{\partial\hspace{-.3em}\sp}\Sigma^{1}_\omega(p^2)\right|_{p^2=m_e^2}=
\frac{\aGR}{4\pi}\frac{1}{m_e^{~}}12\(2+1\log{\frac{\lambda_\omega^2}{m_e^{2}}}\)\!,\\
\partial\Sigma^{\gamma}_\omega(m_e^2)&:=
\left.\frac{\partial}{\partial\hspace{-.3em}\sp}\Sigma^{\gamma}_\omega(p^2)\right|_{p^2=m_e^2}=
-\frac{\aGR}{4\pi}\frac{1}{m_e^2}6\(3+1\log{\frac{\lambda_\omega^2}{m_e^{2}}}\)\!.
\end{align*}
where $\lambda_\omega$ is the fictitious spin connection mass.
Consequently, electron field and mass renormalisation constants, respectively, are provided as follows: 
\begin{subequations}
\begin{align}
{Z^{~}_{\psi\omega}}^{\hlf}&=1-\frac{1}{2}\(
\Sigma^{\gamma}_\omega\(m_e^2\)+m_e^{~}\(m_e^{~}\hspace{.1em}\partial\Sigma^{\gamma}_\omega\(m_e^2\)+
\partial\Sigma^{1}_\omega\(m_e^2\)\)
\)\!,\notag\\
&=1-\frac{1}{2}\frac{\aGR}{4\pi}\hspace{.2em}3\(
C_{UV}+4-\log{\frac{m_e^{2}}{\mu_R^2}}+2\log{\frac{\lambda_\omega^2}{m_e^{2}}}\)\label{ZpsiW}
\intertext{and}
\delta{m^{\omega}_{e}}&=m_e^{~}\hspace{.1em}\Sigma^{\gamma}_\omega\(m_e^2\)+\Sigma^{1}_\omega\(m_e^2\)
=-\frac{\aGR}{4\pi}\hspace{.1em}9m_e\!\(
C_{UV}+\frac{4}{3}-\log{\frac{m_e^{2}}{\mu_R^2}}
\)\!.\label{dmeW}
\end{align}
\end{subequations}

We have obtained renormalisation constants (\ref{ZpsiW}) and (\ref {dmeW}) by independent calculations for the electron-spin connection diagram.
In reality, they are factorised as
\begin{align}
\(\eta^\bcdots\eta^\stars-\eta^{\star\bcdot}\eta^{\star\bcdot}\)\eta_\stars=3\eta^\bcdots_{~}&\implies
\text{(\ref{ZpsiW})}=3\times\text{(\ref{ZpsiA})}~~\text{and}~~\text{(\ref {dmeW})}=3\times\text{(\ref {dmeA})}.\label{threeeta}
\end{align}
Whereas the photon polarisation vector has four components, the spin connection has 24 components.
After the convention (\ref{wconst}), it has twelve components.
A factor of three arises from the ratio of the photon's degrees of freedom to the spin connection's one.

\subsubsection{Coupling constant renormalisation}
\begin{figure}[t]
  \begin{minipage}[b]{0.45\linewidth}
    \centering
   \includegraphics[width=7.cm]{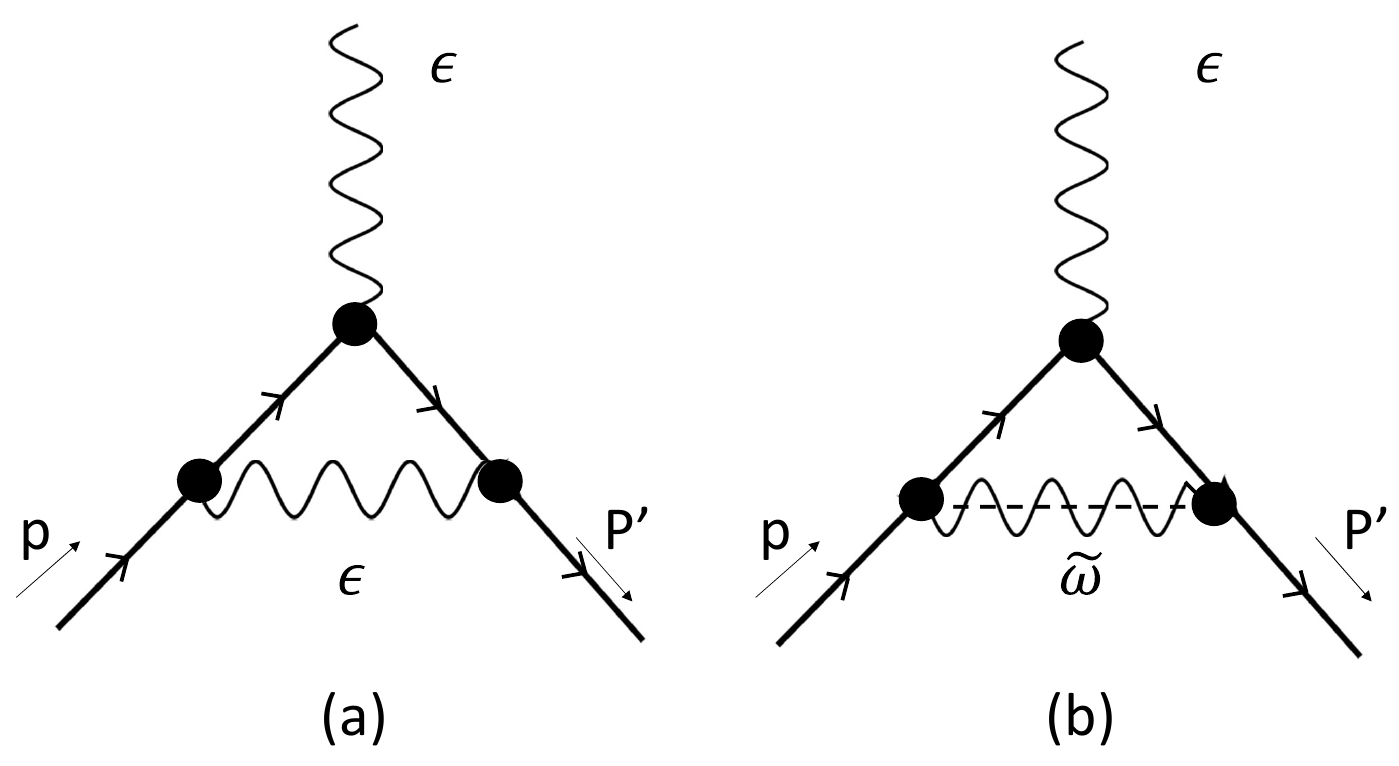}
 \caption{\label{fig7}\small
Vertex correction diagrams for the electron photon vertex
} 
  \end{minipage}
  \begin{minipage}[b]{0.45\linewidth}
    \centering
   \includegraphics[width=7.cm]{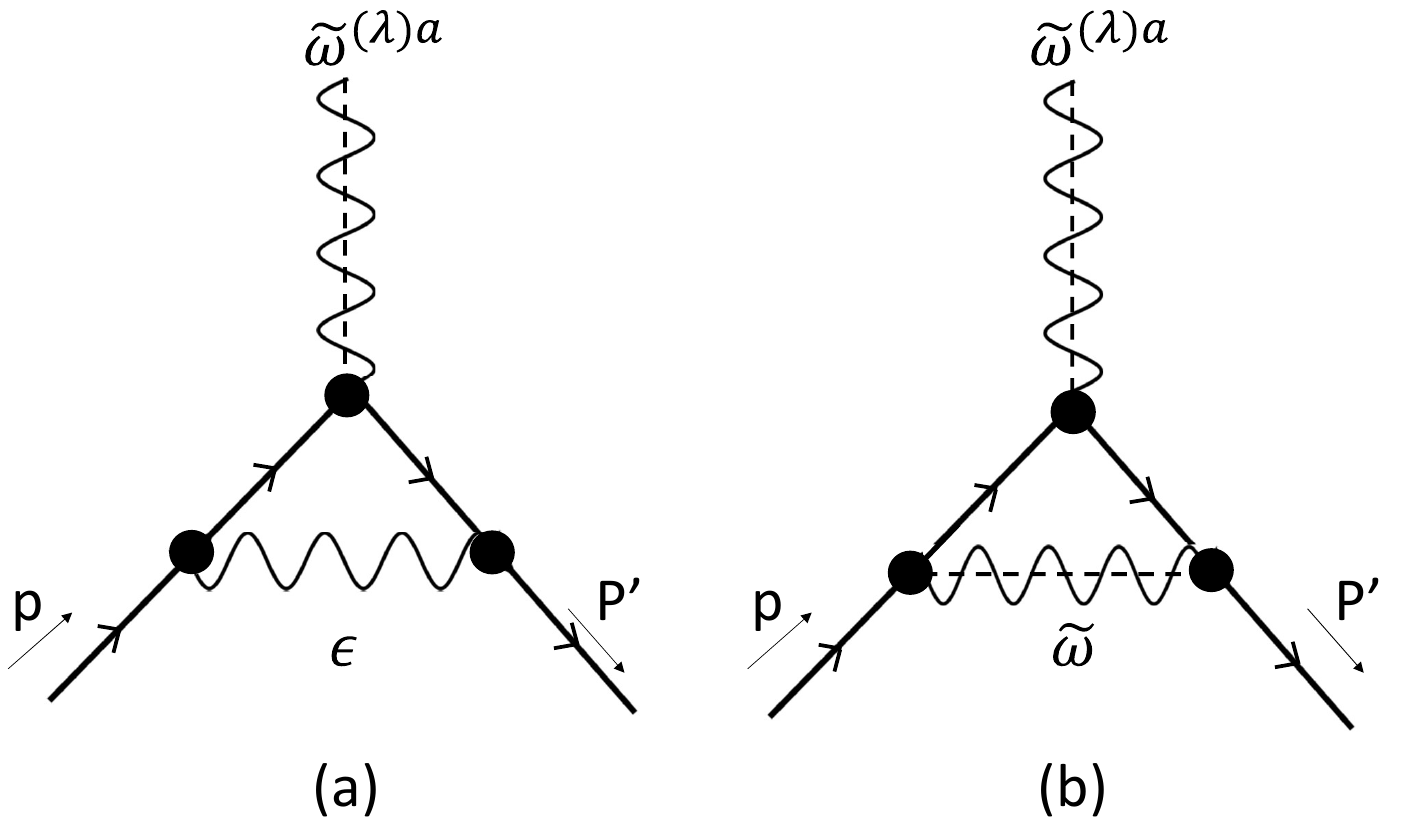}
 \caption{\label{fig8}\small
Vertex correction diagrams for the electron spin connection vertex
} 
  \end{minipage}
\end{figure}

Coupling constant renormalisation constants are provided owing to vertex diagrams in Figures \ref{fig7} and \ref{fig8} together with other renormalisation constants as
\begin{align*}
\text{Figure \ref{fig7}-(a)}=:&Z^{V}_{\hspace{-.2em}\Aa\hspace{-.2em}\Aa}
\rightarrow Y^{~}_{\hspace{-.2em}\Aa\hspace{-.2em}\Aa}=
Z^{V}_{\hspace{-.2em}\Aa\hspace{-.2em}\Aa}\hspace{.1em}
{Z^{~}_{\psi\Aa}}^{\hspace{-.2em}-1}\hspace{.1em}{Z^{~}_{\hspace{-.1em}\Aa}}^{\hspace{-.2em}-\hlf},\\
\text{Figure \ref{fig7}-(b)}=:&Z^{V}_{\hspace{-.2em}\Aa\hspace{-.1em}\omega}
\rightarrow Y^{~}_{\hspace{-.2em}\Aa\hspace{-.1em}\omega}=
Z^{V}_{\hspace{-.2em}\Aa\hspace{-.1em}\omega}\hspace{.1em}
{Z^{~}_{\psi\omega}}^{\hspace{-.2em}-1}\hspace{.1em}
{Z^{~}_{\Aa\hspace{-.1em}\omega}}^{\hspace{-.2em}-\hlf},\\
\text{Figure \ref{fig8}-(a)}=:&Z^{V}_{\hspace{-.2em}\omega\Aa}
\rightarrow Y^{~}_{\hspace{-.2em}\omega\Aa}=
Z^{V}_{\hspace{-.2em}\omega\Aa}\hspace{.1em}
{Z^{~}_{\psi\Aa}}^{\hspace{-.2em}-1}\hspace{.1em}
{Z^{~}_{\Aa\hspace{-.1em}\omega}}^{\hspace{-.2em}-\hlf},\\
\text{Figure \ref{fig8}-(b)}=:&Z^{V}_{\omega\omega}
\rightarrow Y^{~}_{\hspace{-.1em}\omega\omega}=
Z^{V}_{\omega\omega}\hspace{.1em}
{Z^{~}_{\psi\omega}}^{\hspace{-.2em}-1}\hspace{.1em}
{Z^{~}_{\hspace{-.1em}\omega}}^{\hspace{-.2em}-\hlf}.
\end{align*}
For the QED part, current conservation induces the Ward--Takahashi identity\!\cite{PhysRev.78.182,1957NCim....6..371T}, such that
\begin{subequations}
\begin{align}
Z^{V}_{\hspace{-.2em}\Aa\hspace{-.2em}\Aa}&=Z_{\psi\Aa}^{~}.\label{ZVAA}
\end{align} 
For vertices including the spin connection, the Ward--Takahashi identity (or, more precisely,  the Slavnov--Taylor identity\!\cite{TAYLOR1971436,Slavnov:1972fg} for the no-Abelian gauge theory) ensures similar relations when it is maintained.
In reality, the vertex rules for the photon (\ref{v1}) and the spin connection (\ref{v2}) have a common structure; thus, the same identity is maintained for both vertices. 
Consequently, we obtain relations:
\begin{align}
Z^{V}_{\hspace{-.2em}\Aa\hspace{-.1em}\omega}&=Z_{\psi\Aa}^{~},~
Z^{V}_{\hspace{-.2em}\omega\Aa}=Z_{\psi\omega}^{~}\label{ZVAW}
\intertext{and}
Z^{V}_{\omega\omega}&=Z_{\psi\omega}^{~}.\label{ZVWW}
\end{align} 
\end{subequations}
Even though general relativity is a non-Abelian gauge theory, Feynman rules are similar to the Abelian ones since the spin connection self-interaction is eliminated using the torsionless condition.
We discuss this point in detail in \textbf{section \ref{torsion}}.

Owing to the above relations, we immediately obtain the coupling renormalisation as
\begin{subequations}
\begin{align}
Y^{~}_{\hspace{-.2em}\Aa\hspace{-.2em}\Aa}&=
1+\frac{\alpha}{4\pi}\hspace{.2em}
\frac{2}{3}\sum_{i=1}^\Ncf\(C_{UV}-\log{\frac{m_i^2}{\mu^2_R}}\)\!,\label{YAA}\\
Y^{~}_{\hspace{-.2em}\Aa\hspace{-.1em}\omega}&=
1+\frac{\(\alpha\aGR\)^{1/2}}{4\pi}\left\{
\frac{2}{3}\sum_{i=1}^\Ncf\(C_{UV}-\log{\frac{m_e^2}{\mu^2_R}}\)
+2\(C_{UV}+4-\log{\frac{m_e^2}{\mu^2_R}}+2\log{\frac{\lambda^2}{m_e^2}}\)\right\}\!,\label{YAW}\\
Y^{~}_{\hspace{-.2em}\omega\Aa}&=1+\frac{\(\alpha\aGR\)^{1/2}}{4\pi}\left\{
\frac{2}{3}\sum_{i=1}^\Nf\(C_{UV}-\log{\frac{m_e^2}{\mu^2_R}}\)
-2\(C_{UV}+4-\log{\frac{m_e^2}{\mu^2_R}}+2\log{\frac{\lambda^2}{m_e^2}}\)\right\}\!,\label{YWA}\\
Y^{~}_{\hspace{-.1em}\omega\omega}&=
1+\frac{\aGR}{4\pi}\left\{
\frac{2}{3}\sum_{i=1}^\Nf\(C_{UV}-\log{\frac{m_e^2}{\mu^2_R}}\)
-\frac{1}{12}\(C_{UV}+\frac{2}{3}+\log{\frac{\mu^2_R}{\mu^2_{I\hspace{-.2em}R}}}\)\right\}\!.\label{YWW}
\end{align}
\end{subequations}

\paragraph{Tadpole renormalisation:}
At last, we note that the \QGED has no fundamental scalar particle; thus, the tadpole diagram does not appear in the physical processes.
\begin{table}[t]
\begin{center}
\caption{\label{tableRC}\small
Table of renormalisation constants}
\vskip 2mm
\begin{tabular}{lllc}
\multicolumn{2}{c}{renormalisation constants}&order of coupling(s)&equation number\\
\hline
\multirow{6}{*}{Firld renomalisation}
&${Z^{~}_{\hspace{-.1em}\Aa}}^{\hlf}$&$\hspace{3em}\mathcal{O}(e^2)$&(\ref{ZAA})\\
&${Z^{~}_{\omega}}^{\hlf}$&$\hspace{3em}\mathcal{O}(\cGR^{\hspace{-.3em}2})$&(\ref{ZWW})\\
&${Z^{~}_{\Aa\hspace{-.1em}\omega}}^{\hspace{-.2em}\hlf}$&$\hspace{3em}\mathcal{O}(e\hspace{.1em}\cGR)$&(\ref{ZAW})\\
&${Z^{~}_{\hspace{-.1em}\E}}^{\hspace{-.2em}\hlf}$&\hspace{3em}$\mathcal{O}(\cGR^{\hspace{-.3em}2})$&(\ref{ZEE})\\
&${Z^{~}_{\psi\Aa}}^{\hlf}$&\hspace{3em}$\mathcal{O}(\cGR^{\hspace{-.3em}2})$&(\ref{ZpsiA})\\
&${Z^{~}_{\psi\omega}}^{\hlf}$&\hspace{3em}$\mathcal{O}(\cGR^{\hspace{-.3em}2})$&(\ref{ZpsiW})\\
\hline
\multirow{2}{*}{Mass renomalisation}
&$\delta{m^{\Aa}_{e}}$&\hspace{3em}$\mathcal{O}(e^2)$&(\ref{dmeA})\\
&$\delta{m^{\omega}_{e}}$&\hspace{3em}$\mathcal{O}(\cGR^{\hspace{-.3em}2})$&(\ref{dmeW})\\
\hline
\multirow{4}{*}{Vertex renomalisation}
&$Z^{V}_{\hspace{-.2em}\Aa\hspace{-.2em}\Aa}$ &\hspace{3em}$\mathcal{O}(e^2)$&(\ref{ZVAA})\\
&$Z^{V}_{\hspace{-.2em}\Aa\hspace{-.1em}\omega}$ &$\hspace{3em}\mathcal{O}(e\hspace{.1em}\cGR)$&(\ref{ZVAW})\\
&$Z^{V}_{\hspace{-.2em}\omega\Aa}$ &$\hspace{3em}\mathcal{O}(e\hspace{.1em}\cGR)$&(\ref{ZVAW})\\
&$Z^{V}_{\omega\omega}$ &\hspace{3em}$\mathcal{O}(\cGR^{\hspace{-.3em}2})$&(\ref{ZVWW})\\
\hline
\multirow{4}{*}{Coupling constant renomalisation}
&$Y^{~}_{\hspace{-.2em}\Aa\hspace{-.2em}\Aa}$& \hspace{3em}$\mathcal{O}(e^2)$& (\ref{YAA})\\
&$Y^{~}_{\hspace{-.2em}\Aa\hspace{-.1em}\omega}$& \hspace{3em}$\mathcal{O}(e\hspace{.1em}\cGR)$& (\ref{YAW})\\
&$Y^{~}_{\hspace{-.2em}\omega\Aa}$& \hspace{3em}$\mathcal{O}(e\hspace{.1em}\cGR)$& (\ref{YWA})\\
&$Y^{~}_{\omega\omega}$&\hspace{3em}$\mathcal{O}(\cGR^{\hspace{-.3em}2})$& (\ref{YWW})\\
\hline
\end{tabular}
\end{center}
\end{table}
\subsubsection{Determination of coupling constants and electron mass}\label{DCCEM}
We summarise all renormalisation constants appearing in one-loop diagrams in the \QGED in Table \ref{tableRC}.
All divergent loop corrections are encapsulated in three physical parameters in the bare Lagrangian, $e^\bare$, $\cGR^\bare$ and $m^\bare_e$.
When we replace them with measured values, we can numerically evaluate probabilities to observe GraviElectro phenomena.
This section discusses how to measure such parameters experimentally.

\paragraph{Electron charge and mass:}
First, we determine the electromagnetic parameters of electric charge and electron mass. 
As discussed in {\bf section \ref{CC}}, a coupling constant can be absorbed in definitions of a connection and a curvature when we look at a single gauge interaction.
For the pure Yang--Mills theory given by the Lagrangian (\ref{LSUfreeint}), the coupling constant $\cSU$ can be set to unity.
In reality, we have to fix the coupling constant through the interaction Lagrangian (\ref{LMTfreeint}) as a relative value between the coupling constant and an electron mass.

Formally, an electron charge (equivalently, the fine structure constant) is fixed due to a free electron's forward-scattering of photons, i.e., the Thomson scattering.
A ratio of the fine structure constant and an electron mass appears in the Thomson scattering cross section at $\theta=0$, such that
\begin{align*}
\frac{d\sigma^{~}_{\text{Th}}}{d\cos{\theta}}&=\pi\hbar\frac{\alpha^2}{m^{2}_e}\(1+\cos^2{\theta}\)
\xrightarrow{\theta\rightarrow0} 2\pi\hbar\frac{\alpha^2}{m^{2}_e}.
\end{align*}
In reality, the fine structure constant is provided by measuring the Lamb shift of atomic spectra.
An essential relation to determining the electromagnetic parameters is
\begin{align*}
\alpha^2=\frac{R_\infty}{c}\times\frac{m^{~}_{\hspace{-.2em}A}}{m^{~}_e}\times\frac{h}{m^{~}_{\hspace{-.2em}A}},
\end{align*}
where $R_\infty$ is the Rydberg constant and $m^{~}_{\hspace{-.2em}A}$ is an atomic mass used in the experiment.
The Rydberg constant appears in an atom's electromagnetic spectra and has been measured using the optical frequency of the 2S-12D transitions in hydrogen and deuterium\!\cite{PhysRevLett.82.4960}.
This measurement does not depend on the measured value of an electron mass.
A mass ratio between electron and ionized atomic masses is provided by a ratio of the cyclotron frequency of electron and ionized atom as
\begin{align*}
\frac{m^{~}_{\hspace{-.2em}A}}{m^{~}_e}&=\frac{2}{g}\frac{q}{e}\frac{\nu^{~}_{\hspace{-.2em}A}}{\nu^{~}_e},
\end{align*}
where $g$ is a spin $g$-factor (the Land\'{e} $g$-factor) of an electron, $q$ is a charge of an ion, and $\nu^{~}_{\hspace{-.2em}A}(\nu^{~}_e)$ is an ion (electron) cyclotron frequency, respectively.
The most precise measured value of ${m^{~}_{\hspace{-.2em}A}}/{m^{~}_e}$ to date is reported in Ref$.$\!\cite{Sturm_2014}.
This relative atomic mass is a dimensionless observable and independent of the precise value of an electric charge.
We note that a ratio ${q}/{e}$ is integer-valued.
A rubidium atomic mass has been measured using the combination of Bloch oscillations with a Ramsey-Bord\'{e} interferometer\!\cite{Cadoret_2008}, which measures the inertial mass of the atom.
Consequently, the electron inertial mass is provided through the measured value of the relative electron mass.

\paragraph{Gravitational coupling constant:}
Next, we consider how we can fix the gravitational coupling constant experimentally.
For the measurements of the electron mass and charge, we discuss the inertial electron mass in the classical equation of motion and the interaction Lagrangian between an electron and a photon.
On the other hand, when we treat $\widetilde{V}_{\omega\psi}$ defined by (\ref{Vwpsi0}), we may fix the gravitational coupling constant concerning the gravitational mass of the fermion since the Born approximation of the electron scattering with $\widetilde{V}_{\omega\psi}$  provides the amplitude as
\begin{align*}
i\bm\tau&=
\langle f|i\widetilde{V}_{\omega\psi}|i\rangle\simeq
-i\(2m^{~}_e\)\bm\xi_f^\dagger\widetilde{V}_{\omega\psi}\hspace{.1em}\bm\xi^{~}_i,
\end{align*}
where $\bm\xi$ is the two-component Weyl spinor.
The normalisation of the external electron field (\ref{fnorm}) provides the mass in the above amplitude, which must be the gravitational mass.
The weak equivalence principle insists that gravitational and inertial mass are equivalent.
Experimentally, it is confirmed that a proton's gravitational and inertial masses are equivalent (proportional) to each other on the macroscopic scale.
On a microscopic scale, we can rephrase the equivalent principle as \textit{``the pole mass of an electron in the quantum field theory is equivalent to the gravitational mass in general relativity''}.
Experimental results support the microscopic equivalent principle, e.g., the proton pole mass equals its gravitational mass.

We cannot measure the gravitational mass and a coupling constant simply by using Newton's law of universal gravity, since we formulate a perturbative approach in the inertial system, where gravity is eliminated locally.
Consider an experiment measuring the gravitational coupling constant using the Earth's gravity, other than Newton's law of universal gravity.
We propose to utilise the \textit{gravimagnetic effect}\!\cite{de2010classical,pfister2015inertia}.
The spin connection couples to the electron through tensor coupling; thus, the gravitational field interacts directly with an electron spin.
This effect is measurable using existing\!\cite{Muong-2:2021ojo} and future\!\cite{10.1093/ptep/ptz030} experimental apparatus.
In reality, it gives a possible resolution of the muon $\gmt$ anomaly and $\cG$ is consistent with unity\!\cite{Kurihara:2022g-2}.
We discuss the measurement of $\cG$ in detail in \textbf{Appendix \ref{appB}}.

\subsection{Gravitational running coupling and the Landau pole}
This section discusses an energy scale dependence of the effective gravitational coupling using a gravitational beta function due to the renormalisation group equation.
In perturbative quantum field theory, this effect, known as the running coupling constant, gives a stronger coupling at higher energy scales than at lower energies for QED and vice versa for QCD.
Consequently, an effective coupling can diverge at a specific energy, the \textit{Landau pole}.
The Landau pole is the energy at which the perturbative calculations of quantum field theory break down, which does not imply a breakdown of the theory itself.
For example, non-perturbative QCD, such as lattice QCD, still works below the QCD scale parameter $\Lambda^{~}_{\hspace{-.1em}\text{Q}\hspace{-.1em}\text{C}\hspace{-.1em}\text{D}}\simeq200$MeV.
We apply the running coupling to the \QGED to discuss the effective gravitational coupling and the applicability of the perturbative \QGED to higher energies.

The $\beta$-function of the \QGED governs the renormalised gravitational coupling $\aGR\hspace{-.15em}\(\mu^{~}_{\hspace{-.15em}R}\)$ as a function of the renormalisation scale $\mu^{~}_{\hspace{-.15em}R}$ through the renormalisation group equation (see, e.g., Ref$.$\!\cite{ellis2003qcd}) such as 
\begin{align*}
\mu^{2}_{\hspace{-.15em}R}\hspace{.1em}
\frac{\partial~}{\partial\mu^{2}_{\hspace{-.15em}R}}\aGR\hspace{-.15em}\(\mu^{2}_{\hspace{-.15em}R}\)
&=\bGR\(\aGR;\mu^{2}_{\hspace{-.15em}R}\)\!.
\end{align*}
The renormalised gravitational coupling $\aGR\hspace{-.15em}\(\mu^{~}_{\hspace{-.15em}R}\)$ at a one-loop order is provided using the  renormalisation constant $Y^{~}_{\omega\omega}$ in (\ref{YWW}) as
\begin{align}
\frac{\aGR\hspace{-.15em}\(\mu^{2}_{\hspace{-.15em}R}\)}{4\pi}
=2\hspace{.1em}\delta Y^{~}_{\omega\omega},~~~\text{where}~~
\delta Y^{~}_{\omega\omega}:=Y^{~}_{\omega\omega}-1,\label{delYwwrun} 
\end{align}
yielding
\begin{align*}
\bGR\(\aGR;\mu^{2}_{\hspace{-.15em}R}\)&=
-b^{~}_{\hspace{-.1em}g\hspace{-.1em}r}
\hspace{-.15em}\(\mu^{2}_{\hspace{-.15em}R}\)
\aGR^{\hspace{-.4em}2}\hspace{-.15em}\(\mu^{2}_{\hspace{-.15em}R}\)\!,~~\text{with}~~~
b^{~}_{\hspace{-.1em}g\hspace{-.1em}r}\hspace{-.15em}\(\mu^{2}_{\hspace{-.15em}R}\):=-
\frac{8N^{~}_f\hspace{-.15em}\({\mu^{2}_{\hspace{-.15em}R}}\)-1}{24\pi}.
\end{align*}
The running gravitational coupling discussed here is caused by Figure \ref{figvacpol}, Figure \ref{figvacpol2}-(a) and  Figure \ref{figvacpol2}-(b), through (\ref{delYwwrun}).

\begin{figure}[t]
 \begin{center} 
   \includegraphics[width=12.cm]{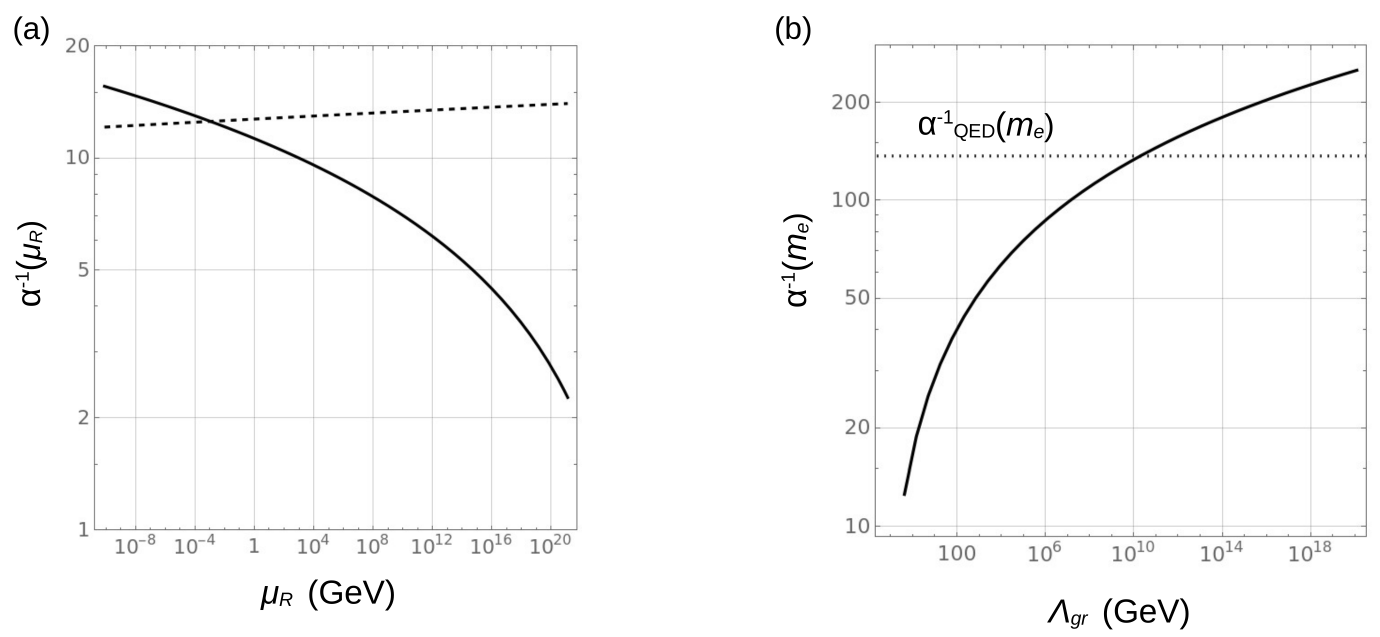}
 \caption{\label{runC}\small
The figure shows effective electric and gravitational coupling constant inverse as a function of the renormalisation scale.
The effective couplings owing to single fermion species and owing to all standard model particles are shown for the gravitational one.
} 
 \end{center} 
\end{figure}

Consequently, we obtain a solution with the boundary condition $\aGR(\LGRt)=\aGRz$ as
\begin{align*}
\aGR\hspace{-.15em}\(\mu^{2}_{\hspace{-.15em}R}\)
&=\frac{\aGRz}{1+\aGRz\hspace{.1em}b^{~}_{\hspace{-.1em}g\hspace{-.1em}r}\hspace{.1em}\log{\({\mu^{2}_{\hspace{-.15em}R}}/{\LGRt}\)}}.
\end{align*}
An energy dependence of the gravitational effective coupling is shown in Figure \ref{runC}-(a).
In the calculation, we include only an electron as a fermion.
Below the electron mass, the vierbein and ghosts contribute to the gravitational $\beta$-function.
Consequently, the effective coupling decreases from the zero-energy limit (a dashed line in the figure) to the threshold of neutrino pair creation. 
Above that energy, it increases with energy owing to the electron loop (a solid line in the figure).
We set the effective coupling constant to $\aGRz=1/4\pi$ at the electron pair-creation threshold.
For the QED case, only charged particles contribute to the running coupling. 
A larger number of particles contributes to the gravitational $\beta$-function in the QGED than in the QED; thus, the gravitational coupling runs faster than the electromagnetic one.
In reality, the renormalisation group equation for the gravitational interaction with all standard model particles has the Landau pole around $\LGR=400$ MeV, and the perturbative \QGED calculation loses its validity above this energy.
This relatively low value of $\LGR$ is due to the large coupling of $\aGRz$ compared with the $\QED$ coupling as
\begin{align*}
\alpha^{0}_\QED=\frac{1}{137.036}=0.00729735\cdots~~\text{and}~~~
\aGRz=\left.\frac{\cG^{\hspace{-.4em}2}}{4\pi}\right|_{\cG=1}=0.0795775\cdots\!.
\end{align*}
When we set the smaller values of the gravitational coupling constant at the electron pair-creation threshold, accordingly $\LGR$ increases as shown in Figure \ref{runC}-(b).

\section{Particle creation in the strong field}
This section treats particle production due to the strong electromagnetic or gravitational fields as an application of the perturbative \QGED. 
The former is the Schwinger effect, and the latter is nothing other than the Hawking radiation\!\cite{1974Natur.248...30H}.

\subsection{Schwinger effect in the QED}
The Schwinger effect, or the Sauter--Schwinger effect, is a phenomenon in which a strong electric field creates a pair of charged particles\!\cite{Sauter:1931zz,Heisenberg:1936nmg,PhysRev.82.664}, e.g., an electron-positron pair.
A strong electric field separates the created pair from each other, preventing the pair from annihilating back into the vacuum.
The Schwinger effect has not been confirmed experimentally to date.
The effective action of the Schwinger effect, $S^{~}_\QED$,  is provided as\!\cite{Heisenberg:1936nmg,Medina:2015qzc}
\begin{align*}
S^{~}_\QED&=\frac{\FF^{2}_e}{\(2\pi\hbar\)^3}
\sum_{n=1}^\infty\frac{1}{n^2}\exp{\(-n\pi\frac{\tilde{m}^{2}_e}{\FF^{~}_e}\)},\\
\intertext{with}
\FF^{~}_e&:=\frac{eE}{\hbar}~~\text{and}~~~\tilde{m}^{~}_e=\frac{m^{~}_e}{\hbar},
\end{align*}
where $E$ is a electric field strength and $\FF^{~}_e$ is an electric force on an electron.
We note on the physical dimension that $[\FF^{~}_e]_\text{pd}=L^{-2}$ and $[\tilde{m}^{~}_e]_\text{pd}=L^{-1}$ in our units, thus the argument of exponential has a null physical dimension, and the action shall be dimensionless after four-dimensional spatial integrate.

Consider the Schwinger effect owing to the Coulomb-type electric field.
The critical electric-field strength to the electron-positron pair creation owing to the Schwinger effect is  $V^{~}_c=1\times10^{18}$ V/m\!\cite{PhysRev.82.664}, yielding the critical radius ($r^{~}_{\hspace{-.1em}c}$) and the critical energy ($\E^{~}_c$) of the Coulomb potential as
\begin{align*}
V^{~}_c=\frac{1}{4\pi\epsilon^{~}_0}\frac{e}{r^2_{\hspace{-.1em}c}}\implies
r^{~}_{\hspace{-.1em}c}\simeq37.9~\text{fm}~~{\implies}~~\E^{~}_c=
\frac{\hbar}{r^{~}_{\hspace{-.1em}c}}\simeq5.19~\text{MeV},
\end{align*}
where critical radius is larger than the proton charge radius.
We note that the critical radius is smaller than the atomic scale and larger than the nuclear scale.
Elementary charged particles cannot cause the Schwinger effect due to energy-momentum conservation. 
However, this simple estimation suggests that the Schwinger effect may occur around a charged composite particle if the composite system can supply enough energy to create an electron-positron pair.

The infinite sum of vacuum-to-vacuum transition diagrams gives the effective action of the Schwinger effect in the QED\!\cite{PhysRev.82.664}.
We recalculate the vacuum-decay rate of the Coulomb electric field at a one-loop order to compare with the Hawking radiation owing to the \QGED.
We write the vacuum-to-vacuum transition amplitude in QED as
\begin{align*}
\langle0_{\text{out}}|0_{\text{in}}\rangle&=
e^{iS^{~}_{\hspace{-.1em}\text{eff}}},~~~\text{yielding}~~
P^{~}_\text{vd}:=\left|\langle0_{\text{out}}|0_{\text{in}}\rangle\right|^2=
e^{-2\hspace{.15em}\text{Im}\hspace{-.1em}\left[S^{~}_{\hspace{-.1em}\text{eff}}\right]},
\end{align*}
where $P^{~}_\text{vd}$ is the vacuum decay probability.
The effective action of the vacuum-to-vacuum transition, denoted as $S^{~}_{\hspace{-.1em}\text{eff}}$, is related to the vacuum polarisation (\ref{PIgeg})$\sim$(\ref{PIgeg3}) such that\!\cite{Dunne:2008kc}
\begin{align}
\Pi^{(\gamma{e}\gamma)}_{ab}(q^2_{i})\delta{q^2_{i}}&=\frac{\delta^2S^{~}_{\hspace{-.1em}\text{eff}}(\Aa)}
{\delta\Aa^a_{~}\hspace{.1em}\delta\Aa^b_{~}},\notag
\intertext{which has a solution}
S^{~}_{\hspace{-.1em}\text{eff}}(\Aa)&=
\(\Aa_{~}^{\bcdot}\hspace{.1em}\Pi^{(\gamma{e}\gamma)}_{\bcdots}\Aa_{~}^\bcdot\)\(q^2_{i}\)\delta{q^2_{i}},
\label{Seff}
\end{align}
where $\delta{q^2_i}$ is the possible energy spectrum due to the Schwinger effect.
We consider the Coulomb electric field that provides the electric potential in the momentum space as
\begin{align}
\Aa_{~}^a&=\(\frac{1}{\tilde{q}^{\hspace{.1em}2}},0,0,0\)\!,\label{Aa}
\end{align}
where $\tilde{q}:=|\vec{q}\hspace{.1em}|$ is a virtual photon three-momentum.
We exploit an ansatz for the virtual photon four-vector such that $\eta_\bcdots^{~}q^\bcdot_{~}q^\bcdot_{~}=q^2=\tilde{q}^2$, yielding
\begin{align}
q^a=\(\sqrt{2}\hspace{.1em}\tilde{q},0,0,\tilde{q}\)\!.\label{ansatz}
\end{align}
Consequently, we obtain the effective action as
\begin{align*}
-2\hspace{.15em}
\text{Im}\hspace{-.2em}\left[S^{~}_{\hspace{-.1em}\text{eff}}(\Aa)\right]&=
-2\times(4\pi)\sum_i\text{Im}
\left[\(\Aa_{~}^{\bcdot}\hspace{.1em}\Pi^{(\gamma{e}\gamma)}_{\bcdots}
\Aa_{~}^\bcdot\)\(q^{2}_{i}\)\right]{\delta q^2_{i}},\\
&\simeq-2\times(4\pi)\int_{4m^{2}_e}^{E^2_{\text{Max}}}\text{Im}
\left[\(\Aa_{~}^{\bcdot}\hspace{.1em}\Pi^{(\gamma{e}\gamma)}_{\bcdots}
\Aa_{~}^\bcdot\)\(q^{2}_{~}\)\right]{dq^2_{~}},\\
&=-\int_{2m^{~}_e}^{E^{~}_{\text{Max}}}\Pi_{\gamma^*\rightarrow e^+e^-}\(\tilde{q}\)d\tilde{q},
\intertext{with}
\Pi_{\gamma^*\rightarrow e^+e^-}\(\tilde{q}\)&:=4\pi\alpha
\frac{\hspace{.1em}\sqrt{\tilde{q}_{~}^2-4m^{2}_e}\(\tilde{q}^2_{~}+2m^2_e\)}{\tilde{q}_{~}^4}.
\end{align*}
\begin{figure}[t]
 \begin{center} 
   \includegraphics[width=4.5cm]{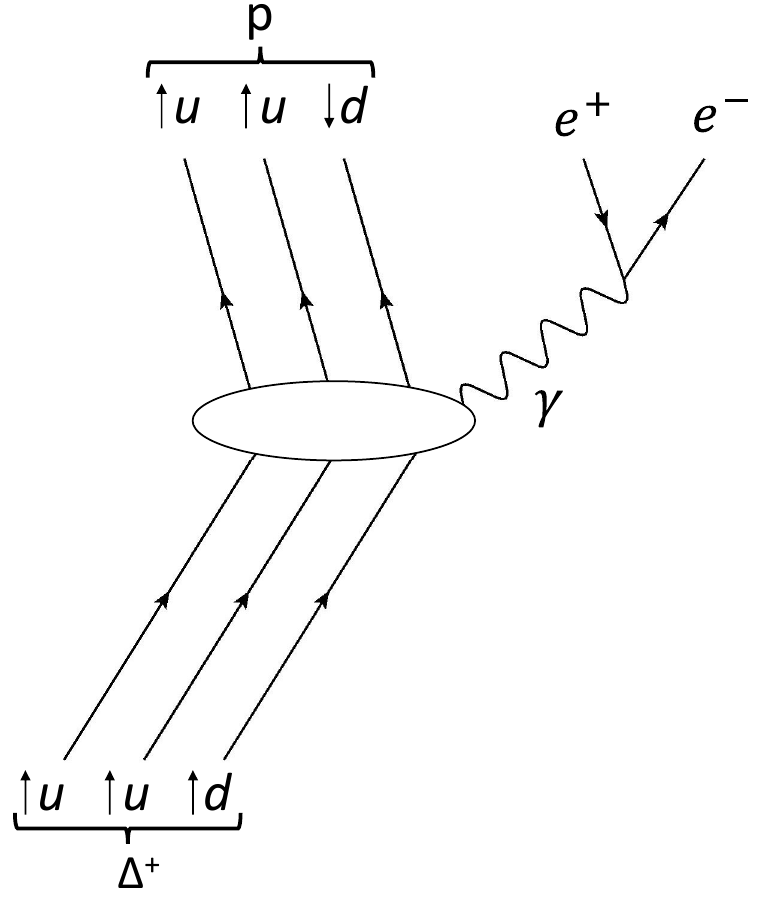}
 \caption{\label{Ddecay}\small
The figure depicts a schematic view of the $\Delta^+\rightarrow p\hspace{.2em}e^+e^-$ decay.
A photon interacts with the hadronic current rather than the individual quarks.
A spin flip can occur in any one of three valence quarks.
} 
 \end{center} 
\end{figure}

We consider the effect on the electromagnetic decay of the $\Delta^{\hspace{-.2em}+}$ baryon in Figure \ref{Ddecay}.
The critical radius is larger than proton and $\Delta^{\hspace{-.2em}+}$ baryon charge radii; thus, the Schwinger effect may occur due to the Coulomb electric field induced by a total charge of the $\Delta^{\hspace{-.2em}+}$ baryon rather than individual quarks.
We estimate the effective action with the $\Delta^+\rightarrow p\hspace{.2em}e^+e^-$ decay parameters as 
\begin{align*}
-2\hspace{.1em}\text{Im}\hspace{-.1em}\left[S_\text{eff}\hspace{-.2em}\(\Delta^+\rightarrow p\hspace{.2em}e^+e^-\)\right]
&=
\int_{2m^{~}_e}^{E^{~}_{\text{Max}}}\Pi_{\gamma^*\rightarrow e^+e^-}\hspace{-.2em}\(\tilde{q}\)d\tilde{q},\\
&=\frac{16\pi\alpha}{3}\left\{
\frac{\sqrt{E^{2}_{\text{Max}}-4m^{2}_e}\(5E^2_\text{Max}+4m^2_e\)}{6E^{3}_{\text{Max}}}
+\log{\(\frac{2m^{~}_e}{E^{~}_\text{Max}+\sqrt{E^2_\text{Max}-4m^2_e}}\)}
\right\},\\
&\simeq6.75\times10^{-1},
\end{align*}
where $E^{~}_{\text{Max}}:=m^{~}_{\hspace{-.1em}\Delta}-m^{~}_p-2m^{~}_e$.

A Feynman diagramatic derivation of the $\Delta^+$ decay is as follows:
The decay amplitude of the $\Delta^+\rightarrow p\hspace{.2em}e^+e^-$ decay is
\begin{align*}
\M\(\Delta^+\rightarrow p\hspace{.2em}e^+e^-\)&=
J_{\Delta^+\rightarrow p+\gamma^*}\(q^2_{\gamma^*}\)\frac{1}{q^2_{\gamma^*}}
J_{\gamma^*\rightarrow e^+e^-}\(q^2_{\gamma^*}\)dq^2_{\gamma^*},
\end{align*}
where $J_{\Delta^+\rightarrow p+\gamma^*}(q^2_{\gamma^*})$ and $J_{\gamma^*\rightarrow e^+e^-}(q^2_{\gamma^*})$ are hadronic and electromagnetic currents such that
\begin{align*}
J_{\Delta^+\rightarrow p+\gamma^*}&:=
e \hspace{.1em}\bar\psi^{~}_p\Gamma_\mu\psi^{~}_\Delta\hspace{.1em}\epsilon^\mu_{\gamma^*},
~~\text{and}~~~
J_{\gamma^*\rightarrow e^+e^-}:=e\hspace{.1em}\bar\psi^{~}_e\gamma_\mu\psi^{~}_e\hspace{.1em}\epsilon^\mu_{\gamma^*},
\end{align*}
respectively, where $\bar\psi^{~}_p\Gamma_\mu\psi^{~}_\Delta$ is the hadron current including an electric form-factor.
They have physical dimension of $[J]_\text{pd}=E$.
We obtain the decay width using the optical theorem and the decoupling approximation such that 
\begin{align}
\Gamma\(\Delta^+\rightarrow p\hspace{.2em}e^+e^-\)&=\frac{1}{2m^{~}_{\hspace{-.1em}\Delta}}\int
\left|\M\(\Delta^+\rightarrow p\hspace{.2em}e^+e^-\)\right|^2d\Omega,\notag \\
&\simeq\frac{1}{2m^{~}_{\hspace{-.1em}\Delta}}
\left|J_{\Delta^+\rightarrow p+\gamma^*}\hspace{.1em}dq^2_{\gamma^*}\right|^2
2\hspace{.1em}\text{Im}\hspace{-.1em}\left[S_\text{eff}\hspace{-.2em}\(\gamma^*\rightarrow e^+e^-\)\right],
\label{G2pee}
\end{align}
where $d\Omega$ is a phase space integration of the three body decay.
$q^{-2}_{\gamma^*}$ is absorbed by $\Aa^2$ (see (\ref{Aa})) in $S_\text{eff}$.
For the hadronic part, we do not have any detailed information.
We treat the hadronic current as a source of particle pair creation owing to the strong electric field; thus, we replace it simply with the critical energy of the Schwinger effect of the Coulomb electric field, such that:
\begin{align*}
\left|J_{\Delta^+\rightarrow p+\gamma^*}\hspace{.1em}dq^2_{\gamma^*}\right|^2\simeq
\E^{2}_c,
\end{align*}
due to the typical energy scale of the $e^+e^-$ pair creation; thus, we obtain
\begin{align*}
\text{(\ref{G2pee})}&\simeq
\frac{1}{2m^{~}_{\hspace{-.1em}\Delta}}\E^{2}_c
\hspace{.2em}\text{Im}\hspace{-.1em}\left[S_\text{eff}\hspace{-.2em}\(\gamma^*\rightarrow e^+e^-\)\right]
\simeq7.39\times10^{-3}~\text{MeV},
\end{align*}
In reality, a partial width of this decay branch is measured as\!\cite{10.1093/ptep/ptac097} 
\begin{align*}
\Gamma^{\text{exp}}_{~}\(\Delta^{\hspace{-.2em}+}\rightarrow
p\hspace{.2em}e^+e^-\)=\(4.91\pm0.82\)\times10^{-3}~\text{MeV}.
\end{align*}
Consequently, we obtain a numerical value consistent with the experimental measurement.
We can interpret the electromagnetic $\Delta^+$ decay as the Schwinger effect due to the Coulomb-type electric field.

\subsection{Hawking radiation in the \QGED}
The Schwinger effect is a phenomenon where a strong static electric field produces pairs of charged particles, which are then separated by the same strong electric field against the attraction between them.
On the other hand, when a strong gravitational field produces a pair of charged particles, both particles are attracted by the gravitational and electric fields, and the pair annihilates back into the vacuum.
Fortunately, exceptional circumstances can exist at the event horizon of a black hole.
The black hole's strong gravitational field produces a pair of charged particles so close to its event horizon that one of the particles in the pair falls into the hole, allowing the rest to reach the asymptotic observer without annihilation.
This phenomenon is known as \textit{Hawking radiation}.
We estimate the thermodynamic temperature of the Hawking radiation from the Schwarzschild black hole using the same method as for the Schwinger effect.

Start from the effective action (\ref{Seff}) for the spin connection vacuum polarisation, which is 
\begin{align*}
S^{~}_{\hspace{-.1em}\text{eff}}(\omega)&=
\(\tilde\omega_{~}^{\bcdot}\hspace{.1em}
\Pi^{(\omega{e}\omega)}_{\bcdots}\tilde\omega_{~}^\bcdot\)\(q^2_{i}\)\delta{q^2_{i}}\!,
\intertext{where}
\Pi^{(\omega{e}\omega)}_{ab}(q^2_{~})&=\aGR
\frac{\hspace{.1em}\sqrt{{q}_{~}^2-4m^{2}_e}\({q}^2_{~}+2m^2_e\)}{3\({q}^{2}\)^\hlf}
\(\eta^{~}_\bcdots-\frac{q^{~}_\bcdot q^{~}_\bcdot}{q^2}\)
\end{align*}
is the amplitude of Figure \ref{figvacpol}-(c) and $\tilde\omega_{~}^a$ is a polarisation vector of the spin connection in the momentum space, which for the Schwarzschild metric in the frame fixed on a hole are given by (\ref{WSE}) and (\ref{WSx}) in \textbf{Appendix \ref{appB}}.
We set the spin connection and the momentum vector to
\begin{align*}
\left.{\tilde\omega}^{\text{S}}_a(\tilde{q})\right|_{\text{lab}}&=
\(\left.{\tilde\omega}^{\text{S}}_E(\tilde{q})\right|_{\text{lab}},\left.{\tilde\omega}^{\text{S}}_x(\tilde{q})\right|_{\text{lab}},0,0\)\!,\\
q^a&=\(\sqrt{2}\hspace{.1em}\tilde{q},\tilde{q}\sin{\theta}\cos{\phi},\tilde{q}\sin{\theta}\sin{\phi},\tilde{q}\cos{\theta}\)\!,
\end{align*}
with the laboratory frame defined as in \textbf{Appendix \ref{appB}} for the black hole instead of the Earth.
We follow the same ansatz as (\ref{ansatz}) for a momentum transfer. 
Hawking radiation is only possible at the event horizon, so the momentum must be
\begin{align*}
\tilde{q}^\text{S}=\frac{\hbar}{R_\bullet^\text{S}}~~\text{with}~~~
 R_\bullet^\text{S}=\frac{\kE M^{~}_\bullet}{\hbar},
 \end{align*}
where $R_\bullet^\text{S}$ is the Schwarzschild radius and $M^{~}_\bullet$ is the mass of the hole.
So the effective action can be written as
\begin{align*}
-2\hspace{.15em}
\text{Im}\hspace{-.2em}\left[S^{~}_{\hspace{-.1em}\text{eff}}(\omega)\right]&=
-2\int_{-1}^{1}d\cos{\theta}\int_{0}^{2\pi}\hspace{-.3em}d\phi\hspace{.3em}2\hspace{.1em}\tilde{q}
\left.\text{Im}\left[\omega_{~}^{\bcdot}\hspace{.1em}\Pi^{(\omega{e}\omega)}_{\bcdots}
\omega_{~}^\bcdot\right]\right|_{\tilde{q}=\tilde{q}^\text{S}}\delta\tilde{q},
\intertext{with}
\text{Im}\left[\omega_{~}^{\bcdot}\hspace{.1em}\Pi^{(\omega{e}\omega)}_{\bcdots}
\omega_{~}^\bcdot\right]&=
{\aGR}
\frac{\hspace{.1em}\sqrt{q_{~}^2-4m^{2}_e}\(q^2_{~}+2m^2_e\)}{3\(q^2\)^{1/2}}\(
\left.{\tilde\omega}^{\text{S}}_\bcdot(\tilde{q})\right|_{\text{lab}}
\(\eta^{~}_\bcdots-\frac{q^{~}_\bcdot q^{~}_\bcdot}{q^2}\)
\left.{\tilde\omega}^{\text{S}}_\bcdot(\tilde{q})\right|_{\text{lab}}\)\!.
\end{align*}
After the angular integration with a massless approximation $m^{~}_e=0$, we obtain that
\begin{align*}
\text{Im}\hspace{-.2em}\left[S^{~}_{\hspace{-.1em}\text{eff}}(\omega)\right]&=
{\aGR}\hspace{.1em}R_\bullet^\text{S}\hspace{.2em}\frac{4+3\pi^2}{9\pi}
\delta\tilde{q}.
\end{align*}
We interpret the result thermodynamically as
\begin{align*}
&-\text{Im}\hspace{-.2em}\left[S^{~}_{\hspace{-.1em}\text{eff}}(\omega)\right]=
-\beta^{~}_\bullet\hspace{.1em}\delta\tilde{q},
\intertext{yielding}
\beta^{-1}_\bullet&=\cGR^{\hspace{-.4em}-1}
\frac{9\pi}{4+3\pi^2}T_\bullet^\text{Hawking}
\quad
\text{with}
\quad
T_\bullet^\text{Hawking}:=\frac{\hbar}{\kE M^{~}_\bullet}={R_\bullet^\text{S}}^{-1},
\end{align*}
where $T_\bullet^\text{Hawking}$ is the Hawking temperature of the Schwarzschild black hole.
{As we expected}, the \QGED provides the hole temperature proportional to the hole mass inverse.
When we set $\cGR=1$ after the coupling constant renormalisation, we obtain the correction factor ${9\pi}/{(4+3\pi^2)}\approx0.841277\cdots$ owing to the \QGED.

\section{Summary}
This report presented a renormalisable quantum theory of gravity (\QGED) based on the standard method used to quantise the Yang--Mills theory.
In our theory, objects quantised are the spin connection ($=$gauge boson) and the vierbein ($=$section) defined in the local inertial manifold, together with the electromagnetic field ($=$gauge boson) and the electron ($=$section) of the $U(1)$ gauge theory.
In classical general relativity, the solution of the Einstein equation provides the metric tensor, which determines the structure of spacetime as a (pseudo-)Riemannian manifold.
In the \QGED, an expected value of quantised vierbein fields provides the metric tensor, instead of a solution of the classical equation of motion. 
The spacetime manifold is not a quantisation target; thus, it is a smooth manifold even after quantisation.
To be brief, we do not quantise spacetime, but we quantise the ruler to measure it.
The author has proposed a space whose metric tensor is given as the expected value of a stochastic process in Ref$.$\!\cite{Kurihara_2018}.

To quantise both general relativity and the Yang--Mills theory simultaneously, the current author has reformulated both theories as the geometrical theory on an equal footing in the previous study\!\cite{Kurihara:2022sso}.
Moreover, the author has been applying the BRST quantisation non-perturbatively for pure gravity in the Heisenberg picture to them\!\cite{doi:10.1140/epjp/s13360-021-01463-3}.
This report took them further and provided a perturbative expansion of quantum gravity.
Although the quantisation method utilised for general relativity in this report is faithful to the standard BRST quantisation of the Yang--Mills theory, it is not the same as traditional methods of quantum general relativity in some aspects; we identify a spin connection, instead of the metric tensor, as a gravitational gauge boson and utilise the covariant differential, including a gravitational coupling constant.
The unconventional gravitational coupling constant provides several benefits to the theory:
\begin{enumerate}
\item We can quantise general relativity entirely parallel to the Yang--Mills theory.
\item The renormalised gravitational coupling constant absorbed the ultraviolet divergence of a scattering vertex.
\item Free gravitational fields with vanishing coupling constant provide a linearised Einstein equation, which is also parallel to the Yang--Mills theory.
\end{enumerate}
The old quantum theoretical consideration provides evidence that the spin connection is the gauge boson
\!\cite{Kurihara_2020}, and a gravitational coupling constant is natural to quantise the gauge theory\!\cite{doi:10.1140/epjp/s13360-021-01463-3}.
Then, we proved all fields are BRST-nilpotent and the Lagrangian is BRST invariant, which ensured the renormalisation with ghosts is anomaly-free.
Moreover, after defining the physical states of gravitation appropriately\!\cite{doi:10.1140/epjp/s13360-021-01463-3}, it ensures the unitarity of the physical amplitude.

We extracted a set of Feynman rules from the \QGED Lagrangian, including gauge-fixing and ghost Lagrangian.
Propagators in the momentum space for the curved spacetime are not trivial because the Fourier transformation kernel includes the metric tensor.
This report defines all Feynman rules in the local inertial frame in which the gravity is eliminated locally; thus, the transformation kernel has a flat metric, and the Fourier transformation is well-defined\!\cite{Kurihara:2022green}.
Utilising Feynman rules of the \QGED prepared here, we provided all renormalisation constants and showed that the theory is perturbatively renormalisable at a one-loop level.
In the renormalisation theory, we must replace infinite-valued bare objects with experimentally measured ones.
We showed that the gravitational coupling constant is measurable experimentally\!\cite{Kurihara:2022g-2}.

Above all, the existence of the gravitational coupling constant allows us to discuss its running effect.
The Einstein (Newtonian) gravitational constant is a fundamental constant of nature that does not change depending on the energy scale we are examining.
On the other hand, the effective coupling can depend on the energy scale of the observables through the renormalisation group equation.
Since a boson loop gives an opposite sign to a fermion loop, a boson loop reduces the effective coupling as the energy scale increases, while the fermion loop does the opposite.
In QED, only charged fermions contribute to the beta function and provide an increasing effective coupling with increasing energy scale. 
In general, the number of fermions contributing to the gravitational $\beta$-function is larger than that for the electromagnetic one, so the gravitational coupling runs faster than the electromagnetic one.

The gravitational Landau pole of the \QGED with one fermion species is above the Planck energy, and for all Standard Model particles it is around $400$ MeV.
Although the perturbative \QGED calculation loses its validity above the Landau pole, it does not imply a brake-down of the \QGED itself.
For the QCD, even though the perturbative QCD fails under the $\Lambda_\text{QCD}$, non-perturbative treatment works appropriately, i.e., the lattice QCD.
One of the primary candidates of non-perturbative quantum theories of gravity is a loop quantum gravity (see, e.g., Refs.\!\cite{rovelli_vidotto_2014,Ashtekar:2021kfp}).
Although the loop quantum gravity has common aspects with the lattice QCD, one of the main differences is that the former does not assume the existence of a smooth manifold as a background.
In contrast, the latter takes the continuum limit of the lattice space to simulate a smooth spacetime.
The loop quantum gravity keeps a size parameter finite, which protects the theory from UV divergence.
On the other hand, the lattice QCD provides finite results at the short-distance limit since QCD is asymptotically free and non-perturbative QCD is well-defined at zero distance.
The lattice approximation of the \QGED, if any, differs from both theories.
For a lattice approximation of the \QGED, we must take a short-distance limit to obtain realistic physical quantities since the \QGED theory is constructed based on the smooth manifold.
However, the \QGED is not an asymptotic-free theory; thus, it may have a divergence.
We need yet another formulation of the non-perturbative quantum theory that utilises the discrete spacetime.

This report also provided a perturbative estimation of the Hawking radiation using the \QGED, with reference to the Schwinger effect of the QED.
The Hawking radiation and the Schwinger effect are particle-pair creations owing to strong static fields; the former is due to the gravitational field, and the latter is due to the electric field.
The Schwinger effect, owing to the electric force, has yet to be observed experimentally since the critical field strength is enormous.
On the other hand, a quark pair creation owing to the strong force is an indispensable part of the hadronisation models in jet simulation programs, e.g., PYTHIA\!\cite{Sjostrand:2014zea} and HERWIG\!\cite{Bahr:2008pv}.
These hadronisation models can describe a jet structure quite well. Thus, we can state that the Schwinger effect, owing to QCD, has experimental support.
We provided a possible interpretation of the $\Delta^+\rightarrow{p}\hspace{.1em}e^+\hspace{-.1em}e^-$ decay owing to the Schwinger effect.
In contrast with the electric force, which has attractive and repulsive forces, the gravitational force is attractive only.
Consequently, though a strong electric field separates a created charged particle pair, they re-combine immediately in the gravitational field case.
One possible loophole in this problem is considering the black hole's event horizon, i.e., Hawking radiation.
The perturbative \QGED shows that the hole temperature is proportional to the hole mass for the Schwarzschild black hole, consistently with the semi-classical derivation of the hole temperature.

Finally, we stress that the proposed perturbative \QGED is an experimentally testable theory, e.g. future high-precision measurements of the anomalous magnetic moment of the muon\!\cite{10.1093/ptep/ptz030}.
Other applications, such as state-of-the-art measurements of the atomic energy spectrum, the Rydberg constant and others, are also possible test benches for the \QGED.
A search for the non-perturbative approach of the \QGED is also an essential direction of future studies.

%
%
\section*{Acknowledgements}
I would like to thank Dr Y$.$ Sugiyama, Prof$.$ J$.$ Fujimoto and Prof$.$ T$.$ Ueda for their continuous encouragement and fruitful discussions.

\numberwithin{equation}{section}
\vskip 2cm
\noindent
{\Large\bf Appendix:}\\
\vspace{-0.8cm}
\begin{appendix}
\section{Proof of nilpotent}\label{app1}
This section summarizes proof of nilpotent for the gravitational BRST transformation..\\

\noindent\fbox{{\bf Coordinate vector}}\\
The coordinate vectors are fundamental vectors on $T\MM$, which is nilpotent as
\begin{align*}
\delBRST^\GR\left[\delBRST^\GR\left[x^{\mu}\right]\right]
&=\delBRST^\GR\left[\chi^\mu\right]=0.
\end{align*}
\noindent
\fbox{{\bf Ghost field}}\\
Ghost  $\chi^\mu$ is trivially nilpotent.
For $\chi^a_{~b}:=\eta_{b\bcdot}\chi^{a\bcdot}$, we obtain
\begin{align*}
\delBRST^\GR\left[\delBRST^\GR\left[\chi^a_{~b}\right]\right]&=
\delBRST^\GR\left[\chi^a_{~c}\chi^{c}_{~b}\right]
=
\chi^a_{~c_2}\chi^{c_2}_{~~c_1}\chi^{c_1}_{~~b}-\chi^a_{~c_1}\chi^{c_1}_{~~c_2}\chi^{c_2}_{~~b}=0,
\end{align*}
due to anticommutativity of the ghost field. A tensor $\partial^{~}_{\mu}\chi^{\nu}$ is also nilpotent as
\begin{align*}
\delBRST^\GR\left[\delBRST^\GR\left[\partial^{~}_{\mu}\chi^{\nu}\right]\right]&=
-\delBRST^\GR\left[\partial^{~}_{\mu}\chi^{\rho}\partial^{~}_{\rho}\chi^{\nu}\right]=
-\partial^{~}_{\mu}\chi^{\rho_1}\partial^{~}_{\rho_1}\chi^{\rho_2}\partial^{~}_{\rho_2}\chi^{\nu}
+\partial^{~}_{\mu}\chi^{\rho_1}\partial^{~}_{\rho_1}\chi^{\rho_2}\partial^{~}_{\rho_2}\chi^{\nu}=0.
\end{align*}

\noindent
\fbox{{\bf Vierbein form}}
\begin{align*}
\delBRST^\GR\left[\delBRST^\GR\left[\eee^a\right]\right]&=
-\delBRST^\GR\left[\eee^b\chi^a_{~b}\right]=\(
\eee^{b_1}\chi^{b_2}_{~~b_1}\chi^a_{~b_2}+\eee^{b_2}\chi^a_{~b_1}\chi^{b_1}_{~~b_2}\)=0.
\end{align*}

\noindent
\fbox{{\bf Spin form}}\\
\begin{align*}
\delBRST^\GR\left[\delBRST^\GR\left[\www^{ab}\right]\right]&=
\delBRST^\GR\left[
d\chi^{ab}
-\(\www^{a}_{~\bcdot}\hspace{.1em}\chi^{\bcdot b}
+\www^{b}_{\hspace{.3em}\bcdot}\hspace{.1em}\chi^{a \bcdot}\)
\right],\\
&=\cG\hspace{.3em}d\(\chi^a_{\hspace{.3em}\bcdot}\chi^{\bcdot b}\)
-\cG\left\{
\(d\chi^a_{\hspace{.3em}\bcdot}\)\chi^{\bcdot b}+
\(d\chi^b_{\hspace{.3em}\bcdot}\)\chi^{a\bcdot}
-\cG\www^\bcdot_{\hspace{.3em}\star}
\(
\chi^{a\star}\chi_\bcdot^{\hspace{.3em}b}+
\chi^{b\star}\chi_{\hspace{.3em}\star}^a
\)
\right\},\\&=0.
\end{align*}

\noindent
\fbox{{\bf Surface form}}\\
Applying the BRST-transformation on it again, one can get
\begin{align*}
\delBRST^\GR\left[\delBRST^\GR\left[\SSS_{ab}\right]\right]&=
\hspace{.1em}\epsilon_{abc_1c_2}\delBRST^\GR\left[\chi^{c_1}_{~c_3}\eee^{c_3}\wedge\eee^{c_2}\right],\\
&=\hspace{.1em}\epsilon_{abc_1c_2}\Bigl\{
\chi^{c_1}_{~c_4}\chi^{c_4}_{~c_3}\eee^{c_3}\wedge\eee^{c_2}-
\chi^{c_1}_{~c_3}\chi^{c_3}_{~c_4}\eee^{c_4}\wedge\eee^{c_2}-
\chi^{c_1}_{~c_3}\chi^{c_2}_{~c_4}\eee^{c_3}\wedge\eee^{c_4}
\Bigr\},\\&=0,
\end{align*}
because first term is the same as the second term and the third term is symmetric with $c_1$ and $c_2$ exchange.\\

\noindent
\fbox{{\bf Volume form}}\\
The volume form is global scalar and their BRST transformation is expected to vanish, which can be confirmed as
\begin{align*}
\delBRST^\GR\left[\vvv\right]&=\frac{1}{4!}\epsilon_{\bcdott}
\delBRST^\GR\left[\eee^\bcdot\wedge\eee^\bcdot\wedge\eee^\bcdot\wedge\eee^\bcdot\right]=
\frac{1}{3!}\epsilon_{a_1\bcdot\bcdots}
\delBRST^\GR\left[\chi^{a_1}_{~~a_2}\eee^{a_2}\wedge\eee^\bcdot\wedge\eee^\bcdot\wedge\eee^\bcdot\right]=0,
\end{align*}
due to $\eee^\bcdot\wedge\eee^\bcdot\wedge\eee^\bcdot\wedge\eee^\bcdot\propto\epsilon^{\bcdott}$ and $\chi^{a_1}_{~~a_2}=0$ when $a_1=a_2.$\\

\noindent
\fbox{{\bf ghost forms}}\\
The BRST transformation of $\ccc^a$ is given by
\begin{align*}
\delBRST^\GR\left[\ccc^a\right]&=
\delBRST^\GR\left[\chi^a_{~b}~\E^b_\mu~dx^\mu\right]
=\chi^a_{~b_1}\chi^{b_1}_{~b_2}\E^{b_2}_\mu dx^\mu-
\chi^a_{~b_1}~\E^{b_2}_\mu \chi^{b_2}_{~b_1}dx^\mu
+\chi^a_{~b}~\left(\partial^{~}_{\!\mu}\chi^\nu\right)\E^b_\nu~dx^\mu
-\chi^a_{~b}~\E^b_\mu d\chi^\mu=0,
\end{align*} 

\noindent
\fbox{{\bf Other forms}}\\
Nilpotent of other forms are trivial and the proof is omitted here.\\

\vskip 2mm
\noindent
\fbox{{\bf Gravitational Lagrangian}}\\
The quantum Lagrangian must be the BRST-null.
Gauge-fixing and Fadeef--Popov Lagrangians are constructed to satisfy the BRST-null condition in section {6-2}.
Nilpotent of only the gravitational Lagrangian is given here.
The BRST transformation of the gravitational Lagrangian is provided as
\begin{align*}
\delBRST^\GR\left[\LLL_G\right]&=
\frac{1}{2}\delBRST^\GR\left[\left(
d\www^\bcdots
+\hspace{.1em}\www^\bcdot_{~\star}\wedge\www^{\star\bcdot}\right)\wedge{\SSS}_\bcdots
-\frac{\Lambda}{3!}\vvv\right].
\end{align*}
The BRST transformation for the volume form is vanished by itself. 
For the derivative term,
\begin{align*}
\delBRST^\GR\left[
d\www^\bcdots\wedge{\SSS}_\bcdots
\right]&=
\epsilon_{abc_2c_3}\chi_{~c_1}^{b}d\www^{ac_1}\!\wedge\eee^{c_2}\wedge\eee^{c_3}+
\epsilon_{abc_2c_3}\!\www^{ac_1}\wedge d\chi_{~c_1}^{b}\!\wedge\eee^{c_2}\wedge\eee^{c_3}+
\epsilon_{abc_1c_2}\chi^{c_1}_{~c_3}d\www^{ab}\!\wedge\eee^{c_3}\wedge\eee^{c_2},\\
&=2\www^{ac_1}\!\wedge d\chi^{b}_{~c_1}\wedge\SSS_{ab},
\end{align*}
where first- and third-terms are cancelled each other.
Remnant term is transformed as
\begin{align*}
\delBRST^\GR\left[\www^\bcdot_{~\star}\wedge\www^{\star\bcdot}\wedge{\SSS}_\bcdots\right]&=
\epsilon_{abc_2c_3}\chi^{c_2}_{~c_4}\www^{ac_1}\wedge\www_{c_1}^{~~b}\wedge\eee^{c_4}\wedge\eee^{c_3}+
\epsilon_{abc_3c_4}\chi^{c_2}_{~c_1}\www^{ac_1}\wedge\www_{c_2}^{~~b}\wedge\eee^{c_3}\wedge\eee^{c_4}\\&~+
\epsilon_{abc_3c_4}\chi^{b}_{~c_2}\www^{ac_1}\wedge\www_{c_1}^{~~c2}\wedge\eee^{c_3}\wedge\eee^{c_4}-2\www^{ac_1}\wedge d\chi^b_{~c_1}\wedge\SSS_{ab},\\
&=-2\www^{ac_1}\wedge d\chi^b_{~c_1}\wedge\SSS_{ab}.
\end{align*}
In the r.h.s of the first line, the second term is zero as itself, and first- and third-terms are cancelled each other.
Therefore one can confirmed $\delBRST^\GR\left[\LLL_G\right]=0$ after summing up all terms.

If we use a following remake, we can give simpler proofs for above forms.\\
\noindent
\fbox{{\bf Remark}}\\
If both of two fields, $\alpha$ and $\beta$, are nilpotent, $\alpha\beta$ is also nilpotent.\\
\noindent
{\it Proof:}\\
If a field $X$ is nilpotent, signatures of the Leibniz rule satisfy $\epsilon_{X}=-\epsilon_{\delta X}$ due to $\delBRST^\GR[\delBRST^\GR[X]]=0$ and (\ref{Leib}), where
$\epsilon_{X}$ ($\epsilon_{\delta X}$) is a signature of $X$ ($\delBRST^\GR[X]$), respectively.
Therefore
\begin{eqnarray*}
\delBRST^\GR\left[\delBRST^\GR\left[\alpha\beta\right]\right]&=&
\epsilon_{\alpha}\delBRST^\GR\left[\alpha\right]\delBRST^\GR\left[\beta\right]+
\epsilon_{\delta\alpha}\delBRST^\GR\left[a\right]\delBRST^\GR\left[\beta\right]
~=~0.
\end{eqnarray*} 

\section{Measurement of the gravitational coupling constant}\label{appB}
\begin{figure}[t]
 \begin{center} 
   \includegraphics[width=6.cm]{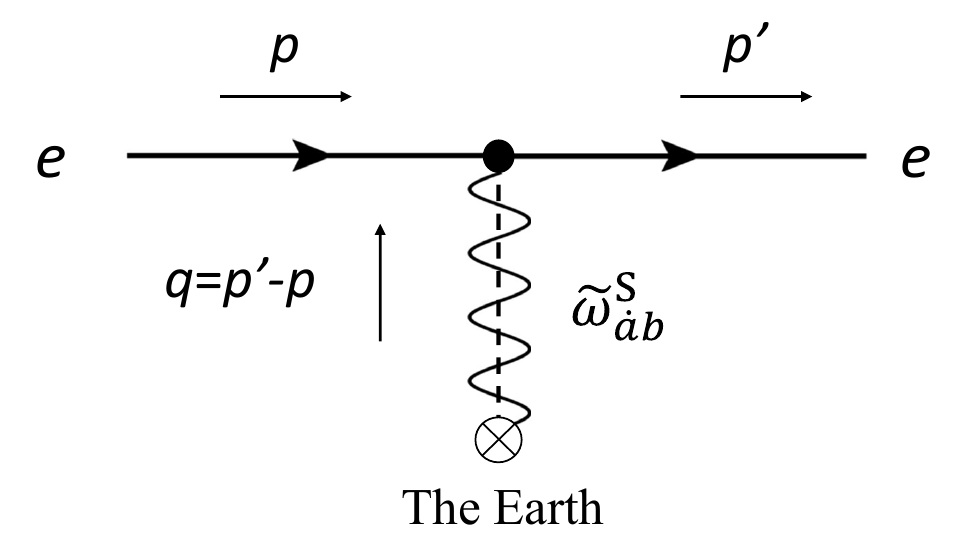}
 \caption{\label{figahc}\small
The figure depicts a Feynman diagram of a gravitational interaction of an electron with the Earth's gravity.
} 
 \end{center} 
\end{figure}

We consider an electron scattering with the Earth's gravity in Figure \ref{figahc} to measure the gravitational coupling constant.

Morishima, Futamase and Shimizu\!\cite{Morishima_2018} proposed a possible resolution to the muon  $\gmt$ anomaly owing to the static gravitational potential of the Earth using the post-Newtonian approximation.
Visser\!\cite{Visser:2018omi} immediately criticised their proposal as contradicting Einstein's equivalence principle and pointed out also that they missed counting the effect of the Sun's and the Galaxy's static potential, which are much stronger than the Earth's.
However, a frame fixed on the Earth is the inertial system concerning the Sun's and the Galaxy's gravity.
Due to Einstein's equivalence principle, the static gravitational potential does not contribute to any local observable obtained in the inertial system.
On the other hand, experimental apparatus on the Earth is not in an inertial system but in an accelerating system concerning the Earth's gravity.
A finite effect on the muon anomalous coupling owing to the Earth's gravity is measurable\!\cite{Kurihara:2022g-2}. 

\begin{figure}[t]
 \begin{center} 
   \includegraphics[width=5.0cm]{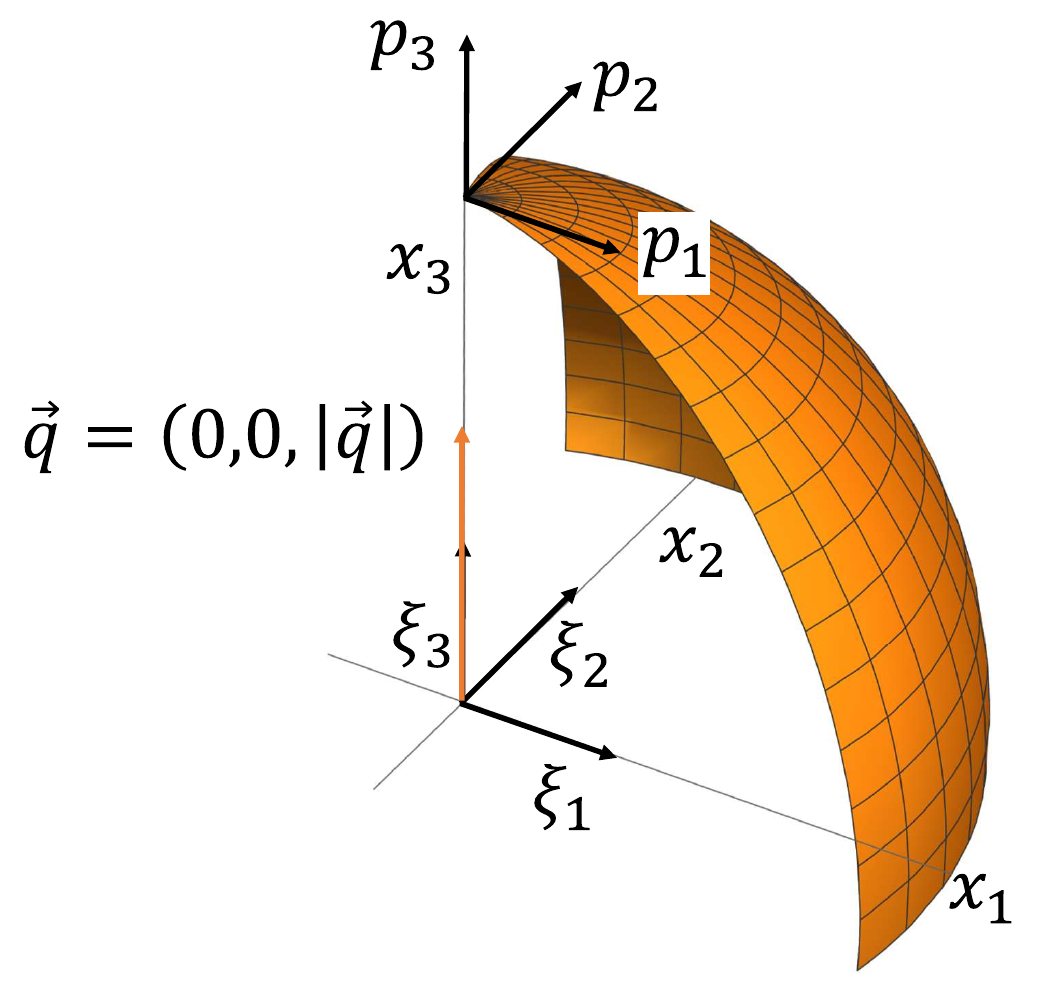}
 \caption{\label{sphare}\small
The standard coordinate frames in the local manifold.
A sphere shows the Earth's surface.
} 
 \end{center} 
\end{figure}

\paragraph{Schwarzscild spin connection:}
The Schwarzschild solution provides a gravitational field induced by the Earth. 
We first calculate a classical spin connection of the Schwarzschild solution in the inertial space.
We set the Earth at rest in the global manifold and the origin of the standard basis at the Earth's centre, as shown in Figure  \ref{sphare}.
We denote the standard basis in the global manifold as $dx^\mu=(dt,dr,d\theta,d\phi)$ in the polar coordinate.
We equate the configuration space to the inertial space and utilise the polar coordinate of $\eee^a=(d\tau,d\rho,d\vartheta,d\varphi)$ as the local standard basis.  
The local inertial manifold located at the Earth's surface at time $t=\tau=0$. 
It has a standard basis whose spatial axes are parallel to the global ones.
At first, we set the origin of the local standard basis at the Earth's centre.
After the Fourier transformation, we set the origin of the momentum coordinate at the Earth's surface. 
(See Figure \ref{sphare}.)
The Schwarzscild spin connection in the global manifold is given as\!\cite{fre2012gravity}
\begin{align*}
{\omega}^{\text{S}\hspace{.2em}tr}_{\hspace{.5em}t}&=
-{\omega}^{\text{S}\hspace{.2em}rt}_{\hspace{.5em}t}\hspace{.3em}=-\frac{{\Rs}}{2r^2},\hspace{4em}
{\omega}^{\text{S}\hspace{.2em}r\theta}_{\hspace{.5em}\theta}=
-{\omega}^{\text{S}\hspace{.2em}\theta r}_{\hspace{.5em}\vartheta}=\fsch(r),\\
{\omega}^{\text{S}\hspace{.2em}r\phi}_{\hspace{.5em}\phi}&=
-{\omega}^{\text{S}\hspace{.2em}\phi r}_{\hspace{.5em}\phi}=\fsch(\rho)\sin{\theta},\hspace{2.2em}
{\omega}^{\text{S}\hspace{.2em}\theta\phi}_{\hspace{.5em}\phi}=
-{\omega}^{\text{S}\hspace{.2em}\phi\theta}_{\hspace{.5em}\phi}=\cos{\theta},
\intertext{otherwise zero, where}
\fsch(\rho)&:=\sqrt{1-\frac{\Rs}{\rho}},\hspace{3.em}
\Rs:={2\GN M_\Earth},
\end{align*}
and $M_\Earth$ and $\Rs$ are a mass and the Schwarzschild radius of the Earth, respectively.
The spin connection in the inertial space is provided using $\omega^{\text{S}\hspace{.2em}ab}_{\hspace{.2em}c}(\xi)=\omega^{\text{S}\hspace{.2em}ab}_{\hspace{.2em}\mu}(x(\xi))\E^{\text{S}\hspace{.1em}\mu}_{\hspace{.2em}c}(x(\xi))$, such that:
\begin{align*}
{\omega}^{\text{S}\hspace{.2em}\tau\rho}_{\hspace{.5em}\tau}&=
-{\omega}^{\text{S}\hspace{.2em}\rho\tau}_{\hspace{.5em}\tau}\hspace{.3em}=-\frac{{\Rs}}{2\fsch(\rho)\rho^2},\hspace{2em}
{\omega}^{\text{S}\hspace{.2em}\rho\vartheta}_{\hspace{.5em}\vartheta}=
-{\omega}^{\text{S}\hspace{.2em}\vartheta\rho}_{\hspace{.5em}\vartheta}=\frac{\fsch(\rho)}{\rho},\\
{\omega}^{\text{S}\hspace{.2em}\rho\varphi}_{\hspace{.5em}\varphi}&=
-{\omega}^{\text{S}\hspace{.2em}\varphi\rho}_{\hspace{.5em}\varphi}=\frac{\fsch(\rho)}{\rho},\hspace{4.2em}
{\omega}^{\text{S}\hspace{.2em}\vartheta \varphi}_{\hspace{.5em}\varphi}=
-{\omega}^{\text{S}\hspace{.2em}\varphi\vartheta}_{\hspace{.5em}\varphi}=\frac{\cos\vartheta}{\rho\sin{\vartheta}},
\end{align*}
Hereafter, the superscript ``S'' on ${\omega}^\text{S}$ is omitted for simplicity.
After the three-dimensional spacial Fourier transformation, we obtain the Schwarzscild spin connection in the momentum space as\!\cite{Kurihara:2022green}
\begin{subequations}
\begin{align}
{\tilde{\omega}^{\hspace{.3em}\rho\tau}_{\tau}(\pb)}
&\simeq\frac{\pi}{2}\Rs^2\(
\frac{\pi}{\pb}-2\log{\frac{\pb}{2}}-2\gamma^{~}_E-2i\pi+{\cal O}\(\pb\)
\)\!,\label{wb}\\
\tilde{\omega}^{\hspace{.3em}\rho\vartheta}_{\vartheta}(\pb)=
{\tilde{\omega}^{\hspace{.3em}\rho\varphi}_{\varphi}(\pb)}
&\simeq
\frac{\pi}{2}\Rs^2\(
\frac{2}{\pb^2}-\frac{\pi}{\pb}
+\log{\frac{\pb}{2}}+\gamma^{~}_E-\frac{1}{2}-i\pi
+{\cal O}\(\pb\)\)\!,\label{wd}
\intertext{and otherwise zero, where}
\pb&:=\frac{\Rs}{2\hbar}|\vec{q}\hspace{.08em}|,\notag
\end{align}
\end{subequations}
and $q^a_{~}=(E,\vec{q})$ is a momentum of a spin connection.

As shown in Figure \ref{sphare}, we set a local coordinate system such that the origin is at the Earth's centre, the test particle is on the $\xi^{~}_3$-axis, and its three-momentum $\vp$ lies in a $\xi_1$-$\xi_3$ plane.
We refer to it as the laboratory frame.
We set the $\xi_1$-axis to a beam momentum $p^{~}_e$ in the laboratory frame for experiments using an electron beam\footnote{Here, ``electron'' includes other leptons used in experiments, such as a muon.}; thus,  on-shell beam momenta are
\begin{align*}
p^{~}_{\text{lab}}&=\(E^{~}_{e},p^{~}_{e},0,-\frac{q^{~}_e}{2}\) ,~~~~
p'_{\text{lab}}=\(E'_{e},p^{~}_{e},0,\frac{q^{~}_e}{2}\)\!,
\intertext{yielding $p^{2}_{\text{lab}}=p'^{2}_{\text{lab}}=m^2_e$; and the spin connection momentum is}
q^{~}_\text{lab}&=p'_{\text{lab}}-p^{~}_{\text{lab}}=\(E'_{e}-E^{~}_{e},0,0,q^{~}_e\)\!,
\end{align*}
where $q^{~}_e$ is momentum transfer between an electron and the Earth through the spin connection.
After changing the polar coordinate to the Cartesian coordinate, we obtain the Schwarzschild spin connection as
\begin{subequations}
\begin{align}
{\tilde\omega}^{a}_{~}(\pbL)&=\(
\hspace{.1em}\tilde{\omega}^{\hspace{.3em}\rho\tau}_{\tau}(\pbL),
\hspace{.1em}\pbL\hspace{.2em}\tilde{\omega}^{\hspace{.3em}\rho\vartheta}_{\vartheta}(\pbL)
,0,0\)\!,\notag
\intertext{with}
{\tilde\omega}_{~}^0(\pbL)&=\frac{\pi}{2}\(\frac{\Rs}{\hbar}\)^2
\(\frac{\pi}{\pbL}-2\log{\frac{\pbL}{2}}-2\gamma^{~}_E+{\cal O}\(\pbL\)\)\!,\label{WSE}\\
{\tilde\omega}_{~}^1(\pbL)&=\frac{\pi}{2}\(\frac{\Rs}{\hbar}\)^2
\(\frac{2}{\pbL}+\pbL\log{\frac{\pbL}{2}}-\pi+{\cal O}\(\pbL\)\)\!.\label{WSx}
\end{align}
\end{subequations}
We set them to have an inverse energy-squared dimension, such as $[{\tilde\omega}^{0}_{~}]_\text{pd}=[{\tilde\omega}^{1}_{~}]_\text{pd}=E^{-2}$, which is the same as the spin connection propagator.
The Schwarzschild spin connection has only an index type
 $\tilde{\omega}^{\hspace{.3em}\rho\bullet}_{\bullet}$ yielding a non-zero value, consistent with the observation in (\ref{wpolvec}); thus, the spin connection has a vector representation, even if it is a tensor object.

To estimate the gravitmagnetic effect quantitatively for the $\cG$ measurement, we utilise the Breit frame in which we have $p^{~}_{\text{Br}}+p'_{\text{Br}}=(2m^{~}_e,0,0,0)$ with neglecting $\OO(q^{~}_e/m^{~}_e)$, such that:
\begin{align*}
p^{~}_{\text{Br}}&=
\(m^{~}_e-\frac{\sqrt{E^{2}_e-m^2_e}}{2m^{~}_2}q^{~}_e,\hspace{.6em}\frac{E^2_e-m^2_e}{2E^{~}_em^{~}_e}q^{~}_e,0,-\frac{q^{~}_e}{2}\) ,\\
p'^{~}_{\text{Br}}&=
\(m^{~}_e+\frac{\sqrt{E^{2}_e-m^2_e}}{2m^{~}_2}q^{~}_e,-\frac{E^2_e-m^2_e}{2E^{~}_em^{~}_e}q^{~}_e,0,\hspace{.5em}\frac{q^{~}_e}{2}\) ,\\
q^{~}_\text{Br}&=
p'_{\text{Br}}-p^{~}_{\text{Br}}.
\end{align*}
We ignore effects due to the centrifugal and Coriolis forces in following calculations.
In the Breit frame, the spin connection of the Schwarzschild solution in the momentum space is
\begin{align*}
{\tilde\omega}^{0}_{~}(\pbB)&=
\gamma^{~}_e\hspace{.1em}{\tilde\omega}^{0}_{~}(\pbL)-
\gamma^{~}_e\hspace{.1em}\beta^{~}_e\hspace{.1em}{\tilde\omega}^{1}_{~}(\pbL),~~~
{\tilde\omega}^{1}_{~}(\pbB)=
\gamma^{~}_e\hspace{.1em}{\tilde\omega}^{1}_{~}(\pbL)-
\gamma^{~}_e\hspace{.1em}\beta^{~}_e\hspace{.1em}{\tilde\omega}^{0}_{~}(\pbL)
\end{align*}
after the Lorentz boost, where $\beta_e={p^{~}_e}/{E^{~}_e}$ and $\gamma_e={E^{~}_e}/{m^{~}_e}$.
In the following, we discuss the scattering amplitude in the Breit frame and omit the subscript ``Br'' for simplicity.

\paragraph{Scattering amplitude:}
We formulate the scattering amplitude by applying the Feynman rule (\ref{v2}) with setting $\varepsilon^{~}_{\hspace{-.1em}U\hspace{-.1em}V}=0$ to Figure \ref{figahc}, such that:
\begin{subequations}
\begin{align}
i{\tau}^{\text{S}}_\Earth&:=\text{Figure \ref{figahc}}
=\cG\tilde\omega^{\bcdot}_{~}(\pb)
\hspace{.1em}\bar{u}(p')\gamma^\bcdot u(p)\label{tauSE0},
\intertext{yielding}
\text{(\ref{tauSE0})}&=
\cG\hspace{.1em}\tilde\omega^{\bcdot}_{~}(\pb)\hspace{.1em}
\bar{u}(p')\(
\frac{p^\bcdot+p'^{\bcdot}}{2m^{~}_e}+
\frac{i\sigma^{\bcdot\star}\(p-p'\)^\star\eta^{~}_\stars}{2m^{~}_e}\)u(p),\notag\\
&=
\cG\hspace{.2em}
\bar{u}(p')\(
\tilde\omega^{0}_{~}(\pb)+iq^{3}
\frac{
\tilde\omega^{0}_{~}(\pb)\sigma^{03}-
\tilde\omega^{1}_{~}(\pb)\sigma^{13}}{2m^{~}_e}\)u(p).\label{tauSE1}
\end{align}
\end{subequations}
using the Gordon identity.
We exploit the Chiral representation for the gamma matrices and obtain a nonrelativistic Dirac spinor as
\begin{align*}
u(p)&=\(
\begin{array}{c}
\sqrt{\eta^{~}_{\bcdots}\hspace{.1em}p^\bcdot\sigma^\bcdot}\hspace{.2em}\bm{\xi}\\
\sqrt{\eta^{~}_{\bcdots}\hspace{.1em}p^\bcdot\bar\sigma^\bcdot}\hspace{.2em}\bm{\xi}
\end{array}
\)\!,
\intertext{where  $\bm\xi$ is a two-component spinor,}
\sigma^a&:=\(I_2,\sigma^1,\sigma^2,\sigma^3\)\!,~~~\bar\sigma^a:=\(I_2,-\sigma^1,-\sigma^2,-\sigma^3\)\!,
\end{align*}
and $I_2$ is a ($\TxT$) unit-matrix.
We expand $u(p^a_{~})$ concerning ${q^{~}_e}/{m^{~}_e}$ up to $\mathcal{O}({q^{2}_e}/{m^{2}_e})$, since $\tilde\omega^{\bullet}_{~}$ has a $\OO(q^{-1}_e)$ term. 
Two-component spinor $\bm{\xi}$ is normalised as $\sum_\lambda{\bm\xi}^\lambda\cdot{\bm\xi}^{\lambda\hspace{.1em}\dagger}=\bm{1}$ owing to (\ref{spnorm}).
Consider the physical interpretation of tensor coupling terms of the amplitude.
The first term of (\ref{tauSE1}) up to $\OO(q^{2}_e/m^{2}_e)$ is
\begin{align*}
\tilde\omega^{0}_{~}(\pb)\hspace{.1em}\bar{u}(p')u(p)&\simeq
\tilde\omega^{0}_{~}(\pb)\(2+\frac{3q^{2}_e}{4m^{2}_e}\)m^{~}_e\hspace{.1em}
\bm\xi'^\dagger\bm\xi,
\end{align*}
which does not couple to the electron spin; thus, we ignore it.
The second term can be written  as
\begin{align}
iq^{~}_e
\bar{u}(p')\frac{
\tilde\omega^{0}_{~}(\pb)\sigma^{03}-
\tilde\omega^{1}_{~}(\pb)\sigma^{13}}{2m^{~}_e}u(p)&\simeq
i\hspace{.1em}\tilde\omega^{x}_{~}(\pb)\(2+\frac{q^{2}_e}{m^{2}_e}\)
\bm\xi'^\dagger\frac{\sigma^2}{2}\bm\xi,\label{iquSu}
\end{align}
yielding high energy limit $\gamma^{~}_e\rightarrow\infty$ as
\begin{align}
\text{(\ref{iquSu})}&\simeq-\pi\(4-\pi\)
\(\bm\xi'^\dagger\frac{\sigma^2}{2}\bm\xi\)\gamma^{~2}_e\beta^{~}_e\frac{\Rs}{2\hbar}.\label{iquSu2}
\end{align}
Electron current (\ref{iquSu2})  consists of dimensional part $[\Rs/(2\hbar)]_\text{pd}=E^{-1}$ and a dimensionless factor.
The dimensional part can be interpreted as a static gravitmagnetic field owing to the Earth as
\begin{align*}
\BG&:=-\frac{\Rs}{2\hbar}=-\frac{G^{~}_{\hspace{-.1em}\text{N}}M_\Earth}{\hbar}.
\end{align*}
Lorentz factor $\gamma^{~2}_e\beta^{~}_e$ comes from the Lorentz boost from the laboratory frame to the Breit frame.
Thus, we obtain a high energy limit of the amplitude as
\begin{align}
i{\tau}^{\text{S}}_\Earth&\Rightarrow-i(2m^{~}_e)
\cG\left[
-\frac{1}{2m^{~}_e}
\bm\xi'^\dagger\frac{\sigma^2}{2}\bm\xi\right]f^{~}_\text{h.e.l.}
\BG+\OO\(\frac{q^{2}}{m^{2}_e}\)\label{tauSE}
\intertext{with}
f^{~}_\text{h.e.l.}&:=\pi\(4-\pi\)\gamma^{~2}_e\beta^{~}_e,\notag
\end{align}
where $f^{~}_\text{h.e.l.}$ is a factor under a high energy limit, and we estimate its realistic value later, owing to the existing experiment.

\paragraph{Gravimagnetic-moment:}
We define a gravimagnetic-moment vector of an electron as
\begin{align*}
\bmuG:=\frac{\cG}{m^{~}_e}\gG\vec{s}
~~~\text{with}~ ~~\vec{s}=\hspace{.2em}\bm\xi'^\dagger\frac{\vec{\sigma}}{2}\hspace{.1em}\bm\xi,
\end{align*}
where $\vec{s}$ is a spin angular-momentum vector and $\gG$ is a \textit{gravitmagnetic $g$-factor}.
We put constants $\cG$ and $\gG$ in the definition by analogy of the magnetic moment of an electron.
The Hamiltonian of an electron in a gravitational field $\bmBG$ is 
\begin{align}
\VG(r)&=-\bmuG\cdot\bmBG=-\frac{\cG}{m^{~}_e}\gG\hspace{.1em}\vec{s}\cdot\bmBG.\label{Hgr}
\end{align}

We note similarities between electromagnetic and gravitmagnetic interactions of an electron.
An electromagnetic field is not a Lorentz vector but a component of an electromagnetic tensor.
However, we can treat an electromagnetic field as a three-dimensional vector.
Similarly, a gravitational field is not a vector but a tensor field in general relativity.
However,  we can treat it as a vector field obtained using the \textit{gravitational vector potential} such as $\omega^a_{~}:=\omega_\bcdot^{\hspace{.3em}\bcdot a}$. 
In an electromagnetic field, a parallel spin electron to an electromagnetic field has a smaller potential energy than that of an anti-parallel one.
Therefore, a small perturbation makes an electron spin-flip from the anti-parallel to parallel by emitting a photon with gap energy of two states.
Similarly, an up-spin electron concerning the Earth's gravitational field has smaller potential energy than the down-spin one owing to the potential (\ref{Hgr}).
Thus, a small perturbation (a small energy transfer from the Earth) may rotate an electron spin to an upper direction in a coordinate frame fixed on the Earth's surface.

Amplitude ${\tau}^{\text{S}}_\Earth$ provides the Born approximation of the scattering of an electron with the potential due to the Earth's gravity:
 \begin{align*}
 i\bm\tau^{~}_{\text{Born}}=-i\(2m^{~}_e\)\bm\xi'^\dagger\tilde{V}(\pb)\bm\xi\xrightarrow{~\text{Figure \ref{figahc}}~}
 i{\tau}^{\text{S}}_\Earth=-i(2m^{~}_e)\bm\xi'^\dagger\tVG(\pb)\bm\xi.
\end{align*}
By comparing (\ref{tauSE}) with (\ref{Hgr}), we obtain that
\begin{align*}
\tVG(0)&=-\frac{\cG}{m^{~}_e}\gG\hspace{.1em}s^y_{~}\BG~~\text{with}~~
\gG=2,
\end{align*}
in the Breit frame.

A coefficient of an angular momentum operator $s^y=\bm\xi'^\dagger(\sigma^2/2)\hspace{.1em}\bm\xi$ gives a rotation angle of an electron spin; thus, the amplitude (\ref{tauSE}) generates a counter-clockwise spin rotation around a $y$-axis such that an initial horizontal (along a $x$-axis) spin vector to a $z$-axis (upward).
The gravitational force $\BG$ induce an orbital motion with the orbital angular frequency $\omega_O^2=|\BG|$.
On the other hand, an angular frequency of a spin rotation $\omega^{~}_S$ is provided using a relation between a rotation energy of an angular momentum and  an angular frequency, such that:
\begin{align*}
\frac{1}{2}|\bmuG|\omega^{2}_S&={\cG}\gG\left|\BG\right|\rightarrow\omega^{}_{S}=\pm\omega^{~}_{O}.
\end{align*}

In a clockwise orbital motion of an electron, it provides a $2\pi$ counter-clockwise spin rotation to a spin in the inertial system during one cycle.  
On the other hand, the electron spin point in the same direction during orbital motion in the global system, which is fixed on the Earth.
The spin rotation in the inertial system is spurious due to the observation of the rotational coordinate system.
The first-order approximation under the weak gravitational field provides $\gG\hspace{-.2em}=\hspace{-.1em}2$ to induce no spin precession.
This first-order result is consistent with a classical spin precession of a gyroscope in the weak Schwarzschild gravitational field owing to the Earth\!\cite{hoyng2007relativistic}.
It is known that the higher order correction induces a geodesic precession\!\cite{hoyng2007relativistic}.

For the finite momentum transfer case with $\pb\neq0$, the second term in (\ref{tauSE1}) proportional to $\bm\xi'^\dagger(\sigma^2/2)\hspace{.1em}\bm\xi$ provides a gravitmagnetic interaction of the Earth's gravity to an electron spin, which gives a gravitmagnetic moment slightly differ from $\gG\hspace{-.2em}=\hspace{-.1em}2$ owing to the gravitmagnetic effect.
More precisely, this term generates a counterclockwise spin rotation around a $y$-axis such that an initial horizontal (along a $x$-axis) spin vector to a negative $z$-axis (upward).
We define the anomalous gravitmagnetic moment as 
\begin{align}
\aG&:=
\frac{\gG-2}{2}=\frac{1}{2}\(\gG\(\pb\)-\gG\(0\)\)\!,\label{gg-2}
~~\text{with}~~
\gG(\pb):=-\(\frac{\cG}{m^{~}_e}\hspace{.1em}s^y_{~}\BG\)^{-1}\tVG(\pb).
\end{align}
The anomalous magnetic-moment measurements may include an effect of the above contribution.

\paragraph{Experimental measurements:}
In the electron anomalous magnetic-moment measurements\!\cite{PhysRevLett.100.120801,Fan:2022eto}, an electron has a small Lorentz factor $\gamma^{~}_e$ and $\gamma^{~}_e\beta^{~}_e$.
Moreover, these experiments utilise free-falling electrons with $\pb\simeq0$.
Thus, the spin precession owing to the gravitmagnetic effect is negligible compared with the magnetic one. 

On the other hand, the BNL--FNAL type $\gmt$ measurement used the \textit{magic momentum}\!\cite{Muong-2:2006rrc,Muong-2:2021ojo} of $p=4.094$GeV  corresponding to Lorentz factors $\gamma=29.4$ and $\beta=0.999421$ to eliminate the spin precession due to the focusing electric field; thus, we can expect a sizable precession owing to the gravitmagnetic moment with the Earth's gravity.
The electrostatic quadrupole (ESQ) covers $13/30$ of the muon storage ring\!\cite{Muong-2:2015xgu}.
Muons in the storage ring are kept horizontally due to the electric field of ESQ on average. 
Therefore, we estimate each muon receives a momentum transfer of $\pb^{~}_\Earth{\simeq}\GN M^{~}_\Earth/r^{~}_\Earth{\simeq}6.95\times10^{-10}$ on average, which induces the spin precession owing to (\ref{gg-2}), such that:
\begin{align*}
\aG/\cG&=257\hspace{.2em}858\times10^{-11}.
\intertext{On the other hand, the muon anomalous magnetic moment is estimated theoretically\!\cite{Aoyama:2020ynm} as}
\aSM&:=\frac{\gSM-2}{2}=
116\hspace{.2em}591\hspace{.2em}810\times10^{-11}.
\end{align*}
A total precesstion effect is estimated as
\begin{align*}
a^{2}_{T}&:=\aSM^2+\(\cG\hspace{.1em}\aG\)^2+
2\cG\hspace{.1em}\aSM\hspace{.1em}\aG\cos{\theta_a},
\end{align*}
where $\theta_a$ is a angle between two precession axes.
The precession owing to the anomalous grativomagnetic moment is around the $y$-axis and that owing to the anomalous magnetic moment is around the $z$-axis in the lab-frame.
Consequently, $\theta_a=\pi/2$ and the grativomagnetic contribution to the anomalous magnetic moment is
\begin{align*}
\delta\aG&:=\(\left.a^{~}_{T}-\aSM\)\right|_{\cG=1}=2.8\times10^{-9}.
\end{align*}
The measured muon anomalous magnetic moment is reported\!\cite{Muong-2:2021ojo} as
\begin{align*}
\aExp&=116\hspace{.2em}592\hspace{.2em}061\times10^{-11},
\intertext{yielding}
%
\delta{a}&:=\aExp-\aSM=\(2.51\pm0.41\pm0.43\)\times10^{-9},
\end{align*}
which suggests the gravitational coupling constant is consistent with unity.
\section{Soldering of the local and global cotangent bundles}\label{Soldering}
This appendix introduces the beneficial relation owing to the torsionless condition, namely the soldering relation based on Ref$.$\!\cite{fre2012gravity} (\textbf{section 5.4}).
In this appendix, we utilise the unscaled spin connection with omitting a \textit{hat} for simplicity.

The vierbein one-form $\eee^a=\E^a_\mu\hspace{.1em}dx^\mu\in\Omega^1(\TsMM){\otimes}V^1(TM)$ is a local tangent vector of a global one-form.
The covariant differential of the vierbein concerning both local and global bundles is defined as
\begin{align}
d_{\www\otimes\Gamma}\hspace{.2em}\eee^a&:=\(\partial_\mu\E^a_\nu+
\omega_\mu^{\hspace{.3em}ab}\hspace{.1em}\E^a_\nu+
\Gamma^\rho_{\hspace{.2em}\mu\nu}\hspace{.1em}\E^a_\rho
\)dx^\mu\!{\wedge}dx^\nu=:
\(\partial^{\www\otimes\Gamma}_\mu\E^a_\nu\)dx^\mu\!{\wedge}dx^\nu\!,
\end{align}
where $\Gamma^\rho_{\hspace{.2em}\mu\nu}$ is the Levi-Civita connection.
Here, we introduce the covariant differential in the tangent bundle, $\partial^{\www\otimes\Gamma}_\mu\!$, as above.
The torsionless equation provides its component representation as
\begin{align}
d_{\www}\hspace{.1em}\eee^a:=\TTT^a=0&\implies
\partial^{~}_\mu\E^a_\nu-\partial^{~}_\nu\E^a_\mu+
\omega_{\mu\hspace{.2em}\bcdot}^{\hspace{.3em}a}\hspace{.1em}\E^\bcdot_\nu-
\omega_{\nu\hspace{.2em}\bcdot}^{\hspace{.3em}a}\hspace{.1em}\E^\bcdot_\mu=0
\implies
\partial^{\www\otimes\Gamma}_\mu\E^a_\nu-\partial^{\www\otimes\Gamma}_\nu\E^a_\mu=0.
\label{trsless}
\end{align}
Here, we have used that the Levi-Civita connection is symmetric concerning two subscripts.
We assume spacetime is torsionless and the metric tensor fulfils the covariant constancy:
\begin{subequations}
\begin{align}
d_{\www\otimes\Gamma}(d\mathbf{s})=0&\implies
d_{\www\otimes\Gamma}\(g_{\mu\nu}\hspace{-.1em}(x)
dx^\mu{\otimes}\hspace{.1em}dx^\nu\)=
{\eta_\bcdots}\hspace{.1em}d_{\www\otimes\Gamma}\(\eee^\bcdot\!\otimes\eee^\bcdot\)=0,\notag\\
&\implies
\(\(\partial^{\www\otimes\Gamma}_\rho\E^\bcdot_\mu\)\E^\bcdot_\nu+
\E^\bcdot_\mu\(\partial^{\www\otimes\Gamma}_\rho\E^\bcdot_\nu\)\)\eta^{~}_\bcdots=0.\label{cc1}
\end{align}
By permuting three Greek indices of (\ref{cc1}), we obtain two more relations as
\begin{align}
\(\(\partial^{\www\otimes\Gamma}_\mu\E^\bcdot_\nu\)\E^\bcdot_\rho+
\E^\bcdot_\nu\(\partial^{\www\otimes\Gamma}_\mu\E^\bcdot_\rho\)\)\eta^{~}_\bcdots&=0,
\label{cc2}\\
\(\(\partial^{\www\otimes\Gamma}_\nu\E^\bcdot_\rho\)\E^\bcdot_\mu+
\E^\bcdot_\rho\(\partial^{\www\otimes\Gamma}_\nu\E^\bcdot_\nu\)\)\eta^{~}_\bcdots&=0.
\label{cc3}
\end{align}
\end{subequations}
By manipulating these three relations under the torsionless condition, we obtain
\begin{align}
\text{(\ref{cc1})}-\text{(\ref{cc2})}-\text{(\ref{cc3})}&\overset{\text{(\ref{trsless})}}{\implies}
\partial^{\www\otimes\Gamma}_\mu\E^a_\nu+
\partial^{\www\otimes\Gamma}_\nu\E^a_\mu=0\implies
\partial^{\www\otimes\Gamma}_\mu\E^a_\nu=0.\label{soldering}
\end{align}
The last equality in (\ref{soldering}) is called the soldering relation in this study.

Suppose $p$-form objects $\aaa$ in the local and global cotangent bundles, relating each other as
\begin{subequations}
\begin{align}
\aaa^{a_1{\cdots}a_p}&=\(\E^{a_1}_{\mu_1}\cdots\E^{a_p}_{\mu_p}\)\aaa^{\mu_1{\cdots}\mu_p},\label{lg}\\
\aaa^{\mu_1{\cdots}\mu_p}&=\(\E_{a_1}^{\mu_1}\cdots\E_{a_p}^{\mu_p}\)\aaa^{a_1{\cdots}a_p}.\label{gl}
\end{align}
\end{subequations}
The soldering relation ensures that the covariant differentials in the local and in the global cotangent bundle have a simple relation such that
\begin{subequations}
\begin{align}
d^{~}_\www \aaa^{a_1{\cdots}a_p}&=
\(\E^{a_1}_{\mu_1}\cdots\E^{a_p}_{\mu_p}\)d^{~}_\Gamma\aaa^{\mu_1{\cdots}\mu_p},\label{dwdG}\\
d^{~}_\Gamma \aaa^{\mu_1{\cdots}\mu_p}&=
\(\E_{a_1}^{\mu_1}\cdots\E_{a_p}^{\mu_p}\)d^{~}_\www\aaa^{a_1{\cdots}a_p}.\label{dGdw}
\end{align}
\end{subequations}
We set
\begin{align}
\bbb^{a_1{\cdots}a_p}:=d^{~}_\www \aaa^{a_1{\cdots}a_p},\label{beqda}
\end{align}
then, we obtain the covariant Laplacian in the local manifold:
\begin{align}
\text{(\ref{beqda})}\rightarrow
d^{~}_\www \bbb^{a_1{\cdots}a_p}&\overset{\text{(\ref{dwdG})}}{=}
\(\E^{a_1}_{\mu_1}\cdots\E^{a_p}_{\mu_p}\)d^{~}_\Gamma\bbb^{\mu_1{\cdots}\mu_p}\overset{\text{(\ref{gl})}}{=}
\(\E^{a_1}_{\mu_1}\cdots\E^{a_p}_{\mu_p}\)d^{~}_\Gamma\left\{
\(\E_{b_1}^{\mu_1}\cdots\E_{b_p}^{\mu_p}\)\bbb^{b_1{\cdots}b_p}\right\},\notag\\&\overset{\text{(\ref{beqda})}}{=}
\(\E^{a_1}_{\mu_1}\cdots\E^{a_p}_{\mu_p}\)d^{~}_\Gamma\left\{
\(\E_{b_1}^{\mu_1}\cdots\E_{b_p}^{\mu_p}\)d^{~}_\www \aaa^{b_1{\cdots}b_p}\right\}
,\notag\\&\overset{\text{(\ref{dwdG})}}{=}
\(\E^{a_1}_{\mu_1}\cdots\E^{a_p}_{\mu_p}\)d^{~}_\Gamma\left\{
\(\E_{b_1}^{\mu_1}\cdots\E_{b_p}^{\mu_p}\)
\(\E^{b_1}_{\nu_1}\cdots\E^{b_p}_{\nu_p}\)d^{~}_\Gamma\aaa^{\nu_1{\cdots}\nu_p}
\right\}=
\(\E^{a_1}_{\mu_1}\cdots\E^{a_p}_{\mu_p}\)d^{~}_\Gamma d^{~}_\Gamma\aaa^{\mu_1{\cdots}\mu_p},\notag\\
\implies&d^{~}_\www d^{~}_\www \aaa^{a_1{\cdots}a_p}=
\(\E^{a_1}_{\mu_1}\cdots\E^{a_p}_{\mu_p}\)d^{~}_\Gamma d^{~}_\Gamma\aaa^{\mu_1{\cdots}\mu_p},\label{LaplaceGw}
\intertext{and similarly in the global manifold,}~~
&d^{~}_\Gamma d^{~}_\Gamma \aaa^{\mu_1{\cdots}\mu_p}=
\(\E_{a_1}^{\mu_1}\cdots\E_{a_p}^{\mu_p}\)d^{~}_\www d^{~}_\www\aaa^{a_1{\cdots}a_p}\!.\label{LaplacewG}
\end{align}
The covariant Laplacian is also soldered in the local and global cotangent bundles.

\section{Dimensional analysis and renormalisability}\label{appC}
We have shown that the power counting analysis of the QEGD shown in \textbf{section 5} suggests that it may be renormalisable at any order of perturbative expansion.
On the other hand, it is known that canonical quantised general relativity with the background field method is not renormalisable over two-loop order\!\cite{GOROFF1986709}, although it is at one-loop order\!\cite{tHooft:1974toh}.
Although, at first glance, our conclusion contradicts the standard analysis, this is not the case.
This appendix explains the essential difference between the QEGD and the background field method and clarifies why the QEGD has a renormalisable expansion.

\paragraph{general relativity:}
The Einstein-Hilbert gravitational action is provided in the global manifold such that
\begin{align}
\I^{~}_{\GR}&:=\int\LLL_\GR=\frac{1}{\lpt}\int {dx^1\wedge dx^0\wedge dx^2\wedge dx^3}\sqrt{-\text{det}[\bm{g}]}\(\frac{R}{2}-\Lambda\).
\end{align}
In the following we omit the cosmological constant $\Lambda$ for simplicity.
First, we apply the dimensional analysis to the action integral.
This report uses the convention $c=1$ and $\kE\neq1\neq\hbar$ as given at the end of \textbf{section 1}; thus the physical dimension of any object can be written in terms of length ($L$) and energy ($E$). 
In this appendix we omit the subscript ``p.d.'' as $[\bullet]_\text{pd}\rightarrow[\bullet]$ for simplicity.

We start with default bases in $\TsMM$ and $\TsM$: we assign the physical dimension as
\begin{align*}
L&=[dx^\mu]=[\eee]=[\E^a_\mu dx^\mu]\implies[\E^a_\mu]=1,
\intertext{and}
L^2&=[ds^2]=[g_{\mu\nu}][dx^\mu]^2=[\eta_{ab}][\eee]^2\implies[g_{\mu\nu}]=[\eta_{ab}]=1.
\end{align*}
We can fix the physical dimension of the spin connection from the definition of the torsion form as 
\begin{align*} 
\text{\ref{torsionFM}}&\implies[\TTT^a]=[d\eee^a]+[\cG][\www^{ab}][\eee^a]\implies
[\cG]=[\www^{ab}]=1,\\
&\implies 1=[\www^{ab}]=[\omega^{\hspace{.3em}ab}_\mu][dx^\mu]\implies[\omega^{\hspace{.3em}ab}_\mu]=L^{-1}.
\end{align*}
We note that the external differential does not change the physical dimension; thus, $[d]$=1.
Thus, the structure equation provides the physical dimension of the curvature form as
\begin{align*} 
\text{\ref{RRR}}&\implies
[\RRR^{ab}]=[d\www^{ab}]+[\cG][\www^{ab}]^2\implies
[\RRR^{ab}]=1,\\&\implies
1=[\RRR^{ab}]=[{\Ri}^{ab}_{\hspace{.7em}\mu\nu}][dx^\mu]^2\implies
[{\Ri}^{ab}_{\hspace{.7em}\mu\nu}]=[R]=L^{-2}.
\end{align*}
As a consequence, we conform the action integral has null physical dimension as
\begin{align}
[\kE\hspace{.1em}\hbar]=L^2,~~[dx]=L,~~[R]=L^{-2},~~[\bm{g}]=1\implies
\left[\I^{~}_{\hspace{-.1em}\GR}\right]=1.
\end{align}
The physical dimensions of the QED particles, electron and photon, are
\begin{align}
[\psi]=L^{-3/2},~~[\Aa]=L^{-1}\implies\left[\I^{~}_{\hspace{-.1em}\QED}\right]=1.
\end{align}

The divergence degree of all \QGED vertices estimated using (\ref{rhoV}) shows they have logarithmic divergence, and only a finite number of diagrams have the UV divergence, as shown in (\ref{numOCT}).
In reality,  all UV divergences are encapsulated in a finite number of renormalisation constants at the one-loop level and can be replaced by the measured values shown in \textbf{section \ref{OLR}}.

\paragraph{background field:}
The background field method provides the metric tensor $g^{~}_{\mu\nu}$ with its perturbative fluctuation $h^{~}_{\mu\nu}$ around the background metric $\eta^{~}_{\mu\nu}$ as 
\begin{align}
g^{~}_{\mu\nu}=\eta^{~}_{\mu\nu}+\lp\hspace{.1em}h^{~}_{\mu\nu}.\label{gyh1}
\end{align}
The expansion coefficient is the  dimensional object as $[\lp]=L\implies[h^{~}_{\mu\nu}]=L^{-1}$.
Suppose a perturbative expansion by a dimensionless field is possible, like
\begin{align}
g^{~}_{\mu\nu}\overset{?}{=}\eta^{~}_{\mu\nu}+c^{~}_\textsc{bf}\hspace{.1em}h^{~}_{\mu\nu}\label{gyh2}
\end{align}
where $c^{~}_\textsc{bf}$ is assumed to be the dimensionless constant.
In this case, the canonical quantisation condition for the dimensionless object $h^{~}_{\mu\nu}$ in momentum space is given by (\ref{CRaEaE}) for the creation and annihilation operators defined in (\ref{FourierE}).
If we set the creation operator to create an energy quantum, the dimensionless fields scale with the inverse length dimension as shown in (\ref{FourierE}).
As a result, $c^{~}_\textsc{bf}$ must have a length dimension, which contradicts the assumption.

The action integral with respect to the $h^{~}_{\mu\nu}$ quantum in (\ref{gyh1}) after the perturbative expansion can have a form such that 
\begin{align}
\delta\I^{~}_{\hspace{-.1em}\textsc{bf}}(h)&=
\int\hspace{.1em}\frac{1}{2}\hspace{-.1em}\(K^{~}_{\hspace{-.1em}\textsc{bf}}-V^{~}_{\!\textsc{bf}}\)\vvv,~\text{with}~
K^{~}_{\hspace{-.1em}\textsc{bf}}=(\partial{h})^2,~~V^{~}_{\!\textsc{bf}}=\sum_{i=1}^\infty\alpha_i\(\lp{h}\)^i(\partial{h})^2.\label{Vbf}
\end{align}
The kinetic term has physical dimension of $[K^{~}_{\hspace{-.1em}\textsc{bf}}]\hspace{-.2em}=\hspace{-.2em}L^{-4}$; thus, it induces the dimensionless action as $[K^{~}_{\hspace{-.1em}\textsc{bf}}\vvv]\hspace{-.2em}=\hspace{-.2em}1$. 
 We interpret the potential function $V^{~}_{\!\textsc{bf}}$ as the interaction Lagrangian, which has the physical dimension as
\begin{align}
L^{-4}=[V^{~}_{\!\textsc{bf}}]=[\alpha_i][\lp h]^i[\partial]^2[{h}]^2&=
[\alpha_i]1^iL^{-2}L^{-2}\implies[\alpha_i]=1.
\end{align}
Thus, the interaction Lagrangian has the desired physical dimension with the dimensionless coefficient $\alpha_i$.
In reality, the perturbation of $V^{~}_{\!\textsc{bf}}$ has a dimensional coefficient as 
\begin{align}
\alpha^\textsc{bf}_i&:=\alpha_i \lp^i~\propto~L^i.
\end{align}
We can estimate the UV-divergent degree for the vertex $V^{~}_{\!\textsc{bf}}$ using (\ref{rhoV}).
The background field vertex (\ref{Vbf}) has the number of particles and derivatives as
\begin{align}
\text{(\ref{Vbf})}&\implies
N^\textsc{bf}_f=0,~~
N^\textsc{bf}_b=i+2,~~
N^\textsc{bf}_\partial=2.\label{NfNb}
\intertext{Thus, it has a divergent degree according to (\ref{rhoV}) as}
\text{(\ref{rhoV})}&\overset{\text{(\ref{NfNb})}}{\implies}
\rho\(V^{~}_{\!\textsc{bf}}\)=i+4-4>0.
\end{align}
Consequently, it has power divergence.

\paragraph{comparison:}
We have seen that \QEGD has no power divergence, but the background method does.
What is the reason for this difference?

First, the field that mediates gravity itself, namely the graviton, is different: the spin connection in \QEGD and the metric tensor (equivalently the vierbein) in the background method.
From the \YMU theoretical point of view, the force field is the curvature two-form, and the potential field to be quantised corresponds to the connection one-form.
In addition, we place the coupling constant $\cG$ at the interaction vertex, as in the Yang--Mills gauge interaction.
When the Lagrangian does not have coupling constant $\cG$ through the $SO(1,3)$ covariant differerntial, divergence owing to vertices Figure \ref{fig8} cannot be absorbed by renormalisation constant $Y^{~}_{\hspace{-.2em}\omega\Aa}$ and $Y^{~}_{\hspace{-.1em}\omega\omega}$.

In this way, the spin connection can be quantised completely in parallel with the Yang--Mills gauge-boson without any additional difficulties.
The quantisation and quantum field theory calculations are carried out in the local inertial manifold, where the metric tensor is Lorentzian and the kernel of the Fourier transform is well defined.
The role of the vierbein in the inertial manifold is to transform the global vector into the local Lorentz vector.
The vierbein transfers the spin connection field as
\begin{align*}
\text{(\ref{wwwTsM})}&\implies\E:
\omega_\mu^{\hspace{.4em}ab}\mapsto
\omega_c^{\hspace{.4em}ab}(\xi):=\omega_\mu^{\hspace{.4em}ab}(\xi)\E^\mu_c(x(\xi)),~~\xi\in \TM.
\end{align*}
We quantise the spin connection field $\omega_c^{\hspace{.4em}ab}(\xi)$ as a single function including the vierbein.
Thus the quantized spin connection is a local object defined in the inertial manifold.
In the internal manifold, the vierbein field appears only in the spin connection self-coupling in $ \LL_{\GR;int}$.
A squared field term usually gives the mass of the field, e.g. the scalar field mass term.
In reality the spin connection as a graviton has no mass.
Using the torsion equation, we convert the two-point vertex of the spin connection to the (spin connection)-(vierbein) vertex.
As a result, we have two two-point diagrams shown in Figures {\ref{figvacpol2}-(a)} and {\ref{figvacpol2}-(b)}: they have no longitudinal component after summing the two diagrams.
Another two-point diagram of the spin connection, Figure \ref{figvacpol}-(c), also has no longitudinal component.
Therefore, as expected, the spin connection has no mass.

The vierbein field is classified in the \textit{section} of the \textit{global manifold} in the \QGED and is in itself a function defined in $\TsMM$.
In the local inertial manifold, a metric tensor is Lorentzian, and the spin connection contains information about how curved the global manifold is.
Consequently, while the spin connection mediates the gravitational force in $\TsM$, the vierbein propagates globally in $\TsMM$.
The classical equation of motion for the vierbein is the torsion equation \ref{torsionFM}, which is the local $\SO(1,3)$-covariant differential of the vierbein.
This story parallels the equation of motion for an electron, the Dirac equation, which is given as the local $\SU\!(\!N\!)$-covariant differential of the spinor field.

The dimensional analysis above suggests that the intrinsic obstacle to the renormalisable perturbation is the quantisation of a dimensionless field.
In the \QGED\!, the spin connection with an inverse length dimension represents a graviton, so renormalisation has no obstacle.
Although the \QGED also has a dimensionless vierbein as a section, it appears only with the spin connection in the \QGED Lagrangian densities, (\ref{LGrfreeint}), (\ref{LMTfreeint}) and (\ref{LSUfreeint}).
Other vierbein fields are absorbed in the integration measure and contribute to preserving the global $GL(4)$ invariance.
The integration measure is the volume form defined in (\ref{vvv}) and has no dynamic degree in the local inertial manifold.
According to the absence of the vierbein self-couplings in the \QGED Lagrangian density, the theory does not have the unrenormalisable vertices.

\section{Ward--Takahashi identity for the spin connection}\label{WT}
A Ward--Takahashi identity for QED provides the relation (\ref{ZVAA}).
It is maintained from identity
\begin{align}
\frac{\partial~}{\partial p^a_{~}}\frac{-1}{\sp-m_e}=
\frac{-1}{\sp-m_e}\gamma_{~}^a\frac{-1}{\sp-m_e}&\implies
\text{Figure \ref{fig7}-(a)}\big|^{~}_{p'\rightarrow p}=:\Gamma_\Aa^a(\sp,\sp)=\frac{\partial~}{\partial p^a_{~}}
\Sigma^{~}_{\Aa}(\sp).
\end{align}
At the one loop, we represent a renormalised field as
\begin{align}
\Sigma^{R}_{\Aa}(\sp)&=\delta{Z^{~}_{\psi\Aa}}
\(\sp-m_e\)-\delta{m_e}+\Sigma^{~}_{\Aa}(\sp).\label{E2}
\end{align}
Our on-shell renormalisation \textit{\`a la} Tyoto\!\cite{10.1143/PTPS.73.1} requires the renormalisation condition
\begin{align}
\Sigma^{R}_{\Aa}(\sp)\big|^{~}_{\sp=m_e}&=
\frac{\partial~}{\partial p^a_{~}}\Sigma^{R}_{\Aa}(\sp)\big|^{~}_{\sp=m_e}=0,\label{E3}
\intertext{yielding}
\text{(\ref{E3})}\implies0=\frac{\partial~}{\partial p^a_{~}}\text{\ref{E2}}\big|^{~}_{\sp=m_e}&=
\gamma^a\delta{Z^{~}_{\psi\Aa}}+
\frac{\partial~}{\partial p^a_{~}}\Sigma^{~}_{\Aa}(\sp)\big|^{~}_{\sp=m_e}
=\gamma^a\delta{Z^{~}_{\psi\Aa}}+\Gamma_\Aa^a(\sp,\sp)\big|^{~}_{\sp=m_e}.\label{E4}
\end{align}
A renormalised QED vertex is defined as
\begin{align}
\Gamma_{\hspace{-.2em}R\Aa}^a(\sp',\sp)&=\Gamma_{\Aa}^a(\sp',\sp)+\gamma^a
\delta{Z^{V}_{\hspace{-.2em}\Aa\hspace{-.2em}\Aa}}\label{GammaR}
\end{align}
with the renormalisation condition
\begin{align}
\Gamma_{\hspace{-.2em}R\Aa}^a(\sp',\sp)\big|_{\sp=\sp'=m_e}&=0,\label{GammaR0}
\end{align}
yielding the observed electric charge at the Tomson limit.
The renormalised vertex provides
\begin{align}
\left.\begin{array}{c}
\text{(\ref{GammaR})}\\
\text{(\ref{GammaR0})}
\end{array}\right\}&\implies
0=\gamma^a\delta{Z^{V}_{\hspace{-.2em}\Aa\hspace{-.2em}\Aa}}+
\Gamma_\Aa^a(\sp,\sp)\big|_{\sp=m};\label{E5}
\intertext{thus, we obtain} 
\left.\begin{array}{c}
\text{(\ref{E4})}\\
\text{(\ref{E5})}
\end{array}\right\}
&\implies\delta{Z^{V}_{\hspace{-.2em}\Aa\hspace{-.2em}\Aa}}=\delta{Z^{~}_{\psi\Aa}}
\implies
{Z^{V}_{\hspace{-.2em}\Aa\hspace{-.2em}\Aa}}={Z^{~}_{\psi\Aa}},
\end{align}
where it is true at all orders of perturbative expansion.

Next, we consider the gravitational vertices in Figures \ref{fig7} and \ref{fig8}.
A vertex in Figure \ref{fig8}-(b) has a spin connection propagator bridging electron lines, which has the same Lorentz structure as that of a photon propagator, differing by a factor of three as shown in (\ref{threeeta}).
Thus, a vertex in Figure \ref{fig7}-(b) induces the same Ward--Takahashi identity as in \ref{fig7}-(a) owing to the above calculations.
The electron-spin connection vertex (\ref{v2}) has the same Lorentz structure as that of the electron-photon (\ref{v1}).
Thus, we can expect that spin connection vertices in Figure \ref{fig8} provide the same Ward--Takahashi identity as photon vertices. 
Here, we confirm the Ward--Takahashi identity of Figure \ref{fig8}-(b) by direct calculations.

A $\SO(1,3)$ symmetry provides  a representation of a vector current, in general, as
\begin{align}
J_{\omega}^a(p',p)&=\cG \bar{u}(p')\left[
F_1^\omega(q^2)\gamma^a+F^\omega_2(q^2)\frac{i}{2m_e}\sigma^{a\bcdot}q^\bcdot\eta_\bcdots
\right]u(p),\notag\\
&=
\cG \bar{u}(p')\left[
\(F_1^\omega\(q^2\)+F_2^\omega\(q^2\)\)\gamma^a-F^\omega_2\(q^2\)\frac{\(p+p'\)^a}{2m_e}
\right]u(p)\label{E9}
\intertext{with}
q^a&:=(p'-p)^a~\text{and}~q^2:=\eta_\bcdots\hspace{.1em}q^\bcdot q^\bcdot<0,
\end{align}
where $p$ and $p'$ are defined as an incoming momentum as in Figure \ref{fig8}-(b).
The spin connection is intrinsically a tensor field with spin-2.
However, a massless field is special according to Wigner's classification\cite{wigner1948}.
A little group of a spin-2 massless field in $\SO(1,3)$ is $\SO(2)$, having two degrees of freedom.
Consequently, a graviton has two helicity states $h=\pm2$ and has the same Lorentz structure as a photon.
Thus, a graviton emitting electron-current has a general form as in  (\ref{E9}).
We calculate an amplitude of Figure \ref{fig8}-(b) to confirm it.

The amplitude is obtained using QGED Feynman rules as
\begin{align}
\text{Figure \ref{fig8}-(b)}&\rightarrow\notag\\
\Gamma_{\omega}^a(p',p)&=\int(dl)\hspace{.2em}\bar{u}(p')\!\left[
\cG\gamma^\bcdot\frac{-1}{\sp'-\slash{l}-m_e}
\cG\gamma^a\frac{-1}{\sp-\slash{l}-m_e}\cG\gamma^\bcdot\right]\!u(p)\hspace{.2em}
\frac{-1}{l^2-\lambda^2_\omega}\(\eta_\bcdots\eta_\stars-\eta_{\star\bcdot}\eta_{\star\bcdot}\)\eta^\stars,\label{E11}
\intertext{where an integration measure of the loop momentum $l$ is given from (\ref{intmeas}) as}
(dl)&:=\(\mu_R^2\)^{\varepsilon^{~}_{\hspace{-.1em}U\hspace{-.1em}V}}\frac{d^Dl}{i(2\pi\hbar)^D }~~\text{with}~~
D:=4-2\epsilon^{~}_{\varepsilon^{~}_{\hspace{-.1em}U\hspace{-.1em}V}}.\notag
\end{align}
Standard calculations provide that
\begin{align}
\text{(\ref{E11})}=&-\frac{\aGR}{4\pi}3\left\{\(C_{UV}-1-\log{\frac{m^2_e}{\mu^{2}_R}}+
\(\frac{11}{3}+2\log{\frac{\lambda^2_\omega}{m^2_e}}\)\frac{p{\cdot}p'}{m^2_e}
+\frac{10}{3}
\)\gamma^a
-2\times\frac{(p+p')^a}{2m_e}
\right\},\notag\\
\xrightarrow{p=p'\!,~p^2=m^2_e}&
-\frac{\aGR}{4\pi}\hspace{.2em}3\left\{\(
C_{UV}+6-\log{\frac{m_e^{2}}{\mu_R^2}}+2\log{\frac{\lambda_\omega^2}{m_e^{2}}}\)\gamma^a
-2\times\frac{(p+p)^a}{2m_e}\right\}.\label{E12}
\end{align}
By comparing (\ref{E12}) with (\ref{E9}), we obtain
\begin{align}
F_1^\omega(0)+F_2^\omega(0)=-\frac{\aGR}{4\pi}\hspace{.2em}3\(
C_{UV}+6-\log{\frac{m_e^{2}}{\mu_R^2}}+2\log{\frac{\lambda_\omega^2}{m_e^{2}}}\)
,~~F_2^\omega(0)=-\frac{\aGR}{4\pi}\hspace{.2em}3\times2.
\end{align}
Thus, we obtain that
\begin{align}
\delta{Z^{V}_{\omega\omega}}&=F_1^\omega(0)=
2\(\text{(\ref{ZpsiW})}-1\)=\delta{Z_{\psi\omega}^{~}}.
\end{align}
As a result, we confirm a Lorentz structure of the graviton emitting electron-current and its Ward--Takahashi identity. 
\end{appendix}
\bibliographystyle{bmc-mathphys} 
\bibliography{QGED}
\end{document}